\documentclass[letterpage]{aa}
\usepackage{amssymb,graphicx,times,multirow,txfonts,lscape}

\begin{document}
           
\title{Quantitative Spectroscopy of BA-type Supergiants
\thanks{Based on observations collected at the European Southern
Ob\-servatory, Chile (ESO N$^{\circ}$ 62.H-0176).
Table~\ref{taba1} is available in electronic form only.}}

\author{N.~Przybilla\inst{1} \and K.~Butler\inst{2} \and
S.R.~Becker\inst{2,}
\and R.P.~Kudritzki\inst{3}}

\offprints{N.~Przybilla (przybilla@sternwarte.uni-erlangen.de)}
\institute{ 
Dr. Remeis-Sternwarte Bamberg, Sternwartstr. 7, D-96049 Bamberg, Germany
\and
Universit\"ats-Sternwarte  M\"unchen, 
 Scheinerstra\ss e 1, D-81679 M\"unchen, Germany
\and
Institute for
Astronomy, University of Hawaii, 2680 Woodlawn Drive, Honolulu, HI 96822}

\date{Received / Accepted}

\abstract{Luminous BA-type supergiants have enormous potential for modern
astrophysics. They allow topics ranging from non-LTE physics
and the evolution of massive stars to the chemical evolution of galaxies and
cosmology to be addressed. 
A hybrid non-LTE technique for the quantitative spectroscopy
of these stars is discussed. Thorough tests and first
applications of the spectrum synthesis method are presented for the bright Galactic
objects $\eta$\,Leo (A0\,Ib), HD\,111613 (A2\,Iabe), HD\,92207 (A0\,Iae) and
$\beta$\,Ori (B8 Iae), based on high-resolution and high-S/N Echelle spectra.
Stellar parameters are derived from spectroscopic indicators, consistently from multiple
non-LTE ionization equilibria and Stark-broadened hydrogen line profiles,
and they are verified by spectrophotometry. The internal accuracy of the method allows 
the 1$\sigma$-uncertainties to be reduced to $\lesssim$1--2\% in $T_{\rm eff}$ and 
to 0.05--0.10\,dex in $\log g$. Elemental abundances are determined for over
20 chemical species, with many of the astrophysically most 
interesting in non-LTE (H, He, C, N, O, Mg, S, Ti, Fe). The non-LTE computations
reduce random errors and remove systematic trends in the analysis. 
Inappropriate LTE analyses tend to systematically underestimate iron group abundances and 
overestimate the light and $\alpha$-process element abundances by up to
factors of two to three on the mean. This is because of the different
responses of these
species to radiative and collisional processes in the microscopic picture, which 
is explained by fundamental differences of their detailed atomic
structure, and not taken into account in LTE.
Contrary to common assumptions, significant non-LTE abundance corrections of
$\sim$0.3\,dex can be found even for the weakest lines
($W_{\lambda}$\,$\lesssim$\,10\,m{\AA}). Non-LTE abundance uncertainties 
amount to typically 0.05--0.10\,dex (random) and $\sim$0.10\,dex 
(systematic 1$\sigma$-errors). Near-solar abundances are derived for the
heavier elements in the sample stars, and patterns indicative of mixing with nuclear-processed
matter for the light elements. These imply a blue-loop scenario for
$\eta$\,Leo because of first dredge-up abundance ratios, while the other three objects 
appear to have evolved directly from the main sequence. In the most
ambitious computations several ten-thousand spectral lines 
are accounted for in the spectrum synthesis, permitting the accurate 
reproduction of the entire observed spectra from the visual to near-IR.
This prerequisite for the quantitative interpretation of  
intermediate-resolution spectra opens up BA-type supergiants as
versatile tools for extragalactic stellar~astronomy beyond the Local Group. 
The technique presented here is also well suited to improve quantitative analyses of less
extreme stars of similar spectral types.
}
\maketitle


\section{Introduction}\label{intro}
Massive supergiants of late B and early A-type (BA-type supergiants, BA-SGs) are among 
the visually brightest normal stars in spiral and irregular galaxies. At absolute
visual magnitudes up to $M_V$\,$\simeq$\,$-$9.5 they can rival with globular
clusters and even dwarf spheroidal galaxies in {\em integrated} light. 
This makes them primary candidates for quantitative spectroscopy at 
large distances and makes them the centre of interest in the era of 
extragalactic stellar astronomy. The present generation of 8--10m telescopes 
and efficient multi-object spectrographs can potentially observe
individual stars in systems out to distances of the Virgo and Fornax cluster 
of galaxies (Kudritzki~\cite{Kudritzki98}, \cite{Kudritzki00}). The
first steps far beyond the Local Group have already been taken (Bresolin et
al.~\cite{Bresolinetal01}, \cite{Bresolinetal02}; Przybilla~\cite{Przybilla02}).

BA-type supergiants pose a considerable challenge for quantitative
spectroscopy because of their complex atmospheric physics. The large energy and
momentum density of the radiation field, in combination with an
extended and tenuous atmosphere, gives rise to departures from
local thermodynamic equilibrium (non-LTE),
and to stellar winds. Naturally, this makes BA-SGs interesting
in terms of stellar atmosphere modelling and non-LTE physics, but there is more to gain from
their study. They can be used as tracers for elemental abundances, as their line
spectra exhibit a wide variety of chemical species, ranging from
the light elements to $\alpha$-process, iron group and s-process elements.
These include, but also extend the species traced by \ion{H}{ii}-regions
and thus they can be used to investigate abundance patterns and gradients in
other galaxies to a far greater extent than from the study of gaseous
nebulae alone. 
In fact, stellar indicators turn out
to be highly useful for independently constraining (Urbaneja et
al.~\cite{Urbanejaetal05}) the recently identified systematic error budget
of strong-line analyses of extragalactic metal-rich \ion{H}{ii} regions 
(Kennicutt et al.~\cite{Kennicuttetal03}; Garnett et
al.~\cite{Garnettetal04}; Bresolin et al.~\cite{Bresolinetal04}),
and its impact on models of galactochemical evolution.
BA-type supergiants in other galaxies allow the
metallicity-dependence of stellar winds and stellar evolution to be studied.
In particular, fundamental stellar parameters and light element abundances 
(He, CNO) help to test the most recent generation of evolution models 
of rotating stars with mass loss 
(Heger \& Langer~\cite{HeLa00}; Meynet \& Maeder~\cite{MeMa00,MeMa03,MeMa05}; Maeder \& 
Meynet~\cite{MaMe01}) and in addition magnetic fields (Heger et
al.~\cite{Hegeretal05}; Maeder \& Meynet~\cite{MaMe05}). These
make predictions about the mixing of the stellar surface 
layers with nuclear processed matter which can be verified observationally. 
Moreover, BA-SGs can act as primary
indicators for the cosmological distance scale by application of the wind 
momentum--luminosity relationship
(WLR, Puls et al.~\cite{Pulsetal96}; Kudritzki et
al.~\cite{Kudritzkietal99}) and by the flux-weighted gravity--luminosity
relationship (FGLR, Kudritzki et al.~\cite{Kudritzkietal03}; Kudritzki \&
Przybilla~\cite{KuPr03}). In addition to the stellar metallicity,
interstellar reddening can also be accurately determined, so that BA-SGs provide
significant advantages compared to classical distance indicators such as
Cepheids and RR\,Lyrae.

Despite this immense potential, {\em quantitative} analyses of BA-SGs are
scarce. Only a few single objects were studied in an early phase;
several bright Galactic supergiants, preferentially among these $\alpha$\,Cyg (A2\,Iae) 
and $\eta$\,Leo 
(Groth~\cite{Groth61}; Przybylski~\cite{Przybylski69}; Wolf~\cite{Wolf71};
Aydin~\cite{Aydin72}) and the visually brightest stars in the Magellanic Clouds
(Przy\-bylski~\cite{Przybylski68}, \cite{Przybylski71}, \cite{Przybylski72};
Wolf~\cite{Wolf72}, \cite{Wolf73}). Near-solar abundances were found in
almost all cases.
Surveys for the most luminous stars in the Local Group galaxies followed
(Humphreys~\cite{Humphreys80}, and references therein), but were not
accompanied by detailed quantitative analyses.
These first quantitative studies were outstanding for their time,
but from the present point of view they were also restricted in
accuracy by oversimplified analysis techniques/model atmospheres, inaccurate 
atomic data and the lower quality of the observational material
(photographic plates).
Non-LTE effects were completely ignored at that time, as appropriate models
were just being developed
(e.g. Mihalas~\cite{Mihalas78}, and references therein; Kudritzki~\cite{Kudritzki73}).

BA-type supergiants have become an active field of research again, 
following the progress made in model atmosphere techniques and detector 
technology (CCDs), and in particular the advent of 8--10m-class telescopes.
In a pioneering study by Venn~(\cite{Venn95a}, \cite{Venn95b}) over twenty
(less-luminous) Galactic A-type supergiants were systematically 
analysed for chemical abundances,
using modern LTE model atmosphere techniques, and non-LTE refinements in
a few cases. These indicated near-solar abundances for the heavier elements and
partial mixing with CN-cycled gas. A conflict with stellar evolution
predictions was noted, as the observed high N/C ratios were realised through
carbon depletion and not via the predicted nitrogen enrichment. Later, this
conflict was largely resolved by Venn \& Przybilla~(\cite{VePr03}). However,  
an analysis of helium was not conducted, and more luminous supergiants, which are 
of special interest for extragalactic studies, were missing in the sample.
Similar applications followed on 
objects in other spiral and dwarf irregular galaxies (dIrrs) of the Local
Group (McCarthy et al.~\cite{McCarthyetal95}; Venn~\cite{Venn99}; 
Venn et al.~\cite{Vennetal00}, \cite{Vennetal01}, \cite{Vennetal03})
and the nearby Antlia-Sextans Group dIrr galaxy Sextans A (Kaufer et
al.~\cite{Kauferetal04}). The primary aim was to obtain first measurments of
heavy element abundances in these galaxies. Good agreement between stellar
oxygen abundances and literature values for nebular abundances was found,
with the exception of the dIrr WLM (Venn et al.~\cite{Vennetal03}). 
The number of objects analysed is small, thus prohibiting the derivation of
statistically significant further conclusions. Moreover, abundances of the mixing 
indicators He, C and N were not determined in almost all cases.

Parallel to this, a sample of Galactic BA-SGs were studied by 
Takeda \& Takada-Hidai (\cite{TaTa00}, and references therein) for 
non-LTE effects on the light element abundance analyses which confirm
the effects of mixing in the course of stellar evolution.
However, a conflict with the predictions of stellar evolution models was found, 
because C depletion was apparently accompanied by He depletion. The authors regarded 
this trend not being real, but indicated potential problems with the He\,{\sc i} line-formation. 
On the other hand, stellar parameters were not independently determined in 
these studies, but estimated, which can result in severe systematic global
errors as will be shown later. Verdugo et al.~(\cite{Verdugoetal99b}) concentrated
on deriving basic stellar parameters for over 30 Galactic A-type supergiants,
assuming the validity of LTE. 
Independently, the Galactic high-luminosity benchmark $\alpha$\,Cyg was
investigated for elemental abundances by Takeda et
al.~(\cite{Takedaetal96}), solving the restricted non-LTE problem, and by
Albayrak~(\cite{Albayrak00}), using a pure LTE approach. Fundamental parameters 
of $\alpha$\,Cyg were determined by Takeda~(\cite{Takeda94}) and by Aufdenberg
et al.~(\cite{Aufdenbergetal02}). At lower luminosity, $\eta$\,Leo was the
subject of a couple of such studies (Lambert et al.~\cite{Lambertetal88}; Lobel
et al.~\cite{Lobeletal92}). The results from these studies mutually agree only if
rather generous error margins are allowed for, implying considerable
systematic uncertainties in the analysis methods.
Among the late B-type supergiants the
investigations focused on $\beta$\,Ori (Takeda~\cite{Takeda94}; Israelian
et al.~\cite{Israelianetal97}). A larger sample of Galactic B-type supergiants, among those
a number of late B-types, was studied by McErlean et
al.~(\cite{McErleanetal99}) for basic stellar parameters and estimating
elemental abundances, on the basis of unblanketed non-LTE model atmospheres and
non-LTE line formation. The chemical analysis recovers values being consistent 
with present-day abundances from unevolved Galactic B-stars, except for the
light elements, which again indicate mixing of the surface layers with
material from the stellar core.

\begin{figure*}
{\centering
\resizebox{0.948\hsize}{!}{\includegraphics{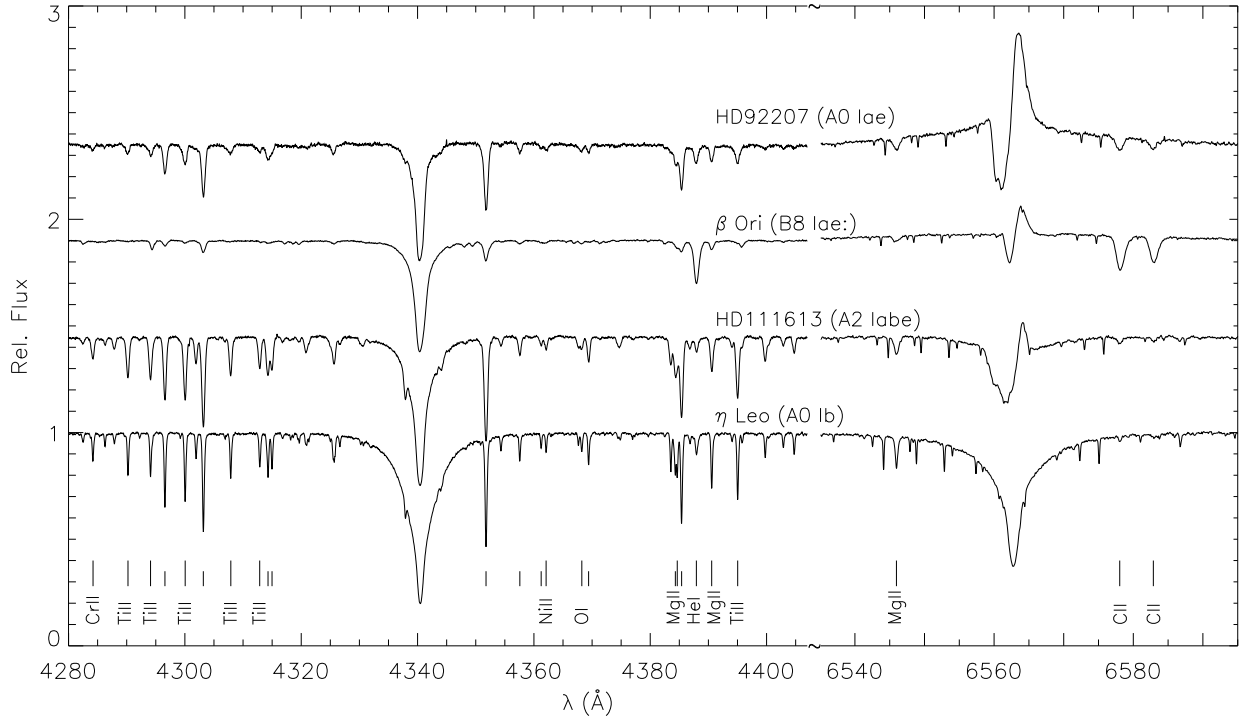}}\\[-2mm]}
\caption[]{Spectra of the sample stars around H$\gamma$ and
H$\alpha$. The major spectral features are identified, short vertical marks 
indicate \ion{Fe}{ii} lines. Luminosity increases from bottom to top.
Likewise, H$\alpha$ develops from a pure absorption line into a 
P-Cygni-profile because of the strengthening stellar wind.
Note the marked incoherent electron scattering wings of H$\alpha$ in
the most luminous object.}
\label{spectra}
\end{figure*}

The basic stellar parameters and abundances from {\em }modern studies of
individual objects can still be discrepant by up to $\sim$20\% in 
$T_{\rm eff}$, $\sim$0.5\,dex in $\log g$ and $\sim$1\,dex in the abundances.
This is most likely to be the result of an interplay of systematic errors and
inconsistencies in the analysis procedure, as we will discuss in the following.
In order to improve on the present status, we introduce a spectrum synthesis approach 
for quantitative analyses of the photospheric spectra of BA-SGs. 
Starting with an overview of the observations of our
sample stars and the data reduction, we continue with a thorough investigation of the
suitability of various present-day model atmospheres for such analyses, and
their limitations. General aspects of our line-formation computations are 
addressed in Sect.~\ref{sectlform}, while the details are summarised in
Appendix~\ref{apa}. Our approach to stellar parameter and abundance
determination is examined and tested on the sample stars in Sects.~\ref{sectparams}
and~\ref{sectabus}. The consequences of these for the evolutionary status 
of our BA-SG sample are briefly discussed in Sect.~\ref{sectevol}.
Finally, the applicability of our technique is tested for intermediate
spectral resolution in Sect.~\ref{sectmedres}. At all stages we put special 
emphasis on identifying and eliminating sources of systematic error, which allows us to
constrain all relevant parameters with unprecedented accuracy. We conclude
this work with a summary of the main results.

We will address the topic of stellar winds in our sample of BA-SGs
separately, completing the discussion on the analysis inventory for this class of
stars. Applications to Galactic and extragalactic BA-SGs in Local Group systems
and beyond will follow. Note that the technique presented and tested here
in the most extreme conditions is also well suited to 
improve quantitative analyses of less luminous stars of similar spectral~classes.


\section{Observations and data reduction}\label{sectobs}

We test our analysis technique on a few bright Galactic supergiants that
roughly sample the parameter space in effective temperature and surface gravity
covered by future applications. We chose the two MK standards $\eta$\,Leo
(\object{HD\,87737}) and $\beta$\,Ori (\object{HD\,34085}), the brightest member
of the rich southern cluster \object{NGC\,4755}, \object{HD\,111613}, and
one of the most luminous Galactic A-type supergiants known to
date,\,\object{HD\,92207},\,for this objective.

For $\eta$\,Leo, HD\,111613 and HD\,92207, Echelle spectra using FEROS
(Kaufer et al.~\cite{Kauferetal99}) at the ESO 1.52m
telescope in La Silla were obtained on January, 21 and 23, 1999. Nearly complete wavelength
coverage between 3\,600 and 9\,200\,\AA\ was achieved, with a resolving power
$R$\,=\,$\lambda/\Delta\lambda$\,$\approx$\,48\,000 (with 2.2 pixels per 
$\Delta\lambda$
resolution element), yielding a S/N of several hundred in $V$ in 120, 600 and
300\,sec exposures. A corresponding spectrum of $\beta$\,Ori was adopted from 
Commissioning II data (\#0783, 20\,sec exposure taken in November 1998). 

Data reduction was performed using the FEROS context in the MIDAS package
(order definition, bias subtraction, subtraction of scattered light, order
extraction, flat-fielding, wavelength calibration, barycentric movement
correction, merging of the orders), as described in the FEROS documentation
(http:\,//www.ls.eso.org/lasilla/Telescopes/2p2T/E1p5M/
FEROS/docu/Ferosdocu.html).
Optimum extraction of the orders with cosmic ray clipping was chosen. In the
spectral region longward of $\sim$8\,900\,{\AA} problems with the optimum extraction
arose due to the faintness of the signal. Standard extraction was therefore
performed in this region. 
No correction for telluric lines was made.
The spectral resolution is sufficiently high to trace the (broadened)
stellar lines in our spectrum synthesis approach even within the
terrestrial O$_2$ and H$_2$O bands in many cases, see e.g. Fig.~9 of
Przybilla et al.~(\cite{Przybillaetal01b}).
The spectra were normalised
by fitting a spline function to carefully selected continuum points. This
suffices to retain the line profiles of the Balmer lines in supergiants as
these are rather weak and sampled by a single Echelle order.
Finally, the spectra were shifted
to the wavelength rest frame by accounting for a radial velocity $v_{\rm rad}$
as determined from cross-correlation with an appropriate synthetic spectrum.
Excellent agreement with $v_\mathrm{rad}$ data from the
literature (see Table~\ref{obj}) was found. Representative parts of the spectra are 
displayed in Fig.~\ref{spectra}.


\begin{figure*}
\resizebox{0.497\hsize}{!}{\includegraphics{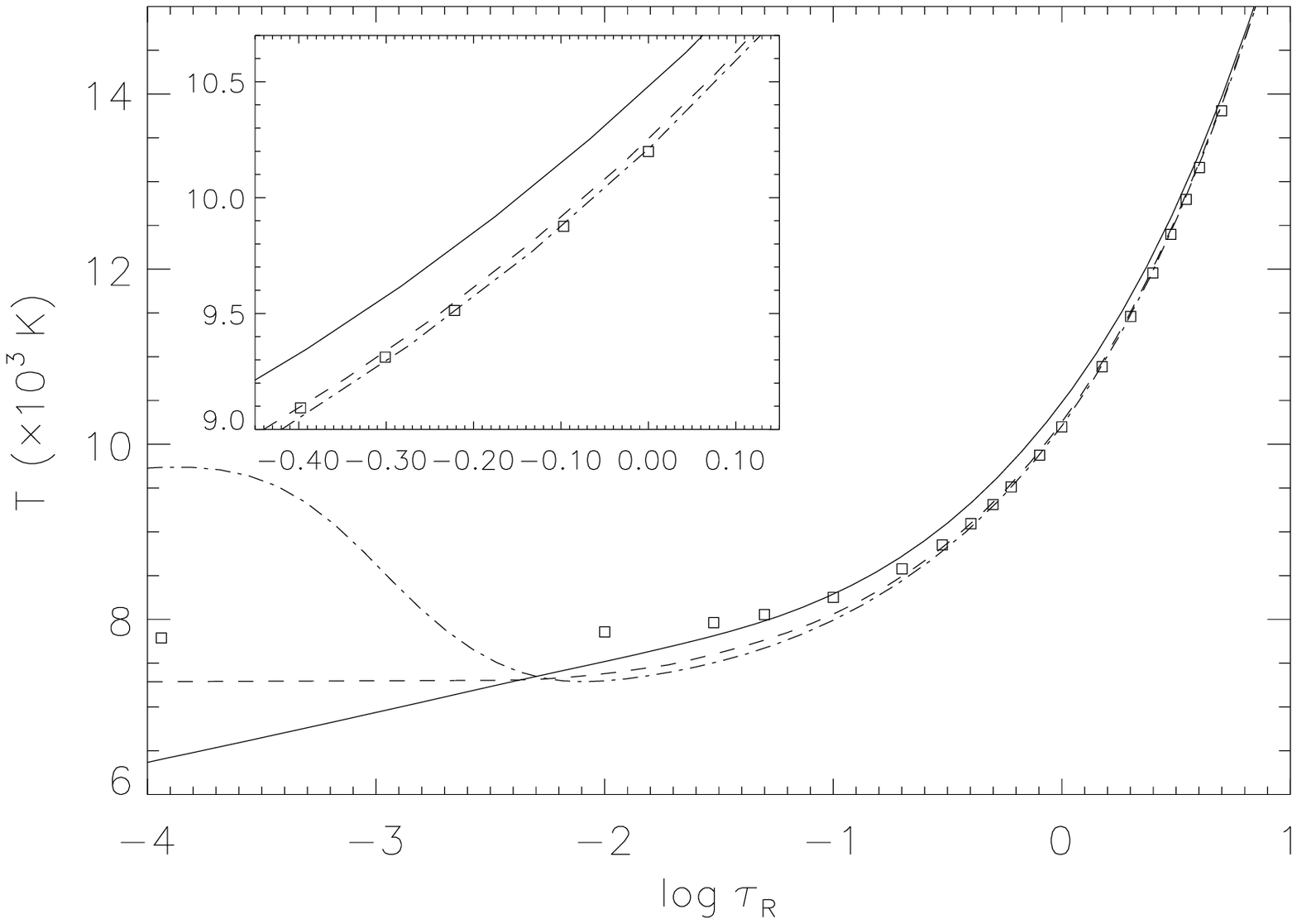}}
\hfill
\resizebox{0.497\hsize}{!}{\includegraphics{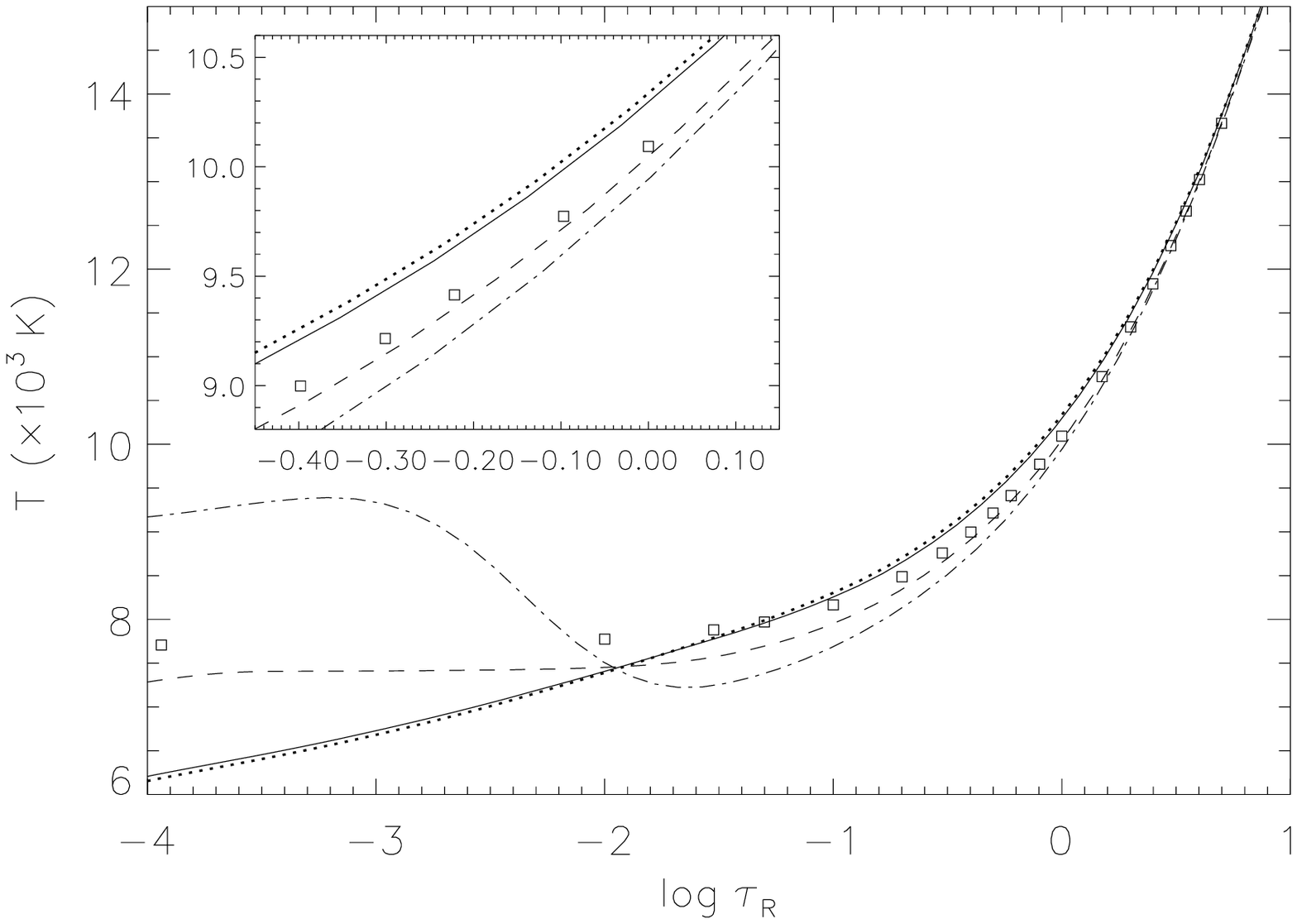}}
\resizebox{0.497\hsize}{!}{\includegraphics{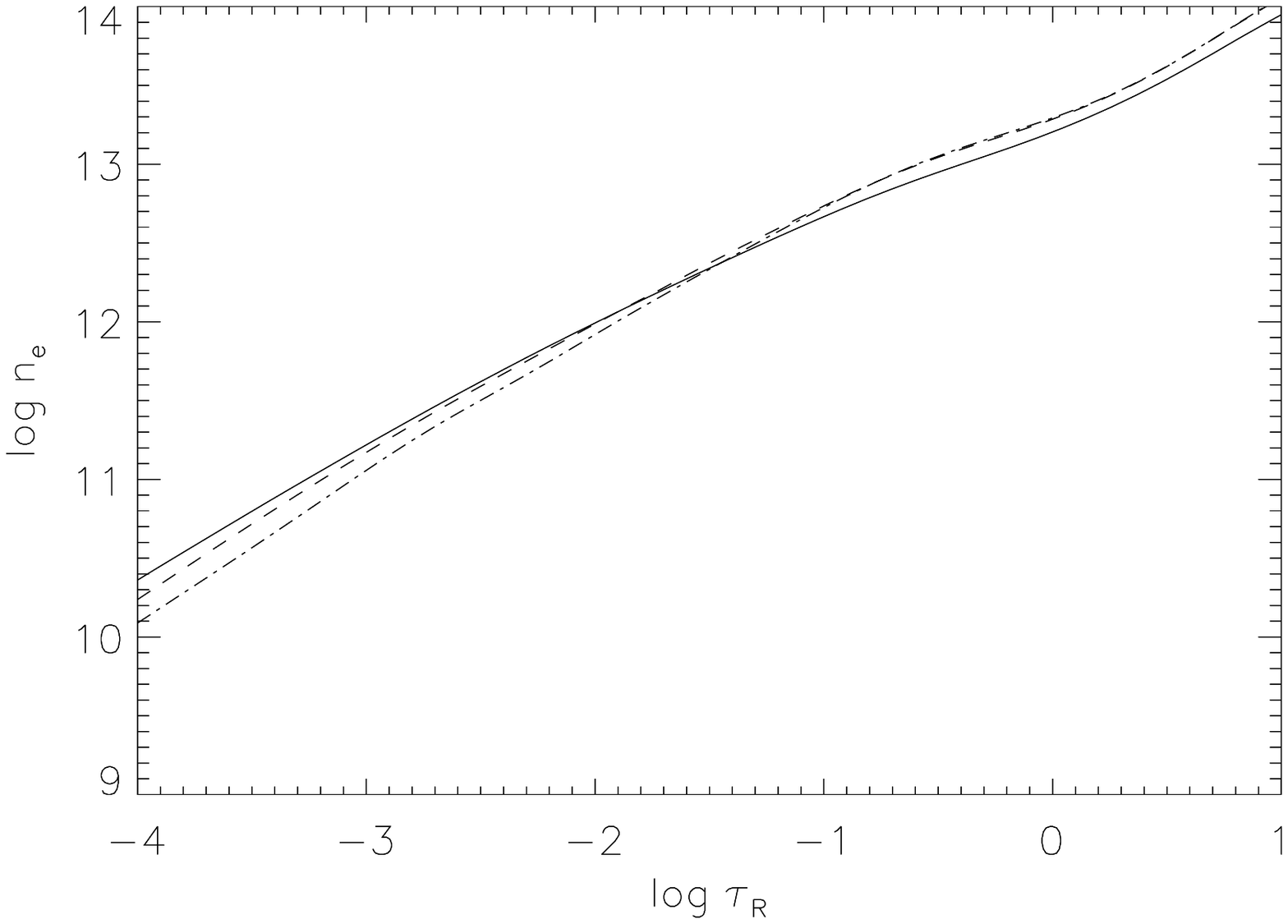}}
\hfill
\resizebox{0.497\hsize}{!}{\includegraphics{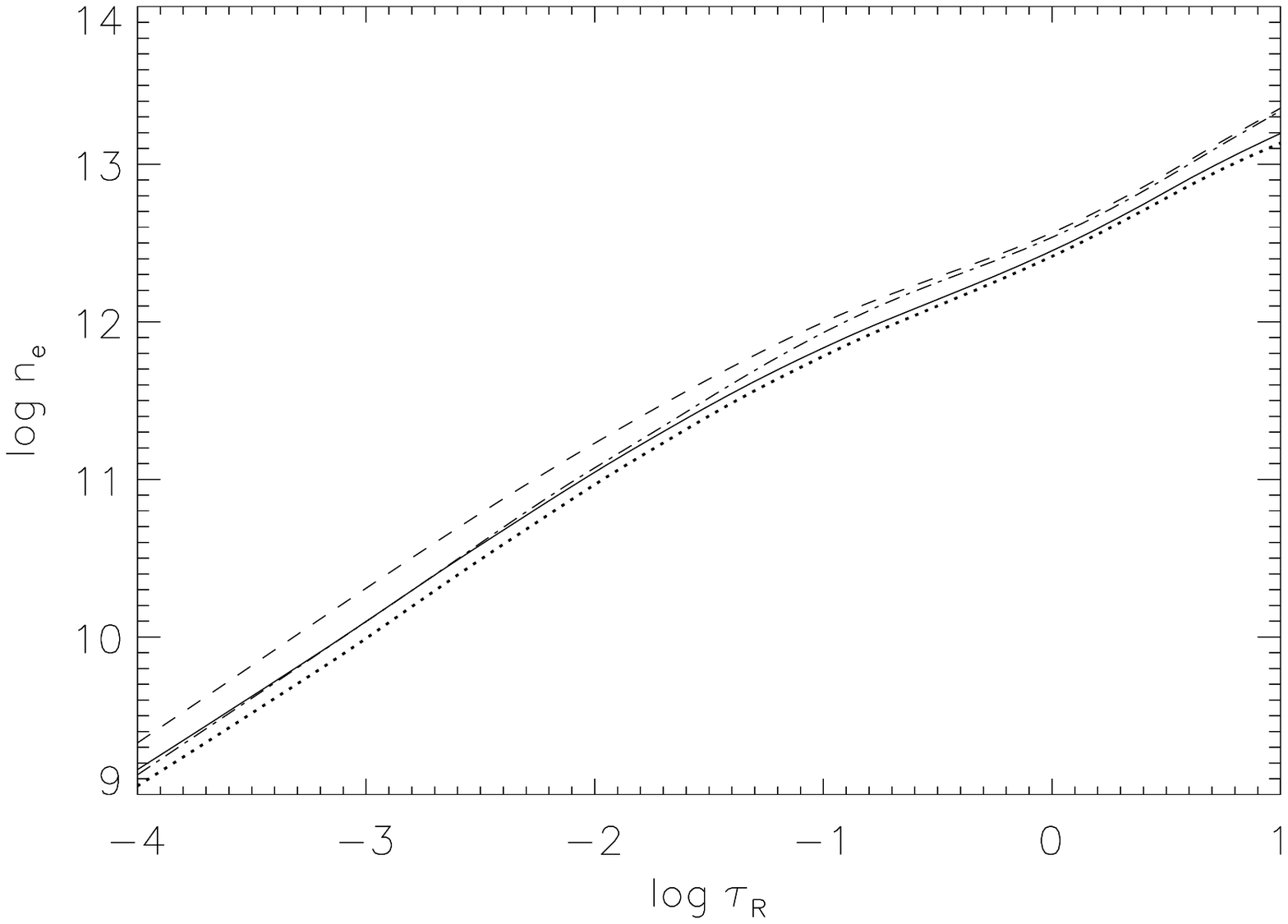}}
\caption[]{Comparison of different model atmospheres:
temperature structure (top) and electron density (bottom) for two supergiant
models as a function of the Rosseland optical depth. On the left for the 
LC Ib with stellar parameters
according to our model for $\eta$\,Leo, on the right for the LC
Iae (HD\,92207), corresponding to our sample objects at lowest and highest
luminosity. Full line:  {\sc Atlas9} line-blanketed LTE model 
(adopted for the analysis), dashed: {\sc Atlas9} H+He LTE model without line
blanketing, dashed-dotted: {\sc Tlusty } H+He non-LTE model without line
blanketing, boxes: grey atmosphere. A second {\sc Atlas9} line-blanketed LTE 
model is displayed in the case of the Iae object (dotted), calculated with
background opacities corresponding to solar metallicity. In our final model 
for HD\,92207 the background opacities are reduced by an empirical factor of two
in metallicity, correcting at least qualitatively for neglected non-LTE
effects. In the inset the formation region of weak lines is displayed enlarged.} 
\label{atmplot}
\end{figure*}

\section{Model atmospheres}\label{sectmodels}
Model atmospheres are a crucial ingredient for solving the {\em inverse
problem} of quantitative spectroscopy. Ideally, BA-SGs should be described by
{\em unified} (spherically extended, hydrodynamical) non-LTE atmospheres, accounting for the
effects of line blanketing. Despite the progress made in the last three
decades, such model atmospheres are still not available for routine
applications. In the following we investigate the suitability of {\em
existing} models for our task, their robustness for the applications and their 
limitations. The criterion applied in the end will be their ability to reproduce
the observations in a consistent way.

\subsection{Comparison of contemporary model atmospheres}
Two kinds of {\em classical} (plane-parallel, hydrostatic and
stationary) model atmospheres are typically applied in the contemporary
literature for analyses of BA-SGs: line-blanketed LTE atmospheres and 
non-LTE H$+$He models (without line-blanketing) in radiative equilibrium.
First, we discuss what effects these different physical assumptions have on
the model stratification and on synthetic profiles of important diagnostic
lines. We therefore include also an LTE H$+$He model without line-blanketing and
additionally a grey stratification in the comparison for two limiting cases: the
least and the most luminous supergiants of our sample, $\eta$\,Leo and
HD\,92207, of luminosity class (LC) Ib and Iae.
The focus is on the {\em photospheric} line-formation depths, where the
classical approximations are rather appropriate -- the velocities in the
plasma remain sub-sonic, the spatial extension of this region is small (only
a few percent) compared to the stellar radius in most cases, and the BA-SGs photospheres retain
their stability over long time scales, in contrast to their cooler progeny, 
the yellow supergiants, which are to be found in the instability strip of the
Hertzsprung-Russell diagram, or the Luminous Blue Variables.

\begin{figure*}
\resizebox{0.495\hsize}{!}{\includegraphics{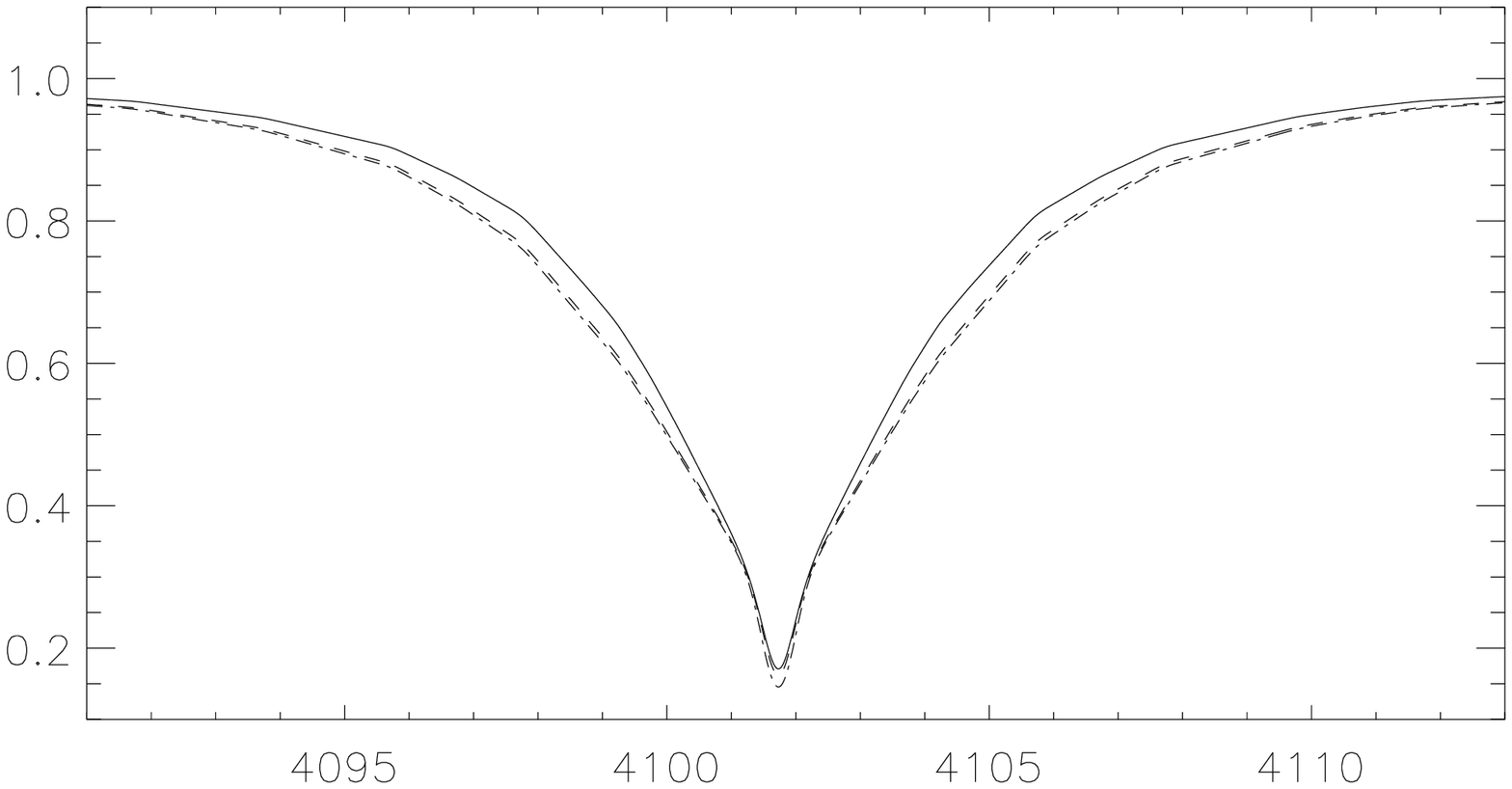}}
\hfill
\resizebox{0.495\hsize}{!}{\includegraphics{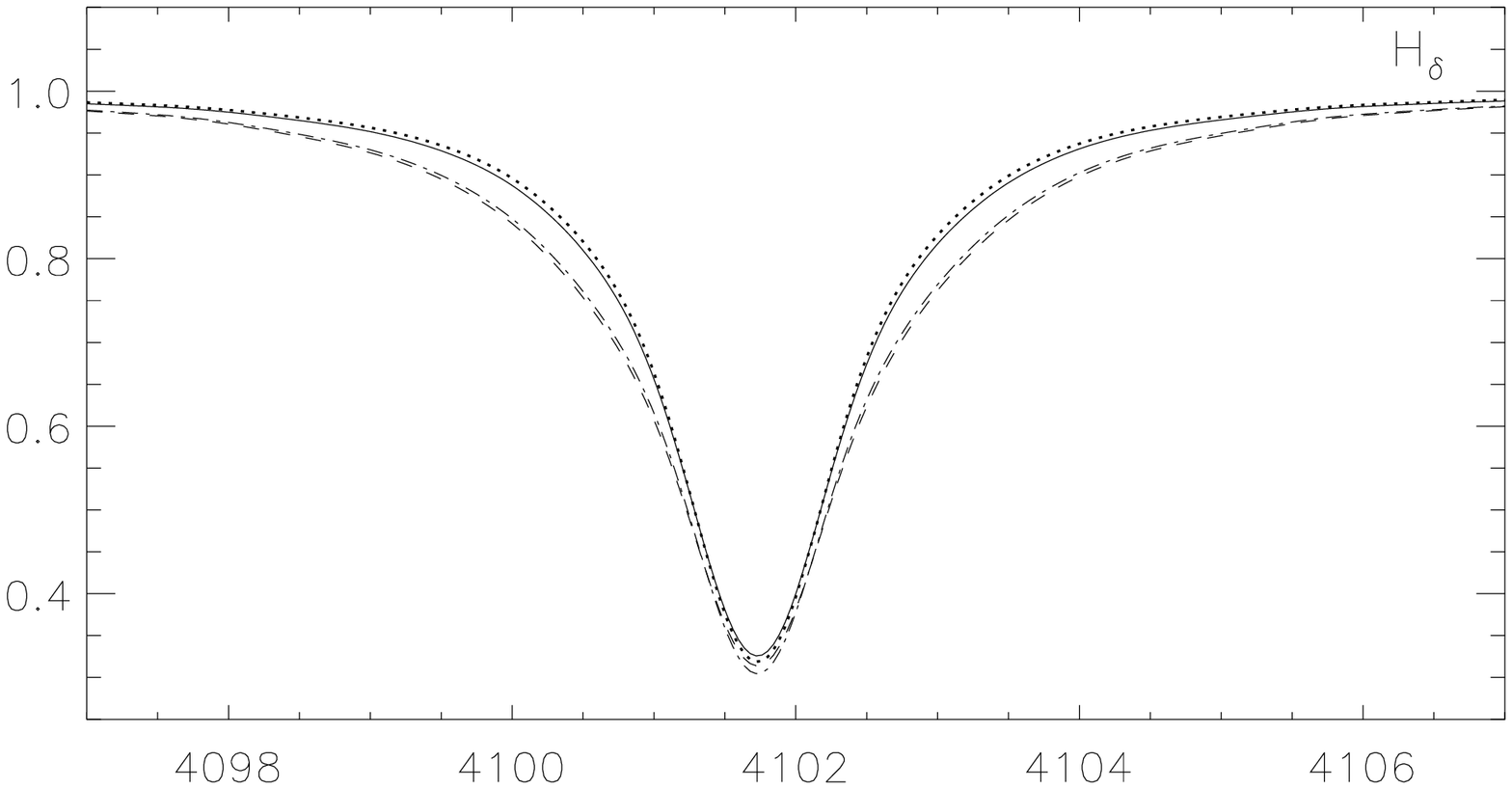}}\\[-2mm]
\resizebox{0.495\hsize}{!}{\includegraphics{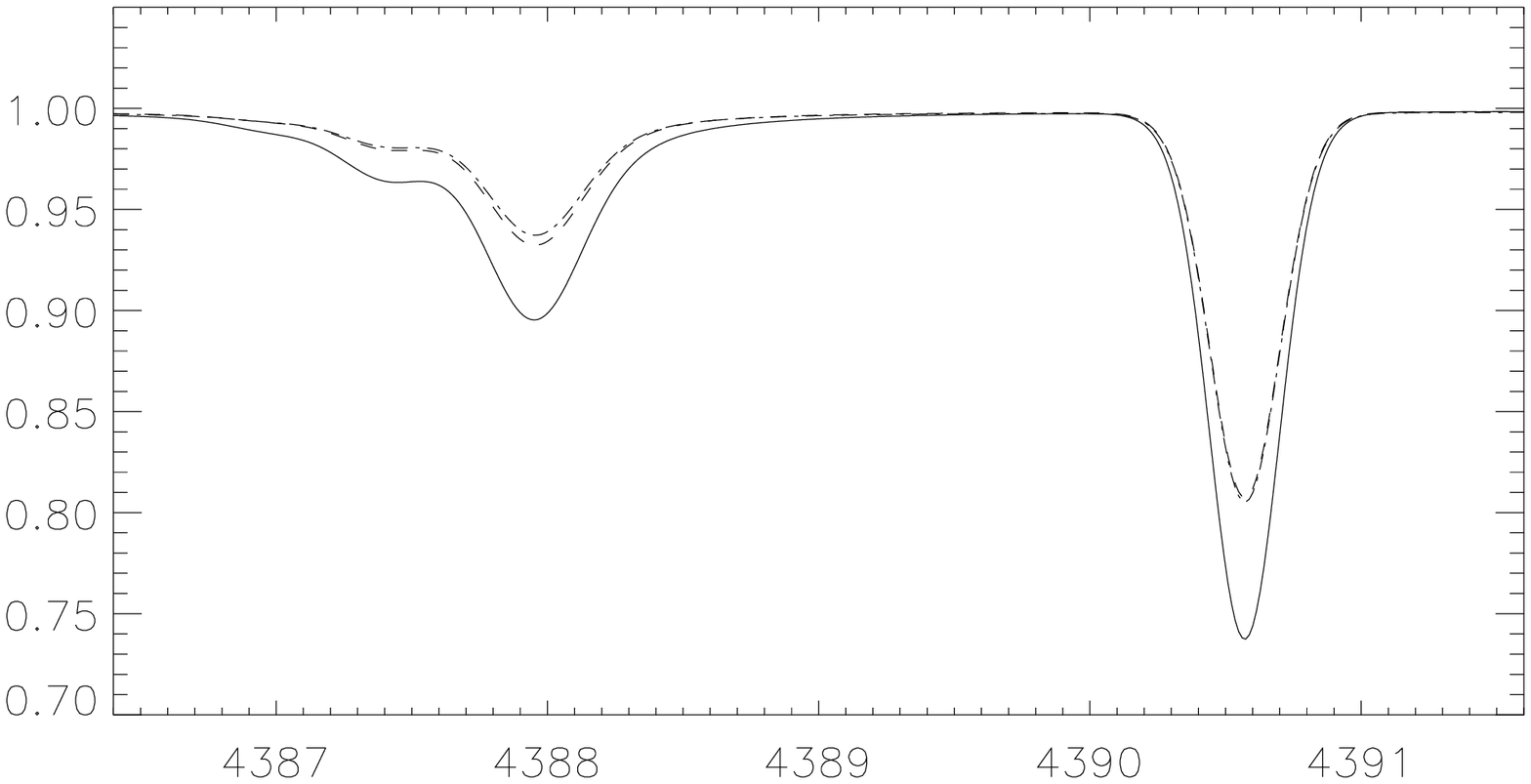}}
\hfill
\resizebox{0.495\hsize}{!}{\includegraphics{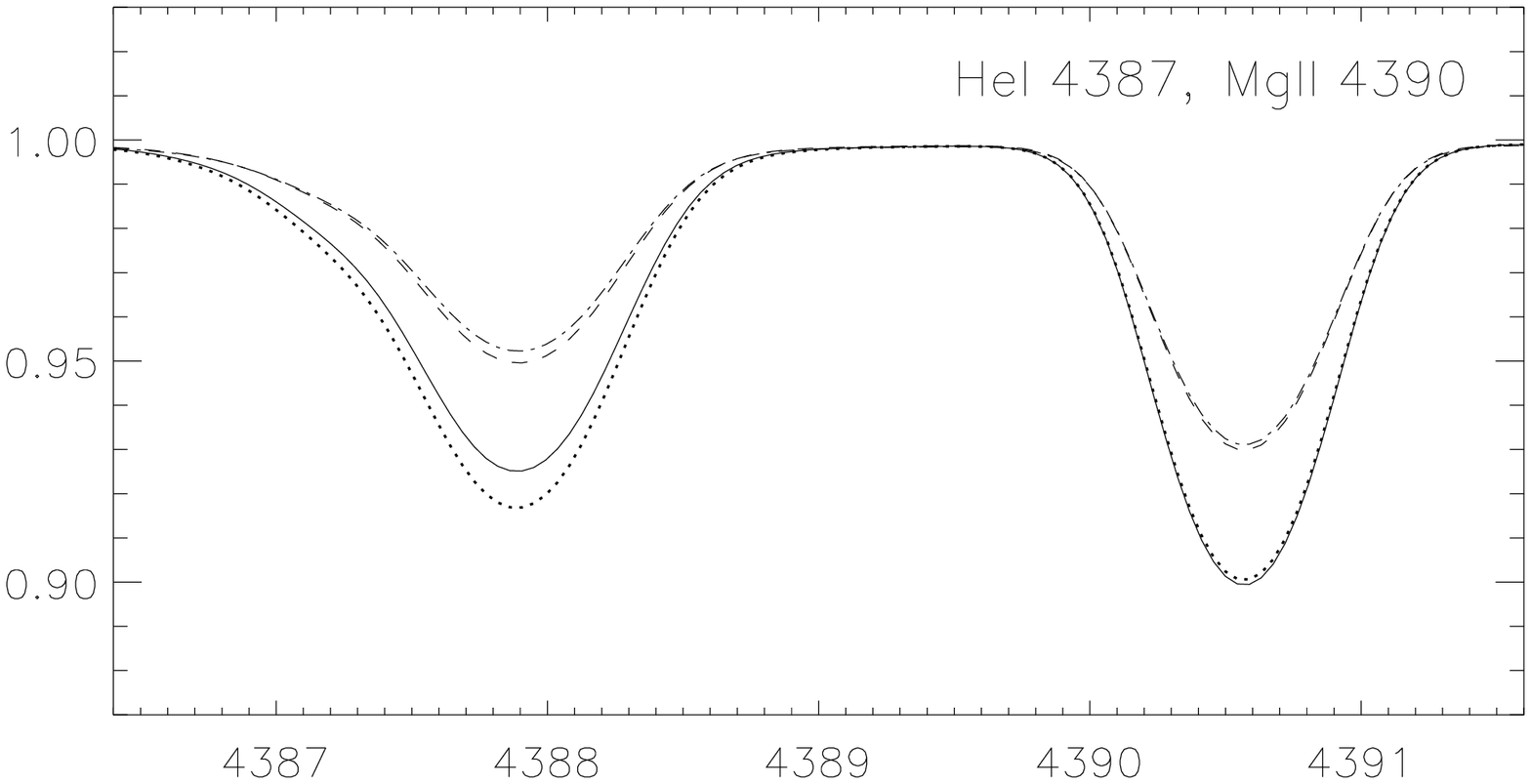}}\\[-2mm]
\resizebox{0.495\hsize}{!}{\includegraphics{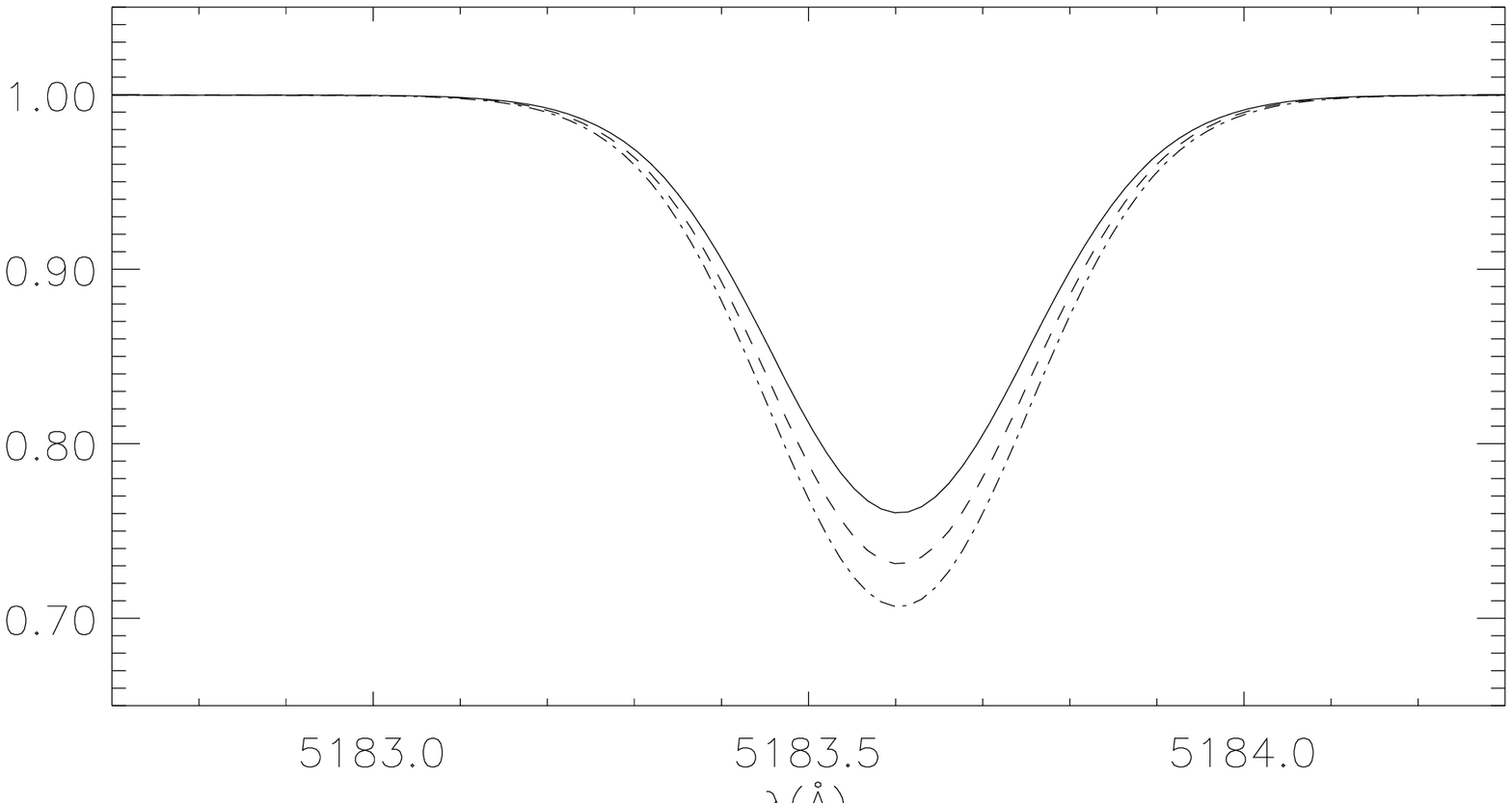}}
\hfill
\resizebox{0.495\hsize}{!}{\includegraphics{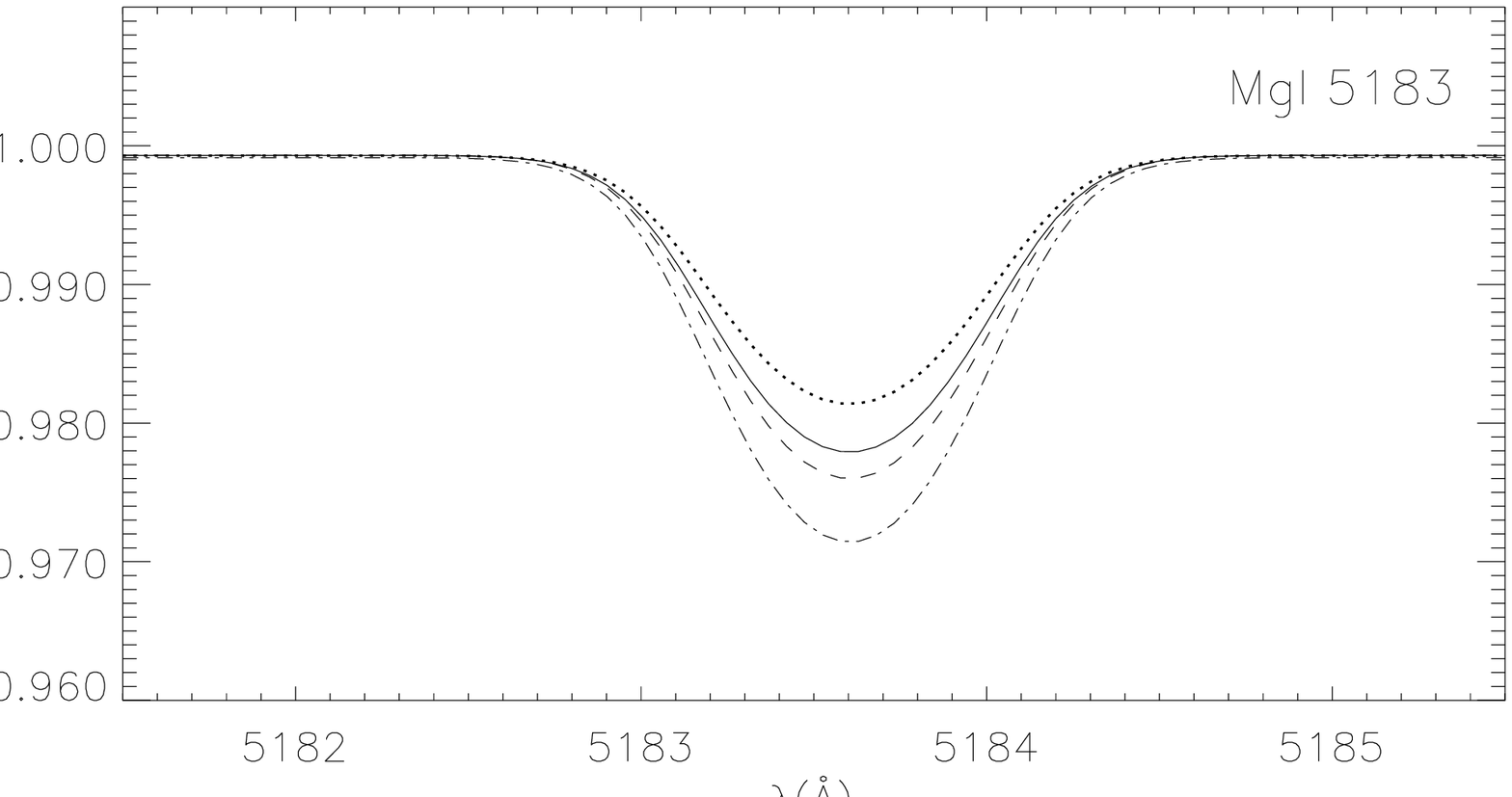}}
\caption[]{Comparison of theoretical non-LTE profiles for several diagnostic
lines, calculated on the basis of different model
atmospheres for A0 supergiants of LC Ib ($\eta$\,Leo, left)
and Iae (HD\,92207, right). Ordinate is normalised flux.
The same line designations as in Fig.~\ref{atmplot} are used, and the profiles are
broadened accounting for instrumental profile, rotation and
macroturbulence (see Table~\ref{obj}). 
Note that the \ion{He}{i} and \ion{Mg}{i} lines react sensitively to the detailed
structure of the models
while for the Balmer lines and the \ion{Mg}{ii} lines only the profiles of the
line-blanketed and the unblanketed models can be discriminated well.}
\label{profplot}
\end{figure*}

The non-LTE models are computed using the code {\sc Tlusty} (Hubeny \&
Lanz~\cite{HuLa95}), the LTE models are calculated with {\sc Atlas9}
(Kurucz~\cite{Kurucz93}), in the version of M.~Lemke, as obtained
from the CCP7~software library, and with further modifications (Przybilla et
al.~\cite{Przybillaetal01b}) where necessary. Line blanketing is accounted
for by using solar metallicity opacity distribution functions (ODFs)
from Kurucz~(\cite{Kurucz92}). For the grey temperature structure,
$T^4$\,$=$\,$\frac{3}{4}\,T_{\rm eff}^4\,[\tau+q(\tau)]$,
exact Hopf parameters $q(\tau)$ (Mihalas~\cite{Mihalas78}, p.~72) are used.

\begin{figure*}
\resizebox{0.497\hsize}{!}{\includegraphics{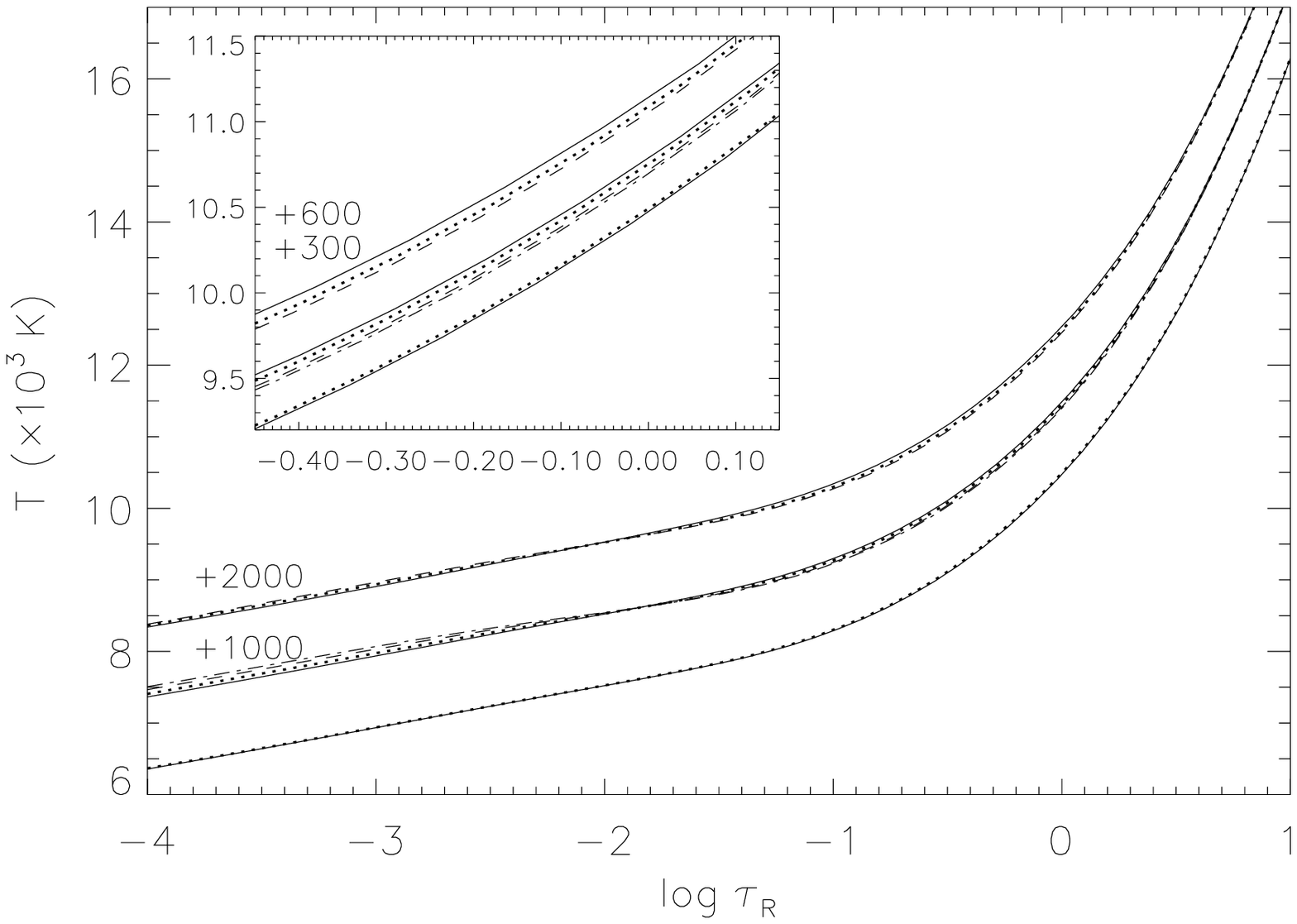}}
\hfill
\resizebox{0.497\hsize}{!}{\includegraphics{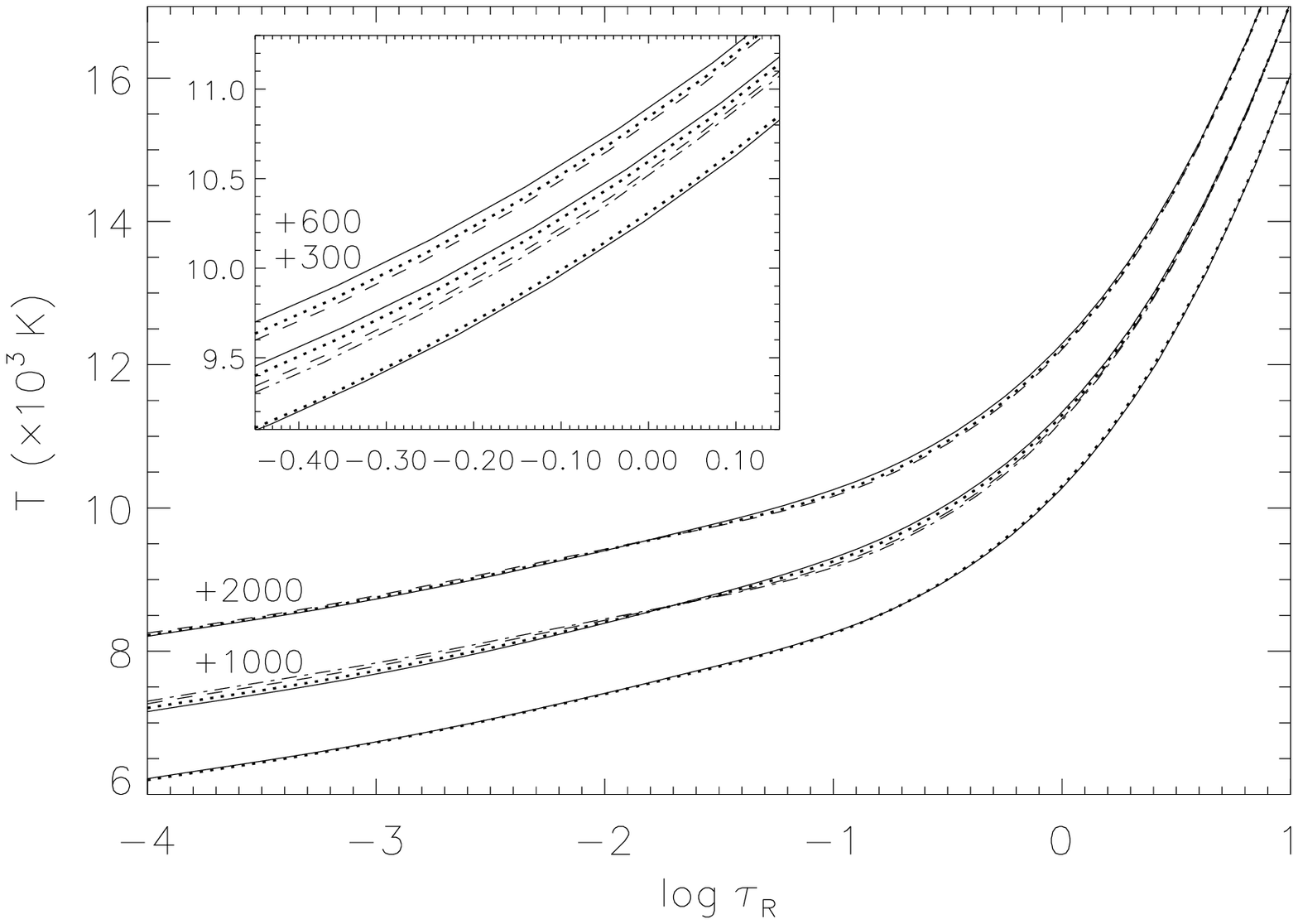}}
\resizebox{0.497\hsize}{!}{\includegraphics{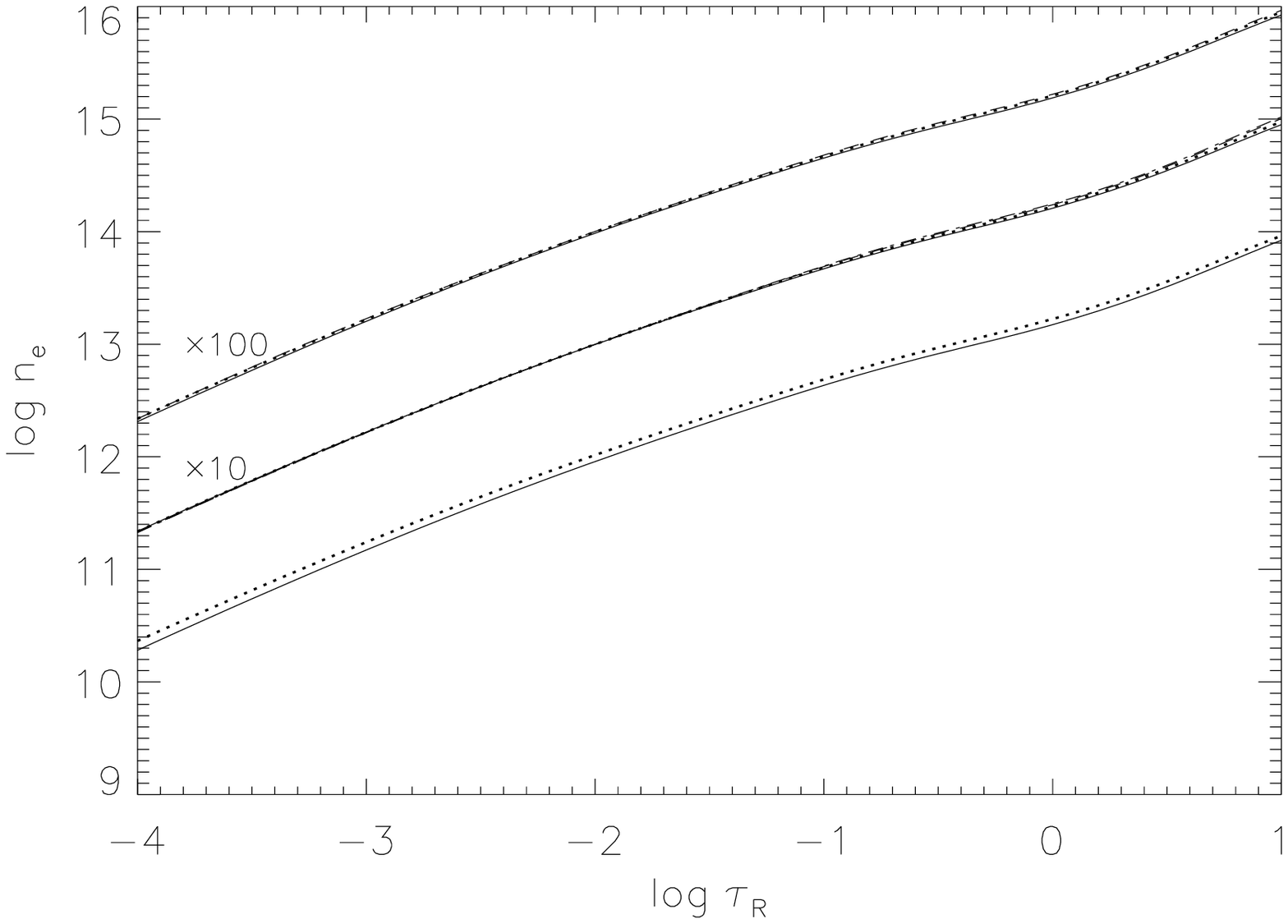}}
\hfill
\resizebox{0.497\hsize}{!}{\includegraphics{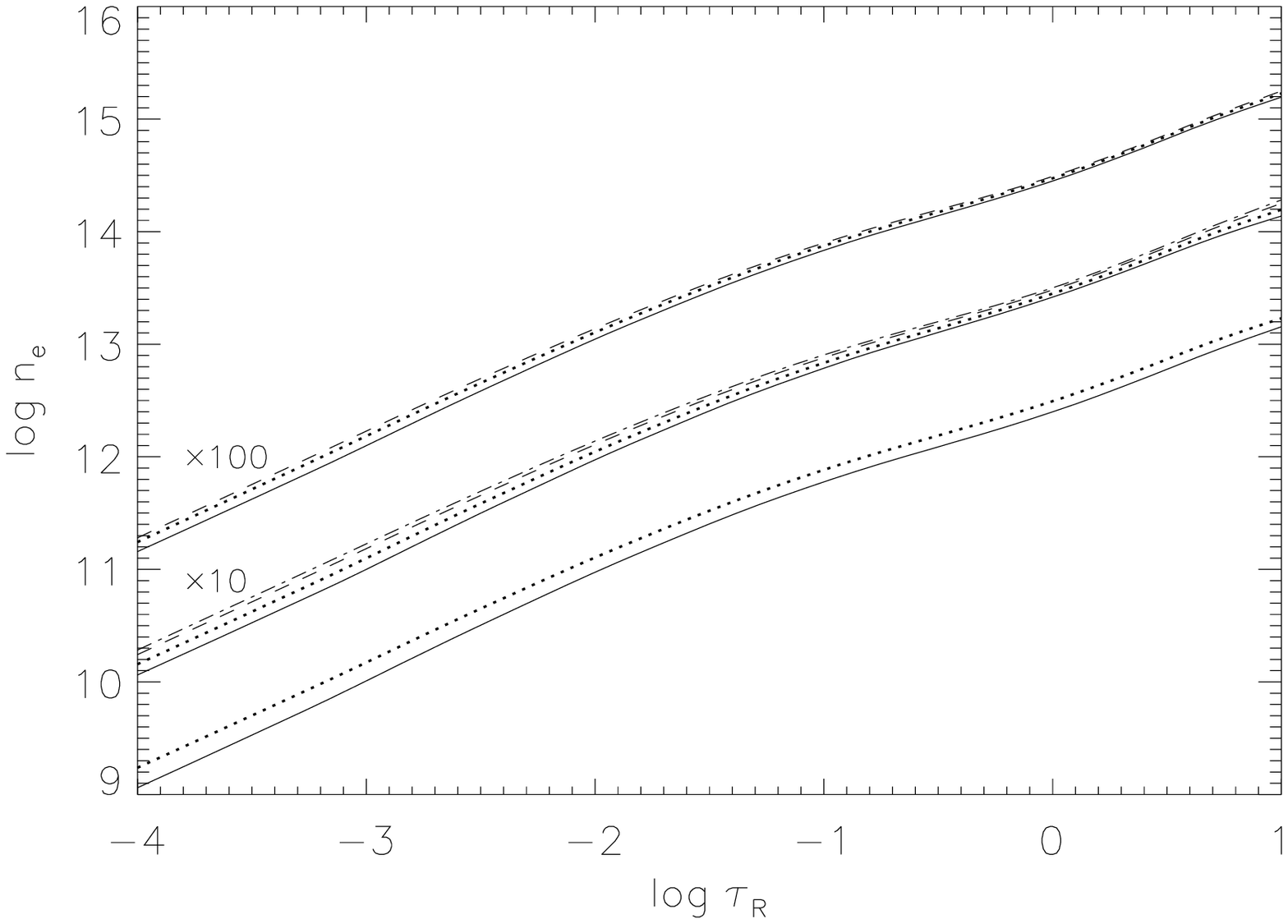}}
\caption[]{Influence of the helium abundance on the atmospheric structure (bottom) 
and effects of metallicity (middle) and microturbulence (top set of curves)
on the atmospheric line blanketing:
temperature structure (upper panels) and electron density (lower panels). On
the left the comparison is made for a LC Ib object with stellar parameters
matching those of our model for $\eta$\,Leo, on the right for the LC Iae (HD\,92207).
Only one parameter is modified each time: atmospheric helium abundance of
$y$\,=\,0.089 (solar value, full line) and $y$\,=\,0.15 (dotted line);
using ODFs with $[$M/H$]$\,=\,0.0, $-$0.3, $-$0.7 and $-1.0$\,dex (full, dotted,
dashed, dashed-dotted lines); using ODFs with $\xi$\,=\,8, 4 and
2\,km\,s$^{-1}$ (full, dotted, dashed lines).
In the inset, the formation region for weak lines is enlarged. For
convenience, the individual sets of curves have been shifted by the factors
indicated.} 
\label{atmplot2}
\end{figure*}

In Fig.~\ref{atmplot} we compare the model atmospheres,
represented by the temperature structure and the run of electron number
density $n_{\rm e}$.
For the less-luminous supergiant marked differences in the line-formation
region (for metal lines, typically between Rosseland optical depth
$-$1\,$\lesssim$\,$\log\,\tau_{\rm R}$\,$\lesssim$\,0) occur only between the line-blanketed
model, which is heated due to the backwarming effect by $\lesssim$200\,K, and the
unblanketed models. In particular, it appears that non-LTE
effects on the atmospheric structure are almost negligible, reducing
the local temperature by $\lesssim$50\,K.
The good agreement of the grey stratification and the unblanketed models
indicates that Thomson scattering is largely dominating the opacity in these cases.
A temperature rise occurs in the non-LTE model because of the recombination of
hydrogen -- it is an artefact from neglecting metal lines.
The gradient of the density rise with atmospheric depth is slightly flatter
in the line-blanketed model.
In the case of the highly luminous supergiant, line-blanketing effects
retain their importance (heating by $\lesssim$300\,K), but non-LTE also becomes
significant (cooling by $\lesssim$200\,K).
From the comparison with observation it is found
empirically, that a model with the metal opacity reduced by a factor of 2
(despite the fact that the line analysis yields near-solar abundances)
gives an overall better agreement. This is slightly cooler than the model
for solar metal abundance and may be interpreted as an {\em empirical
correction} for unaccounted non-LTE effects on the metal blanketing 
in the most luminous objects, i.e. $\beta$\,Ori and HD\,92207 in the
present case. 
The local electron number density in the line-blanketed case is slightly lower
than in the unblanketed models.
We conclude from this comparison that at photospheric line-formation depths 
the importance of line blanketing outweighs non-LTE effects, with the r\^ole
of the latter increasing towards the Eddington limit, as expected.
In the outermost regions the differences between the individual models become more
pronounced. However, none of the stratifications can be expected to give a
realistic description of the real stellar atmosphere there, as spherical extension
and velocity fields are neglected in our approach.

The results of spectral line modelling depend on the details of the
atmospheric structure and different line strengths result from
computations based on the different stratifications. Therefore, non-LTE line profiles for
important stellar parameter indicators are compared in Fig.~\ref{profplot}:
H$\delta$ as a representative for the gravity-sensitive Balmer lines, a
typical \ion{He}{i} line, and features of \ion{Mg}{i/ii}, which
are commonly used for the $T_{\rm eff}$-determination. In the case of the Balmer
and the \ion{Mg}{ii} lines discrimination is only possible between the
line-blanketed and the unblanketed models. Both the LTE and non-LTE
H$+$He model structure without line blanketing result in practically the
same profiles. This changes for the lines of \ion{He}{i} and
\ion{Mg}{i}, which react sensitively to modifications of the
atmospheric structure: all the different models lead to distinguishable line
profiles. These differences become more pronounced at higher luminosity.

This comparison of model structures and theoretical line profiles is
instructive, but the choice of the best suited model
can only be made from a confrontation with observation.
It is shown later that the physically most sophisticated approach 
available at present, LTE with
line blanketing plus non-LTE line formation and appropriately chosen 
parameters, allows highly consistent analyses of BA-SGs to be performed.
This is discussed next. 

\subsection{Effects of helium abundance and line blanketing}\label{heliumetc}
\begin{figure*}
\resizebox{0.497\hsize}{!}{\includegraphics{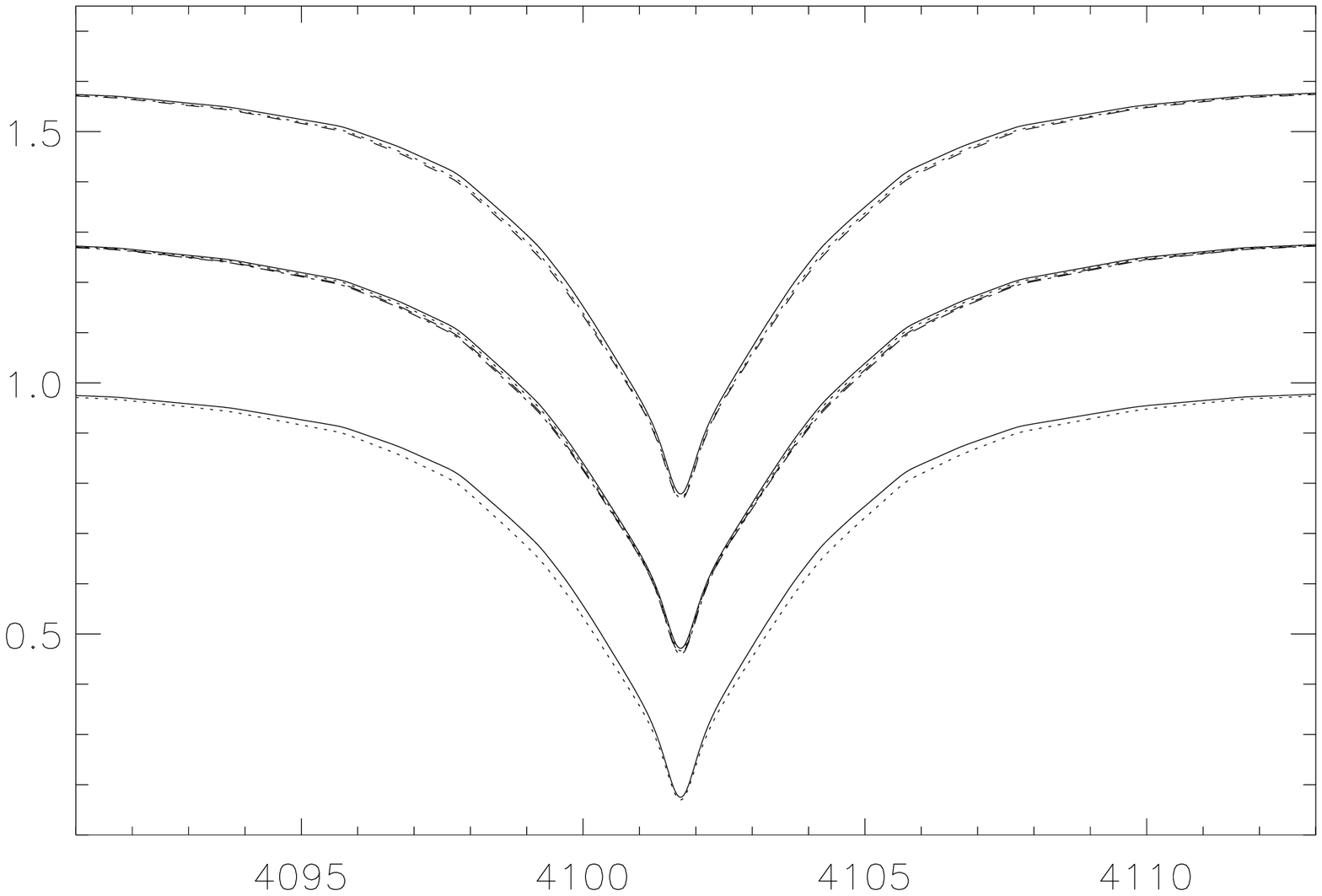}}
\hfill
\resizebox{0.497\hsize}{!}{\includegraphics{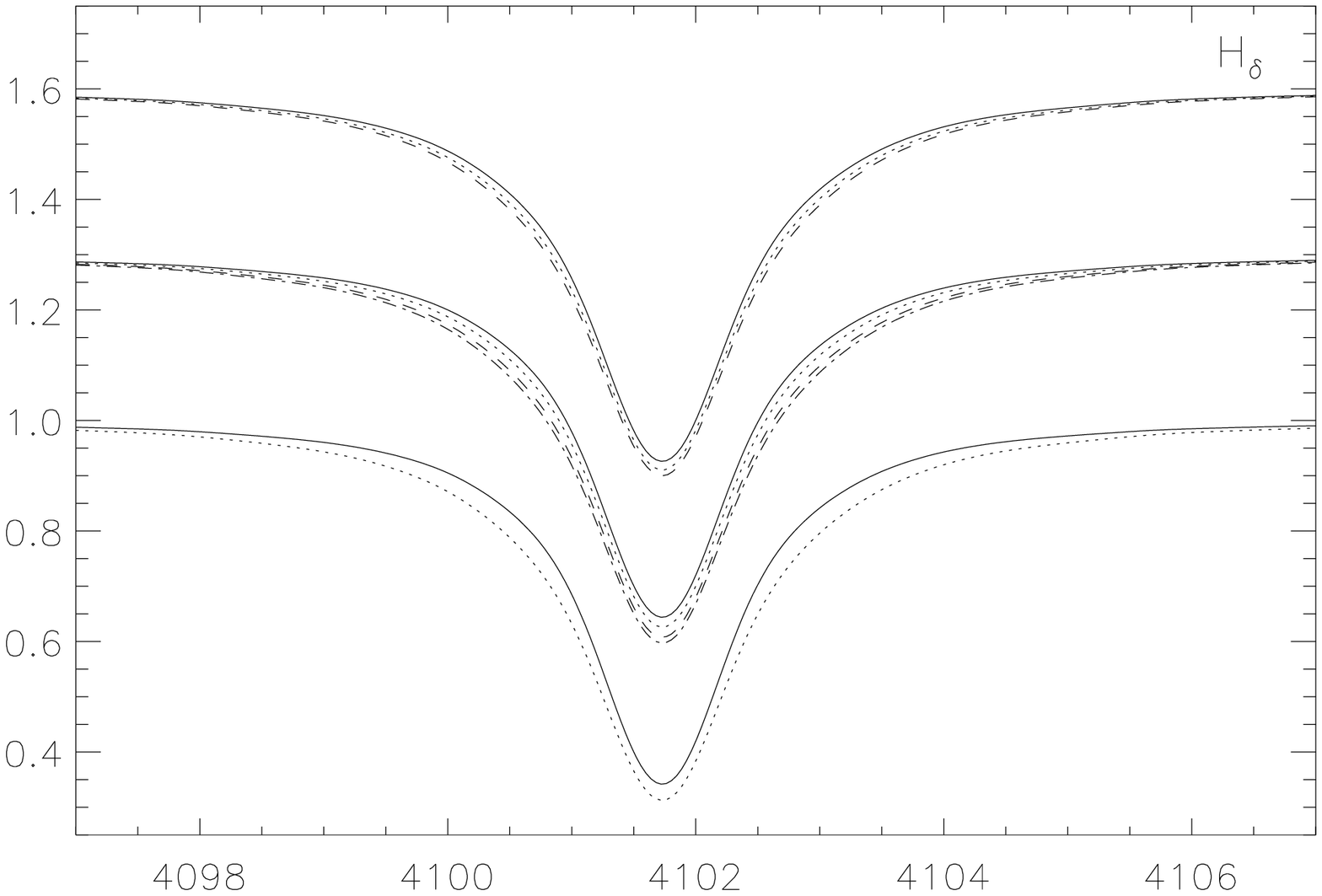}}\\[-4mm]
\resizebox{0.497\hsize}{!}{\includegraphics{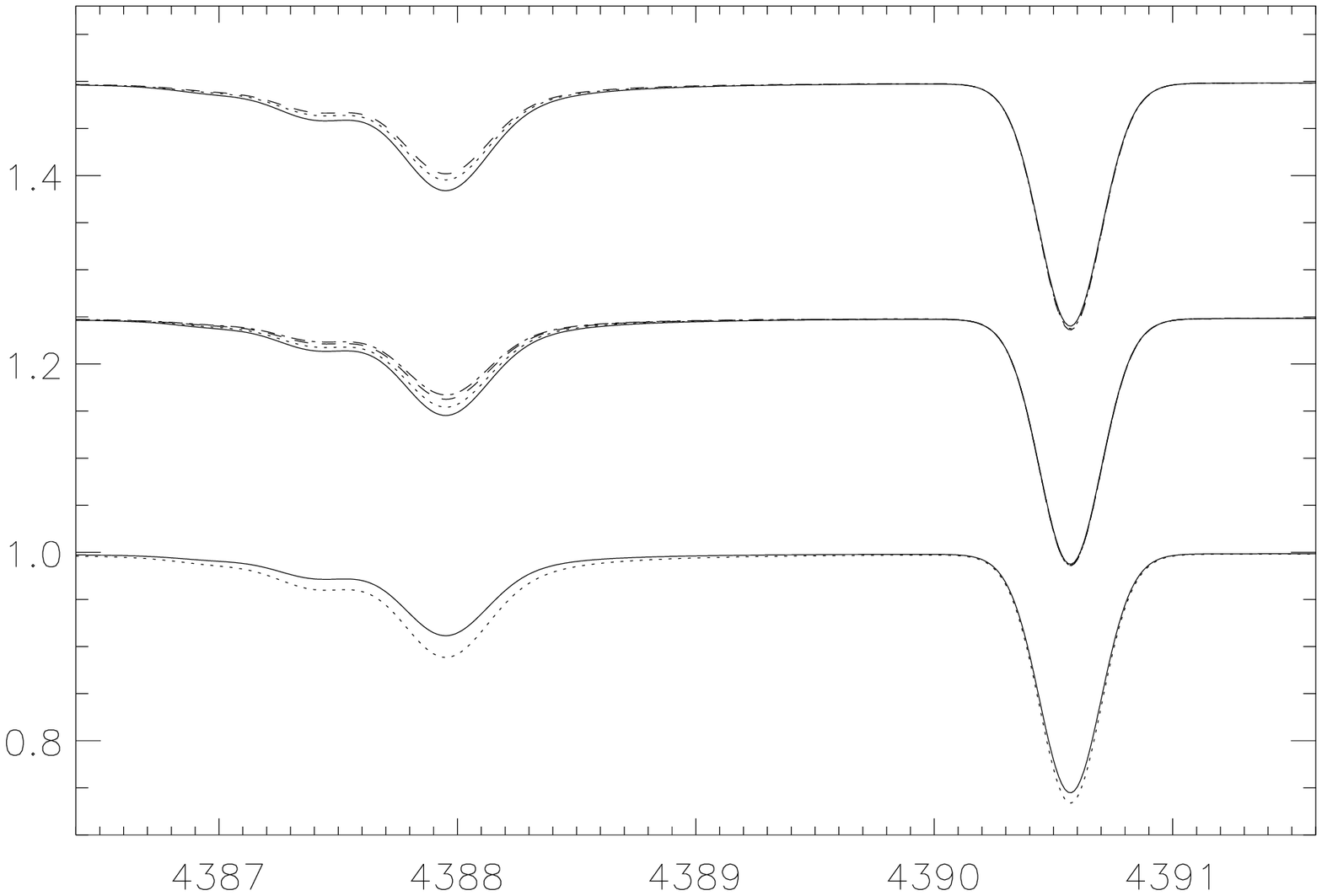}}
\hfill
\resizebox{0.497\hsize}{!}{\includegraphics{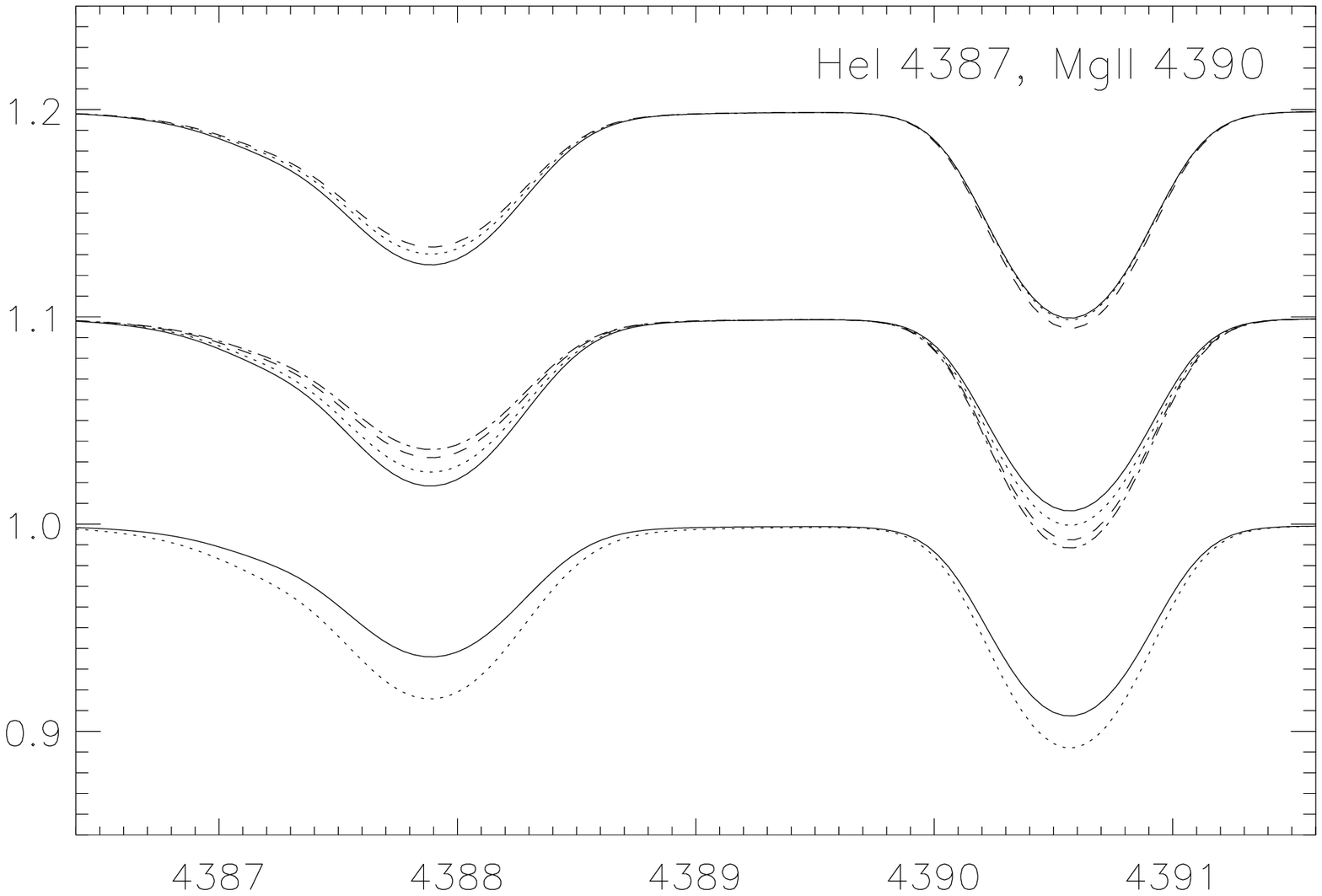}}\\[-4mm]
\resizebox{0.497\hsize}{!}{\includegraphics{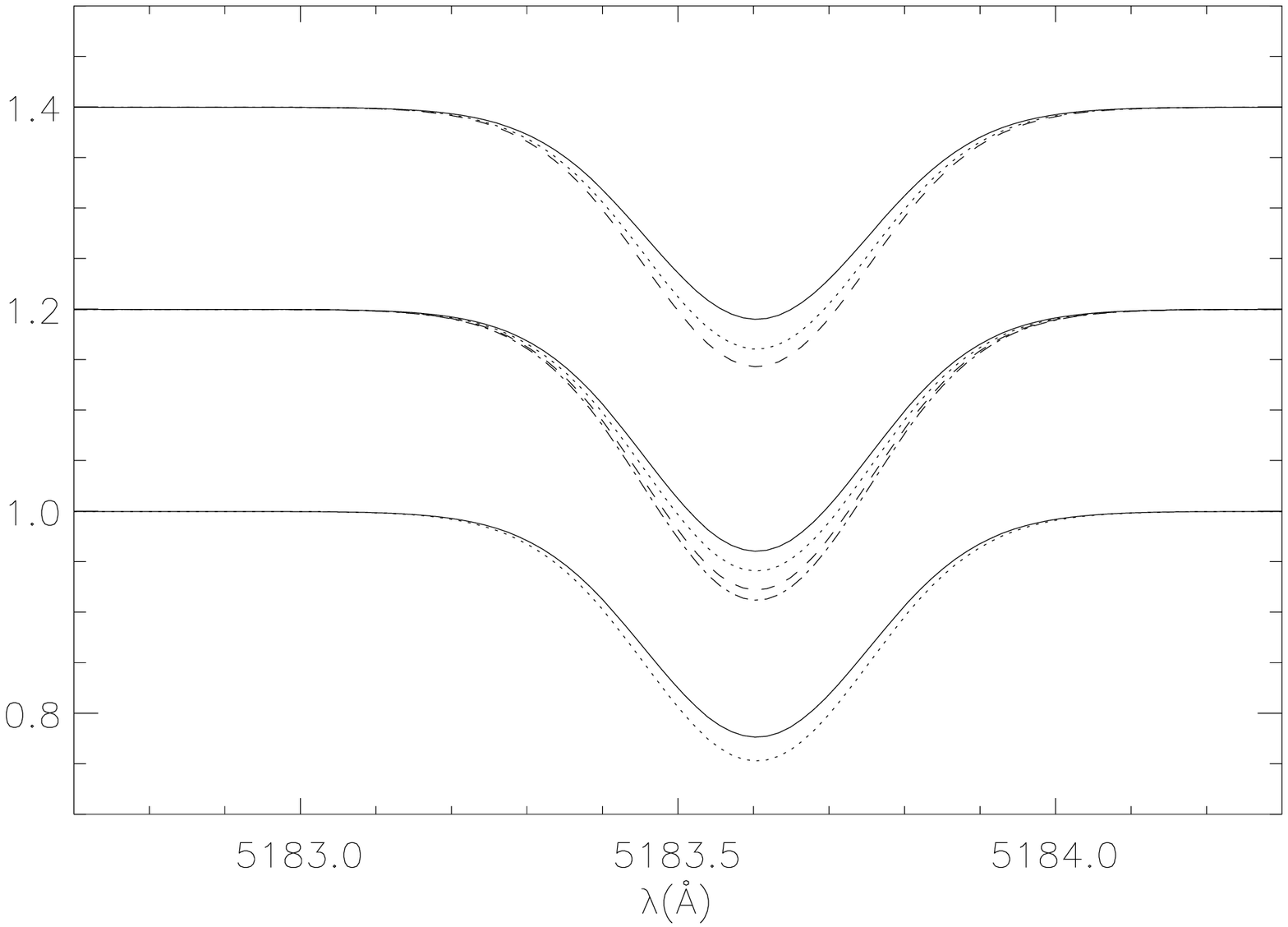}}
\hfill
\resizebox{0.497\hsize}{!}{\includegraphics{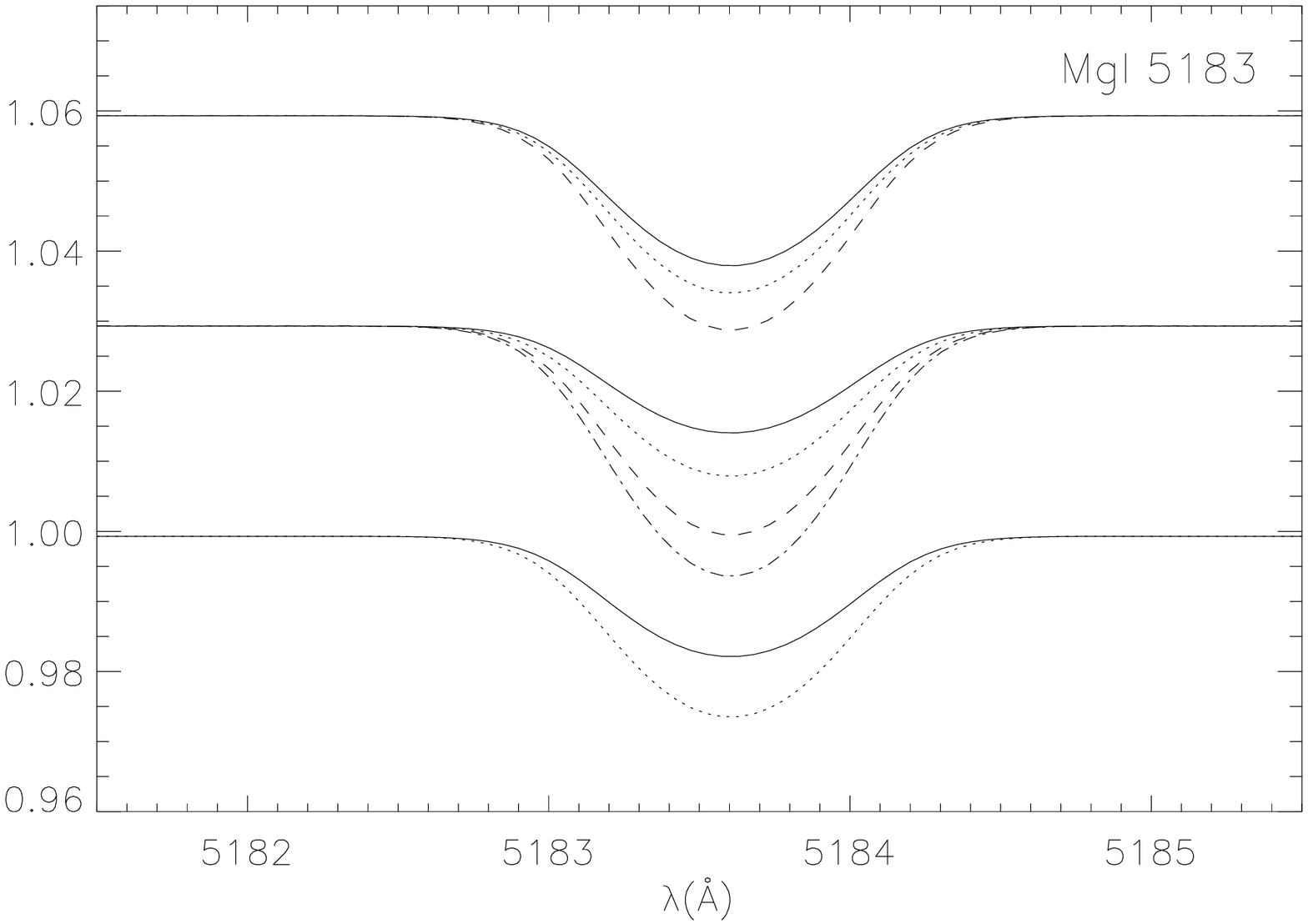}}
\caption[]{Influence of the helium abundance (bottom), metallicity (middle) and 
microturbulence (top set of curves) on the profiles of diagnostic lines for
our models for $\eta$\,Leo and HD\,92207. The
same designations as in Fig.~\ref{atmplot2} are used, and the profiles are
broadened accounting for instrumental profile, rotation and
macroturbulence. Note the strong sensitivity of the lines to these
`secondary' parameters at LC Iae.}
\label{profplot2}
\end{figure*}
Helium lines become visible in main sequence stars at the transition between
A- and B-types at $T_{\rm eff}$\,$\sim$\,10\,000\,K. In supergiants
this boundary is lowered to $\sim$8\,000\,K due to stronger non-LTE effects in
\ion{He}{i} and by the commonly enhanced atmospheric helium abundance in
these stars (see Sect.~\ref{sectabus}). The main effect of an helium enhancement is
the increase of the mean molecular weight of the atmospheric material,
affecting the pressure stratification; the decrease of the opacity and thus
an effect on the temperature structure on the other 
hand is negligible (Kudritzki~\cite{Kudritzki73}). Both effects are
quantified in Fig.~\ref{atmplot2}, where the test is performed for
solar and an enhanced abundance of $y$\,$=$\,0.15, while all other
parameters remain fixed. The relative increase in density strengthens with
decreasing surface gravity. The spectrum analysis of the most luminous
supergiants is distinctly influenced by helium enhancement, while farther
away from the Eddington limit the effects diminish, see Fig.~\ref{profplot2}.
Obviously, the \ion{He}{i} lines are notably strengthened with increasing abundance.
For higher surface gravities, the Balmer lines and the ionization
equilibrium of \ion{Mg}{i/ii} are almost unaffected.
On the other hand, near the Eddington limit the Balmer lines
are noticeably broadened through an increased Stark effect, simulating a
higher surface gravity in less careful analyses. A marked strengthening of the
\ion{Mg}{i} lines is noticed, as the ionization balance is shifted in favour
of the neutral species through the increased electron density. 
In addition, the lines from both
ionic species of magnesium are strengthened by the locally increased absorber
density: the combined effects result in a higher effective
temperature from \ion{Mg}{i/ii}, when helium enhancement is neglected.
All studies of highly
luminous BA-type supergiants (like $\alpha$\,Cyg or $\beta$\,Ori)
to date have neglected atmospheric structure modifications due to
helium enhancements, as these introduce an additional parameter into the
analysis. It is shown that this is not justified: a higher precision in the
stellar parameter determination is obtained in the present work by explicitly 
accounting for this parameter.

Line blanketing is an important factor for atmospheric analyses. However, it
is not only a question of whether line blanketing is considered or not, but
how it is accounted for in detail. The two important parameters to consider
here are metallici\-ty and microturbulence, which both affect the line strengths
and consequently the magnitude of the line blanketing effect. In
studies of supergiants so far this has been neglected.
The introduction of two extra parameters further complicates the analysis
procedure and requires additional iteration steps, but is rewarded in terms
of accuracy.

The influence of the metallicity on the atmospheric line blanketing effect is
displayed in Fig.~\ref{atmplot2}. Here all parameters are kept fixed except the
metallicity of the ODFs used for the model computations. A model sequence
for four metallicities, spanning a range from solar to 0.1\,$\times$\,solar 
abundances, is compared. In the LC Ib supergiant model,
the density structure is hardly affected and the local temperatures in the
line-formation region differ by less than 100\,K for a change of metallicity
within a factor of ten. This difference increases to $\sim$200\,K close to
the Eddington limit: the higher the metallicity, the stronger the
backwarming and the corresponding surface cooling due to line blanketing.
In addition, the density structure is also notably altered,
to a larger extent as in the case of a moderately increased helium abundance.
Radiative acceleration diminishes for decreasing metallicity and thus the
density rises.
The corresponding effects on the line profiles are summarised in
Fig.~\ref{profplot2}. An appreciable effect is noticed only for the highly 
temperature sensitive \ion{He}{i} and
\ion{Mg}{i} lines at LC Ib. Again, at high luminosity all the diagnostic lines are
changed considerably, the extreme case being the \ion{Mg}{i} line which
is strengthened by a factor of almost three. Ignoring the
metallicity effect on the line blanketing in detail will result in
significantly altered stellar parameters from the analysis.

Microturbulence has a similar impact on the line blanketing. 
An increase in the microturbulent velocity $\xi$ strengthens the
backwarming effect as does an increase in the metallicity, since a larger
fraction of the radiative flux is being blocked. The resulting atmospheric
structures from a test with ODFs at three different values for
microturbulence,
for $\xi$\,$=$\,2,\,4 and 8\,km\,s$^{-1}$, and otherwise unchanged parameters
are shown in Fig.~\ref{atmplot2}. In both stellar models the local
temperatures in the line-formation region are increased by $\sim$100\,K when
moving from the lowest to the highest value of $\xi$. A noticeable change of
the density structure is only seen for the LC Iae model.
The corresponding changes in the line profiles are displayed in
Fig.~\ref{profplot2}. Similar effects are found as in the case of
varied metallicity, however they are less pronounced.

The consistent treatment proposed here improves the significance of
analyses, in particular for the most luminous supergiants. This also applies
to the closely related line blocking, which is likewise treated in a
consistent manner. 

\subsection{Neglected effects}
{\em Spherical extension} of the stellar atmosphere is the first of a number of factors 
neglected in the current work. It becomes important in all cases where the atmospheric
thickness is no longer negligible compared to the stellar radius.
Observable quantities like the emergent flux, the colours and line
equivalent widths from extended models will deviate from
plane-parallel results for increasing extension $\eta$ ($=$\,atmospheric
thickness/stellar radius), which can lead to a modified interpretation of
the observed spectra. The expected effects on the line spectrum are (mostly) reduced
equivalent widths due to extra emission from the
extended outer regions and a shift in the ionisation balance. Details on the
differences between spherically extended and plane-parallel hydrostatic 
LTE model atmospheres can be found in Fieldus et al.~(\cite{Fieldusetal90}).
Note that $\eta$ will be on the order of a few percent 
for the photospheres of the objects investigated here (adopting atmospheric
thicknesses from the {\sc Atlas9} models and stellar radii from
Table~\ref{obj}). 

Evidence for macroscopic {\em velocity fields}, i.e. a stellar wind,  
in BA-SG atmospheres is manifold, most obviously from
P-Cygni profiles of strong lines but also from small line asymmetries 
with extra absorption in the blue wing and blue-shifts of the central 
line wavelength in less spectacular cases. 
Kudritzki~(\cite{Kudritzki92}) investigated the influence of
realistic velocity fields from radiation driven winds on the formation of
photospheric lines (in plane-parallel geometry). The subsonic outflow velocity field at the 
base of the stellar wind strengthens lines that are saturated in their cores even for
the moderate mass-loss rates typically observed for BA-SGs.
Desaturation of the lines due to the Doppler shifts experienced by the moving
medium is the driving mechanism for this strengthening.

Velocity fields therefore counteract the proposed line weakening effects of
sphericity, apparently leading to a close net cancellation for the weak line
spectrum. Plane-parallel hydrostatic atmospheres thus seem a good
approximation to spherical and hydrodynamic (unified) atmospheres for the
modelling of BA-SGs, if one concentrates on the photospheric spectrum. In fact, first
results from a comparison of classical LTE with unified non-LTE atmospheres
in the BA-SG regime indicate good agreement (Santolaya-Rey et al.~\cite{SaReetal97}; 
Puls et al.~\cite{Pulsetal05}; J.~Puls, private communication). The
situation appears to be similar in early B-supergiants (Dufton et
al.~\cite{Duftonetal05})

BA-type supergiants have been known as photometric and optical spectrum
variables for a long time. The most comprehensive study to date 
in this context is that of Kaufer et al.~(\cite{Kauferetal96}, \cite{Kauferetal97}).
Additional observational findings in the UV spectral
region, in particular of the \ion{Mg}{ii} and \ion{Fe}{ii} resonance lines,
are presented by Talavera \& Gomez de Castro~(\cite{TaGo87}) and Verdugo et
al.~(\cite{Verdugoetal99a}). 
Kaufer et al.~(\cite{Kauferetal97}) find peak-to-peak amplitude variations 
of the line strength of 29\% on time scales of years.
Unaccounted {\em variability} is therefore a
potential source of inconsistencies when analysing data
from different epochs, see Sect.~\ref{sectparams} for our 
approach to overcome this limitation.

BA-type supergiants are slow rotators with typical observed values of $v\sin i$
between 30 to 50\,km\,s$^{-1}$ (Verdugo et al.~\cite{Verdugoetal99b}).
The modelling can therefore be treated as a {\em 1-D problem}, as rotationally
induced oblateness of the stars is insignificant. Finally, {\em magnetic fields} in BA-SGs 
appear to be too weak to cause atmospheric inhomogeneities. For $\beta$\,Ori
a weak longitudinal magnetic field, on the order of 100\,G
was observed by Severny~(\cite{Severny70}). Little information on other
objects is available.

\subsection{Limits of the analyses}
The spectrum synthesis technique described in the present work is applicable
to a rather wide range of stellar parameters, but nevertheless it is
restricted. Its scope of validity principally concentrates on
BA-SGs and related objects of lower luminosity. This is mainly due to the limits 
posed by the underlying atmospheric models and the atomic models implemented.

From estimates, such as that presented by Kudritzki~(\cite{Kudritzki88},
Fig.\,{\sc iii},\,9), it is inferred that non-LTE effects on the atmospheric
structure, increasing with stellar effective temperature, will inhibit
analyses with the present technique of any supergiants above 
$\sim$20\,000\,K, i.e. in the early B-types. However, main sequence stars and 
even (sub-)giants of such spectral type are analysed with classical atmospheric 
models on a routine basis.
For the less-luminous supergiants of mid B-type the method may still be
applicable. This requires further investigation (including an extension of
the model atom database to doubly-ionized species), but it can be expected to fail 
at higher luminosity. As Dufton et al.~(\cite{Duftonetal05}) indicate that
classical and unified {\em non-LTE} atmosphere analyses give similar results for
early B-SGs, the solution for analyses of highly luminous mid-B supergiants
will be to use non-LTE line-formation computations based on classical line-blanketed 
non-LTE (instead of LTE) model atmospheres.

The lower limit (in $T_{\rm eff}$) for the applicability of the method
is determined by several factors. At $T_{\rm eff}$\,$\simeq$\,8\,000\,K 
helium lines disappear in the spectra of A-supergiants. Thus the helium abundance has to 
remain undetermined, introducing some uncertainties into the analyses. 
Around the same temperature convection can be expected to set in, as shown by
Simon et al.~(\cite{Simonetal02}) for main sequence stars, who place the
boundary line for convection at $T_\mathrm{eff}$\,$\simeq$\,8\,250\,K
(however, no information is available for A-supergiants).
Since the theoretical considerations of Schwarzschild~(\cite{Schwarzschild75}), 
which predict only a small number of giant convection cells scattered over
the stellar surface, atmospheric convection in supergiants has attracted little interest 
until recently. Interferometric observations of the late-type supergiant $\alpha$\,Ori
(Young et al.~\cite{Youngetal00}) can be interpreted in favour of this,
further strengthened by first results from~\mbox{3-D} stellar convection models
(Freytag et al.~\cite{Freytagetal02}). Again, no quantitative information on
this is available for mid and late A-supergiants, resulting in a potential source of
systematic error for model atmosphere and line formation computations. Furthermore, 
convective stellar envelopes give rise to chromospheres (see Dupree et
al.~(\cite{Dupreeetal05}) for observational evidence in the cooler luminous
stars), which introduce an additional
source of UV irradiance, altering the non-LTE populations of the 
{\em photospheric} hydrogen (as compared to models without chromospheres) and
potentially affecting the H$^-$ opacity and thus the stellar continuum 
(Przybilla \& Butler~\cite{PrBu04b}). 

\begin{figure}
\resizebox{\hsize}{!}{\includegraphics{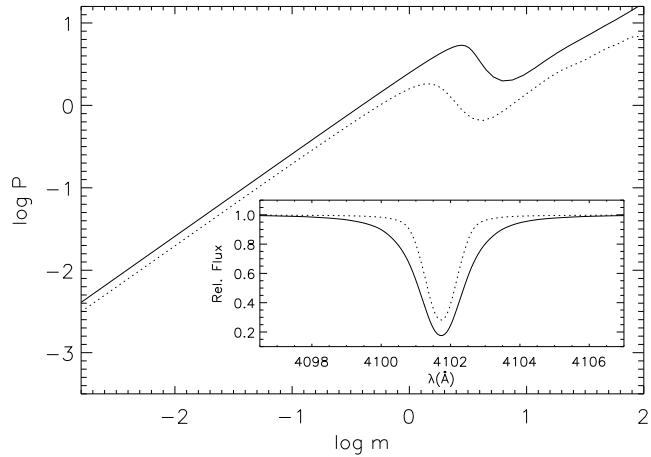}}\\[-6mm]
\caption[]{Example for the occurrence of pressure inversion in supergiants of spectral types
later than $\sim$A4. The pressure stratification as a function of
mass scale is displayed for two {\sc Atlas9} models for $T_\mathrm{eff}$\,$=$\,7\,700\,K
and $\log g$\,$=$\,0.42 (the computed Eddington limit, dotted line) and 
0.47 (full line). The inset compares
the respective H$\delta$ profiles. A change in $\log g$ by 0.05\,dex results
in equivalent widths differing by a factor of 2 -- an artefact of the pressure
bump in the line-formation depth and its impact on the Stark broadening.}
\label{pressureinversion}
\end{figure}

Moreover, supergiant model atmospheres cooler than
$T_\mathrm{eff}$ $\sim$\,8\,200\,K are
characterised by {\em pressure inversion} (and accompanying density inversion).
Pressure inversion can develop close to the Eddington limit when 
radiative acceleration {\em locally} dominates
over gravity because of an opacity bump in the hydrogen ionization zone,
which is located in the photosphere at these effective temperatures. An example of the effect 
is shown in Fig.~\ref{pressureinversion}, where two {\sc Atlas9} models
close to the hotter end of the pressure inversion regime are compared. 
A small change of surface gravity by
$\sim$10\% results in a drastic change of a factor of 2 in equivalent width of the
Balmer lines in this (extreme) case. Despite a general strong sensitivity of 
the hydrogen line equivalent widths to surface gravity close to the Eddington limit 
(see Sect.~\ref{sectparams}) this huge effect is an artefact of the
modelling for the most part. 
Another possible effect of pressure inversion on line-formation computations concerns
deviations from generally inferred trends in the behaviour of
line strengths with stellar parameter variations. At slightly hotter
temperatures the hydrogen line strengths decrease with decreasing surface
gravity. In the pressure inversion regime this trend can be compensated, and even reversed,
by a developing pressure bump. Naturally, pressure inversion affects all spectral features
with line-formation depths coinciding with the pressure bump because of
its effect on absorber densities. Systematic errors for stellar parameters
from ionization equilibria and chemical abundance determinations can be
expected.
Pressure inversion is also discussed in the context of hydrodynamical models 
(Achmad et al.~\cite{Achmadetal97}; Asplund~\cite{Asplund98}).
It is not removed by stationary mass outflow (except for very high 
mass-loss rates not supported by observation). It is not initiating the
stellar wind either.
Abolishing the assumption of {\em stationarity} provides a solution to the problem
as discussed by de Jager~(\cite{deJager98}, and references therein), leading
to pulsations and enhanced mass-loss in the yellow super- and hypergiants, 
which are located in a sparsely populated region of the 
empiric Hertzsprung-Russell diagram. 

From these considerations we restrict ourselves to supergiant analyses at
$T_\mathrm{eff}$\,$\gtrsim$\,8\,000\,K using the methodology presented here.
In our opinion
extension to studies of supergiants at cooler temperatures (of spectral
types mid/late A, F, G, and of Cepheids) requires additional
theoretical efforts in stellar atmosphere modelling, far
beyond the scope of the present work. 
At lower luminosities the problems largely diminish so that the hybrid
non-LTE technique provides an opportunity to improve on the accuracy of
stellar analyses over a large and important part of the Hertzsprung-Russell
diagram. Non-LTE model atoms for many of the astrophysically interesting 
chemical species are already available for studies of stars of later
spectral types, as will be discussed next.


\section{Statistical equilibrium and line formation}\label{sectlform}
\begin{table}
\caption[]{Non-LTE model atoms}
\label{atoms}
\begin{tabular}{ll}
\hline
Ion & Source\\
\hline
\ion{H}{} & Przybilla \& Butler (\cite{PrBu04a})\\
\ion{He}{i} & Husfeld et al. (\cite{Husfeldetal89}), with updated atomic
data\\
\ion{C}{i/ii} & Przybilla et al. (\cite{Przybillaetal01b})\\
\ion{N}{i/ii} & Przybilla \& Butler (\cite{PrBu01})\\
\ion{O}{i/ii} & Przybilla et al. (\cite{Przybillaetal00}) combined with Becker \& \\
& Butler (\cite{BeBu88}), the latter with updated atomic data\\
\ion{Mg}{i/ii} & Przybilla et al. (\cite{Przybillaetal01a})\\
\ion{S}{ii/iii} & Vrancken et al. (\cite{Vranckenetal96}), with updated
atomic data\\
\ion{Ti}{ii} & Becker (\cite{Becker98})\\
\ion{Fe}{ii} & Becker (\cite{Becker98})\\
\hline
\end{tabular}
\end{table}

Detailed quantitative analyses of stellar spectra require another
ingredient besides realistic stellar model atmospheres: an accurate modelling 
of the line-formation process. While our restricted modelling capacities do
not allow the overall problem to be solved in a completely natural fashion,
i.e. simultaneously, 
experience tells us that we can split the task into several steps.
In particular, while non-LTE effects are present in {\em all} cases, as photons are
leaving the stellar atmosphere, they may be of little importance for the
atmospheric structure when the main opacity sources remain close to LTE 
(H and He in early-type stars). Minor species -- this includes trace elements 
as well as high-excitation levels of hydrogen and helium -- behave in this
way. They can be treated in a rather coarse approximation for
atmospheric structure computations, while detailed non-LTE calculations may be required 
for the analysis of their spectra in order to use them as stellar parameter or
abundance indicators. We therefore chose such a hybrid approach for our analyses: 
based on line-blanketed classical model atmospheres we solve the
statistical equilibrium and the radiative transfer problem for
individual species in great detail. The derived level occupations are then used 
in the formal solution to calculate the emergent flux, considering exact line-broadening.  
The last two steps are performed using the non-LTE line-formation package {\sc Detail}
and {\sc Surface} (Giddings~\cite{Giddings81}; Butler \& Giddings~\cite{BuGi85}), 
which has undergone substantial extension and improvement over the years.
In our context the inclusion of an Accelerated Lambda Iteration (ALI) scheme
(Rybicki \& Hummer~\cite{RyHu91}) is of primary interest, as it allows
elaborate non-LTE model atoms to be used while keeping computational expenses
moderate. Line blocking is taken into account via ODFs. 
Special care is required in computations for major line opacity sources, like iron: in order
to prevent counting the line opacity twice (via the ODFs and as radiative
transitions in the statistical equilibrium computations) we adopt ODFs with
a metallicity reduced by a factor 2. This approximately corrects for the
contribution of the element to the total line opacity.

\begin{figure}
\resizebox{\hsize}{!}{\includegraphics{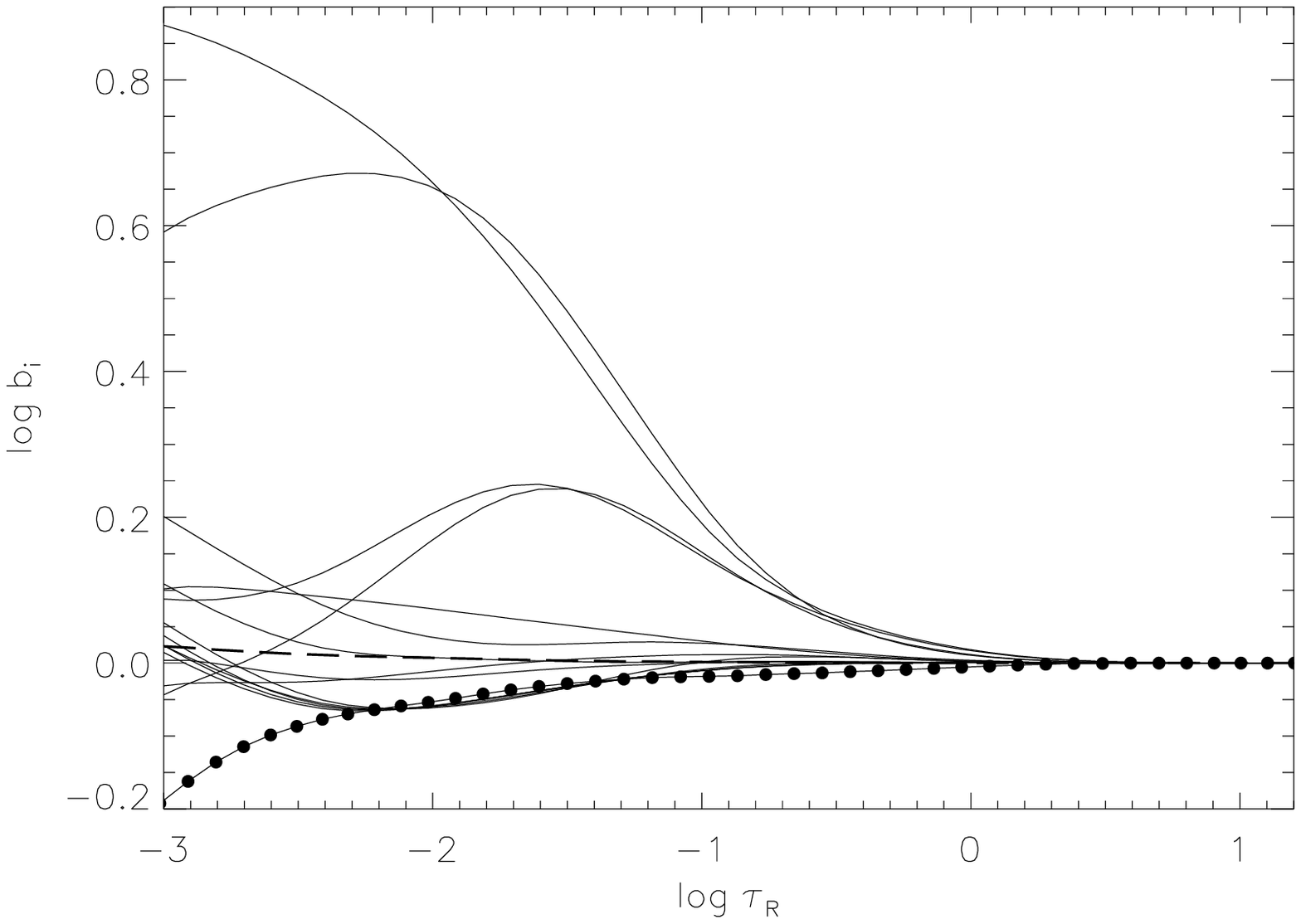}}\\[-.5cm]
\resizebox{\hsize}{!}{\includegraphics{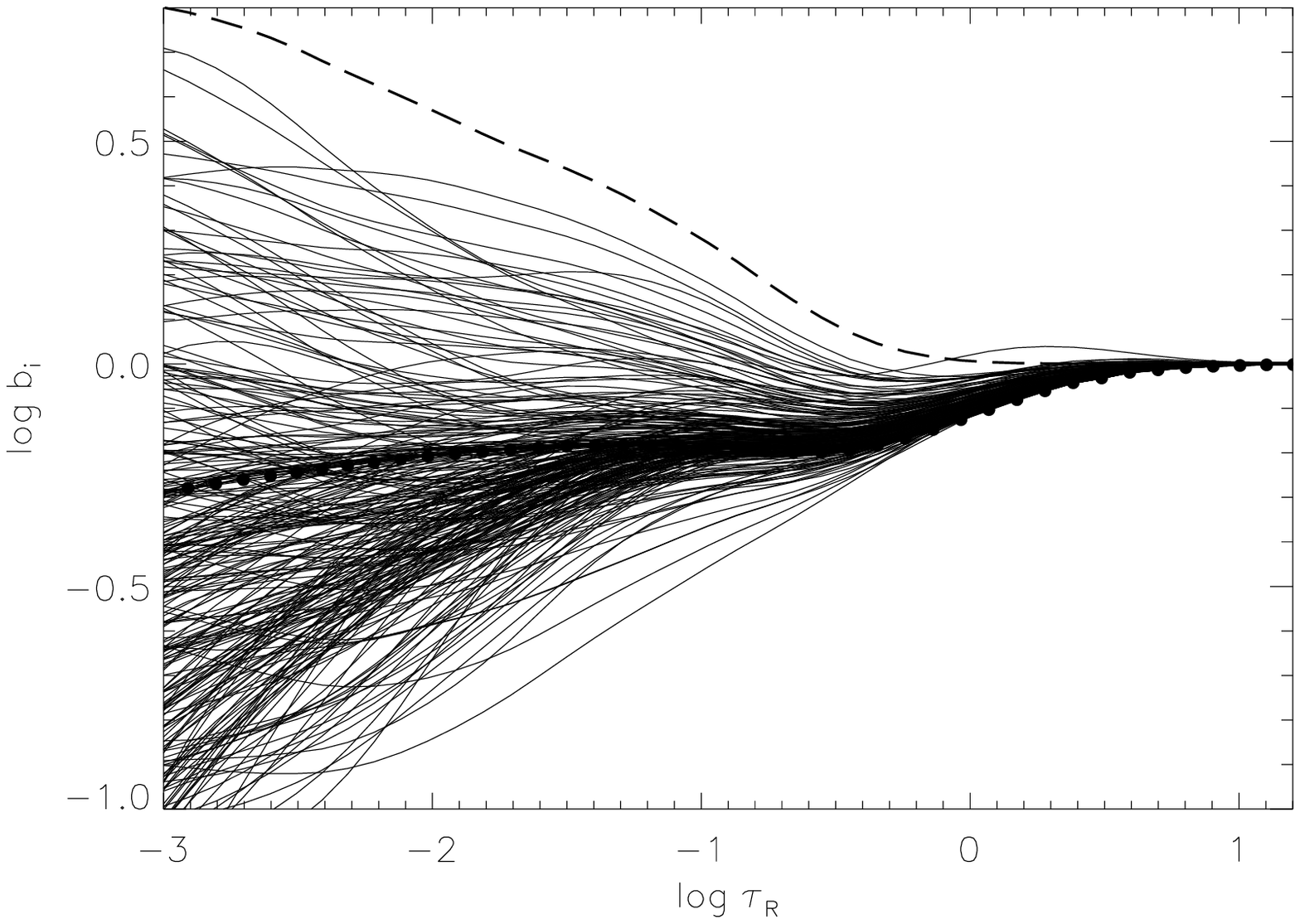}}
\caption[]{Comparison of non-LTE departure coefficients $b_i$ for \ion{O}{i}
(top) and \ion{Fe}{ii} (bottom) in HD\,92207 as a function of 
$\tau_{\rm R}$. Displayed are the departure coefficients of the ground state
(thick-dotted line), of the ground state of the next higher ion (long-dashed
line) and those of the levels from which the diagnostic spectral lines in the 
optical/near-IR arise~(\ion{O}{i}). In the case of \ion{Fe}{ii} the departure coefficients of
all non-LTE levels are shown. The two ionic species are representative for the non-LTE 
behaviour of the light (including the $\alpha$-process elements) and iron 
group elements, see the text for further discussion.\\[-.4cm]}
\label{departures}
\end{figure}

At the centre of our hybrid non-LTE analysis technique stand sophisticated 
model atoms, which comprise many of the most important elements in the astrophysical 
context. An overview is given in Table~\ref{atoms}, summarising the
references where the model atoms and their non-LTE behaviour were
discussed in detail. In brief, they are characterised by the use of accurate
atomic data, replacing approximate data as typically used in such work, by
experimental data (a minority) or data from quantum-mechanical {\em ab-initio}
computations (the bulk). We have profited from the efforts of
the Opacity Project (OP; Seaton et al.~\cite{Seatonetal94}) and the IRON
Project (IP; Hummer et al.~\cite{Hummeretal93}), as well as numerous other
works from the physics literature. In particular, a major difference to
previous efforts is the use of large sets of accurate data for 
excitations via electron collisions. Thus, not only the (non-local) radiative processes
driving the plasma out of LTE are treated realistically
(line blocking is considered via Kurucz~(\cite{Kurucz92}) ODFs), but also the
competing (local) processes of thermalising collisions. These are
essential to
bring line analyses from different spin systems of an atom/ion into agreement.
A few of the older models have been updated/extended with respect to the
original publications. In the case of \ion{S}{ii/iii} the fits to 
the original photoionization cross-sections of the OP data (neglecting resonances due
to autoionizing states) have been replaced by the detailed data, and for
these ions, as well as for \ion{He}{i} and \ion{O}{ii} the features treated
in the line-formation computations have been extended, as well as oscillator strengths 
and broadening parameters updated to more modern values, see
Appendix~\ref{apa} for further detail.
In the following we will discuss some more general conclusions, which can be drawn from 
the analysis of such a comprehensive set of model atoms, while referring the
reader to the original publications (see Table~\ref{atoms}) for the details
of individual atoms/ions.

There is a fundamental difference between the non-LTE behaviour of the light and 
$\alpha$-process elements on the one hand and the iron group elements on the
other. The first group is characterised by a few valence electrons, which
couple to only a few low-excitation states in the ground configuration
that are separated by a large energy gap (several eV) from the higher
excited levels. These excited states in turn show a similar structure on a
smaller scale: a few (pseudo-)metastable levels are detached from the
remainder by a $\sim$2\,eV gap, see e.g. the Grotrian diagrams in the
references given in Table~\ref{atoms}.
Collisional processes are effective in coupling levels 
either below or above the energy gaps, such that these are in or close to LTE {\em
relative} to each other. The levels of highest excitation couple to the ground state
of the next higher ionization stage via collisions. On the other hand, 
only a few electrons in the high-velocity tail of the Maxwell distribution are 
energetic enough in the atmospheres of BA-type stars to pass the first gap, and 
considerably more, but a minority nonetheless, the second gap. This
favours strong non-LTE overpopulations of the excited (pseudo-)metastable states,
as shown exemplarily for \ion{O}{i} in Fig.~\ref{departures}, manifested
in non-LTE departure coefficients $b_i$\,$=$\,$n_i^{\rm NLTE}/n_i^{\rm
LTE}$\,$>$\,1, where the $n_i$ are non-LTE and LTE level occupation numbers,
respectively. Therefore, {\em the diagnostic lines in the optical/near-IR from the light and 
$\alpha$-process elements are typically subject to non-LTE strengthening}.

On the other hand, the electrons in the open 3$d$-shell of the iron group
elements give rise to numerous energetically close levels throughout the
whole atomic structure. Photoionizations from the ground states of the single-ionized
iron group elements are typically not very effective, as the ionization
thresholds fall short of the Lyman limit where the stellar flux is negligible. 
The situation is different for the  well populated levels a few eV above
the ground state. They show the largest non-LTE depopulations (see
Fig.~\ref{departures} for the example of \ion{Fe}{ii}). 
Because of the collisional coupling the ground
states also become depopulated, but to a lower degree. Nonetheless, a net
overionization is established, resulting in a non-LTE overpopulation of 
the main ionization stage (\ion{Fe}{iii} in Fig.~\ref{departures} -- note
that no \ion{Fe}{iii} lines are observed in the optical/near-IR for this star). Again,
the highest excitation levels of the lower ionization stage couple
collisionally to this and are thus also overpopulated. A continuous
distribution of departure coefficients results, with the $b_i$ of
the upper levels of the transitions typically larger than those of the lower
levels. Consequently, {\em the optical/near-IR lines from the iron
group elements experience non-LTE weakening}. Therefore, {\em LTE analyses
of luminous BA-SGs will tend to systematically overestimate abundances of the
light and $\alpha$-process elements, and to underestimate abundances of the
iron group elements}. These effects will be quantified in
Sect.~\ref{sectabus} where the comparison of non-LTE and LTE computations
with observation is made. Note however that non-LTE computations, when
performed with inaccurate atomic data, also bear the risk of introducing
systematic errors to the analysis, see e.g. the discussions in Przybilla \&
Butler~(\cite{PrBu01}, \cite{PrBu04a}). This is because of the nature of
statistical equilibrium, where (de-)population mechanisms couple all energy
levels with each other. The predictive power of non-LTE 
computations is therefore only as good as the models atoms used. 

The spectral lines treated in our non-LTE approach comprise about $\sim$70\% of the
observed optical/near-IR features in BA-SGs. This includes in particular
most of those of large and intermediate line strength. The remainder of the
observed lines is typically weak, except for several \ion{Si}{ii} and \ion{Cr}{ii}
transitions. In order to achieve (near) completeness we incorporate these and another 
ten chemical species into our spectrum synthesis in LTE. It can be expected that this 
approach will introduce some systematic deficiencies, in particular for the
most luminous objects, as will be indicated by the analysis in Sect.~\ref{sectabus}. 
However, the solution for high-resolution observations is to draw no further
conclusions from these species. Interpretation of medium-resolution
spectra (see Sect.~\ref{sectmedres}) on the other hand would suffer more
from their absence than from their less accurate treatment, as they
typically contribute only small blends to the main diagnostic features.

The currently used line lists comprise several ten-thousand transitions, 
covering the classical blue region for spectral analyses
between $\sim$4\,000 and 5\,000\,{\AA} well. At longer wavelengths a number of features
are missing in our spectrum synthesis computations, mostly lines from
highly-excited levels of the iron group elements. However, these are
intrinsically weak (with equivalent widths $W_{\lambda}$\,$\lesssim$\,10\,{m\AA}) 
and typically isolated, so that their absence will hardly be noticed when
compared to intermediate-resolution observations.

\begin{figure*}
\resizebox{0.497\hsize}{!}{\includegraphics{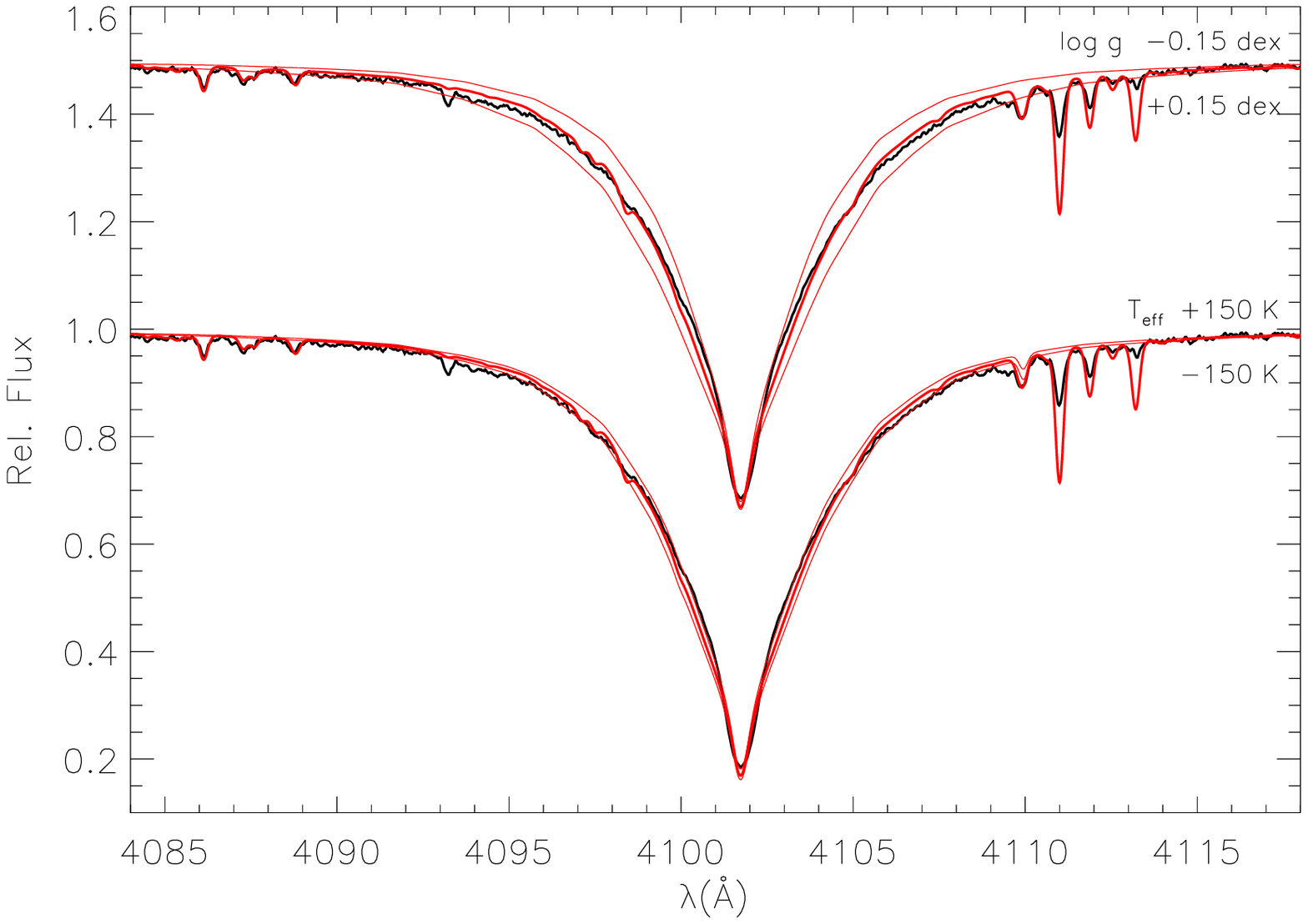}}
\hfill
\resizebox{0.497\hsize}{!}{\includegraphics{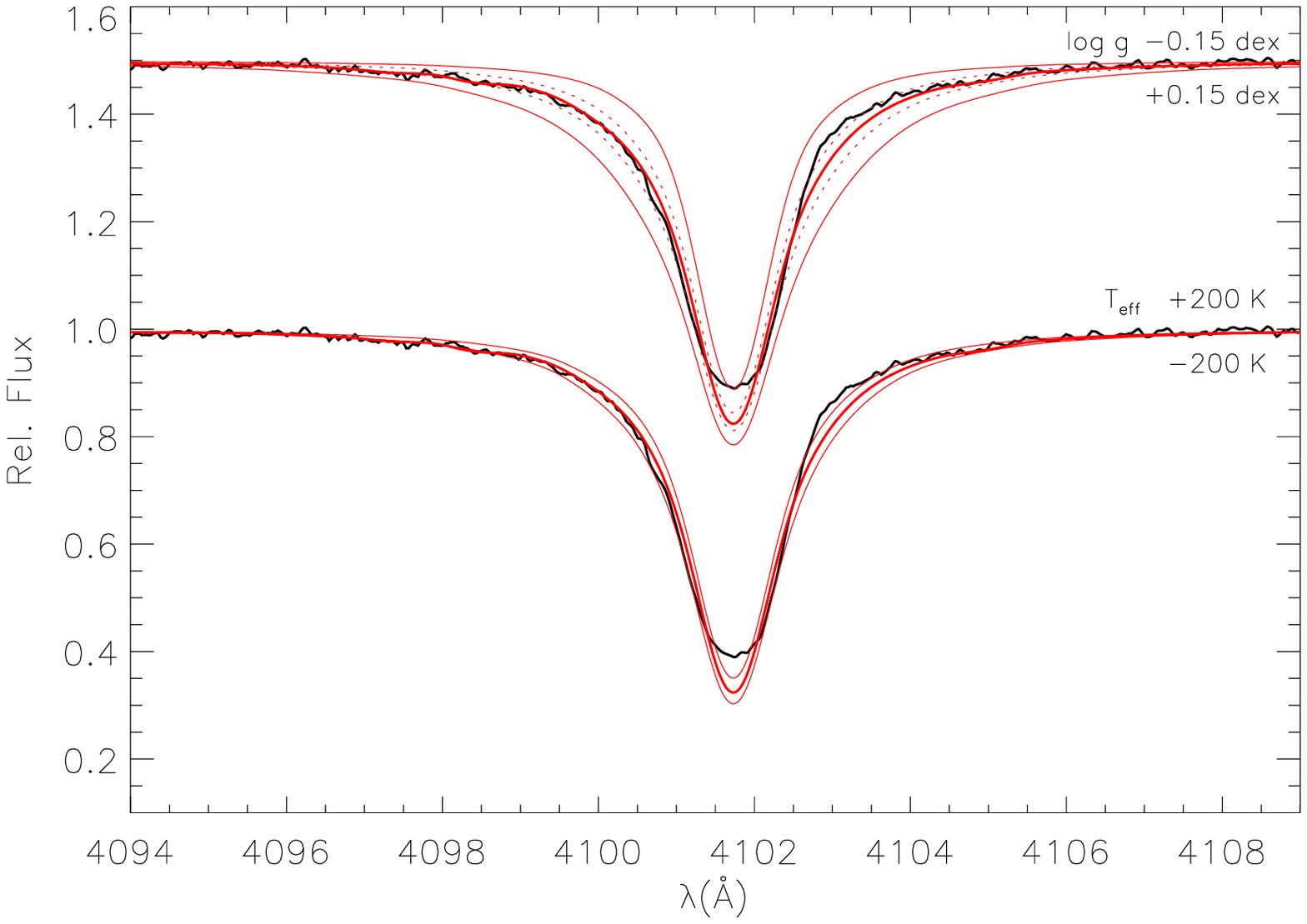}}\\[-8mm]
\caption[]{Impact of stellar parameter variations on non-LTE line profile fits 
for H$\delta$ in $\eta$\,Leo (left) and HD\,92207 (right). Spectrum
synthesis for the adopted parameters (see Table~\ref{obj}, thick red line) and
varied parameters, as indicated (thin lines), are compared to observation. 
A vertical shift by 0.5\,units has been applied to the upper
profiles. Note the decrease of the Balmer line strength in the progression
from LC Ib ($\eta$\,Leo) to Iae (HD\,92207), and the increased sensitivity
to the surface gravity parameter. In fact, the comparison indicates that 
for the more luminous supergiants uncertainties in the surface gravity
determination become as low as 0.05\,dex (dotted lines). The sensitivity to reasonable 
$T_{\rm eff}$-changes is almost negligible. Wind emission slightly contaminates the
red wing of H$\delta$ in HD\,92207.}
\label{loggtest}
\end{figure*}

Where it was once a supercomputing application, comprehensive non-LTE modelling like the
present can now be made on workstations or PCs. Typical running times to
achieve convergence in the statistical equilibrium and radiative transfer 
calculations with {\sc Detail} range from $\sim$10\,min for the most simple 
model atoms to a couple of hours for the \ion{Fe}{ii} model on a 3\,GHz P4 CPU -- 
for one set of parameters. The formal solution for a total of
2--3\,$\times$10$^5$ frequency points with {\sc Surface} requires $\sim$20--30\,CPU min. 
Because of the highly iterative and interactive nature of our analysis
procedure (see next two sections) the total time for a comprehensive study
of {\em one} high-resolution, high-S/N spectrum, assuming excellent wavelength coverage 
from around the 
Balmer jump to $\sim$9\,000\,{\AA} as typically achieved by modern Echelle spectrographs,
amounts to {\em 1--2 weeks} for the experienced user. This is the price to pay for overcoming the
restrictions of contemporary BA-SG abundance studies and advancing them 
from factor 2--3 accuracy astronomy to precision astrophysics.

Future extensions of the present work will aim to provide more non-LTE model atoms, 
but these quite often require the necessary atomic data to be calculated 
first, as many data are still unavailable in present-day literature. In
particular, the status of collisional data has largely to be improved. 
Note that the \ion{Si}{ii} ion in the silicon model atom by Becker \&
Butler~(\cite{BeBu90}) is only rudimentary (though sufficient
for their main purpose, analyses of early B-stars). Several energy
levels involved in the observed transitions in BA-SGs are missing, and test
calculations indicate a wide spread of the non-LTE abundances from the
remaining transitions. On the basis of these findings we refrain from using
the model atom for analyses in the present work, but wish to emphasise that
qualitatively the correct non-LTE behaviour -- line strengthening -- is
predicted.


\section{Determination of stellar parameters}\label{sectparams}
Stellar parameter determination for BA-SGs is complicated because of the
intrinsic variability of these objects. Observational data from different
epochs may therefore reflect different physical states of the stellar atmosphere, 
unless these are obtained (nearly) simultaneously. While we can expect the mass of a
supergiant and its surface abundances to be conserved on human timescales,
other stellar parameters may not. The observed variability patterns of light
curves, radial velocities and line profiles can be
interpreted in favour of a mix of radial and non-radial pulsations (see e.g.
Kaufer et al.~\cite{Kauferetal97}), which may affect $T_{\rm eff}$ and 
$\log g$, that in turn determine colours, the spectral energy distribution and 
the spectra. Even the stellar luminosity may be subject to small changes
(Dorfi \& Gautschy~\cite{DoGa00}). The
parameter variations are not as pronounced as in other variable stars, but
use of information from different epochs can potentially introduce
systematic uncertainties into high-precision analyses. A solution is to derive
the desired quantities from {\em one} set of observational data. Echelle spectra 
are the best choice, containing all information required for the derivation of
stellar parameters and elemental abundances. With few exceptions, use of 
photometric data (from other sources) can thus be avoided. In the following we discuss the
details of our spectroscopic approach for the parameter determination of
BA-SGs. Spectrophotometry is only briefly considered for consistency checks.
Finally, our findings are compared with previous analyses of the two well
studied standards $\eta$\,Leo and $\beta$\,Ori.

\begin{figure*}
\resizebox{0.497\hsize}{!}{\includegraphics{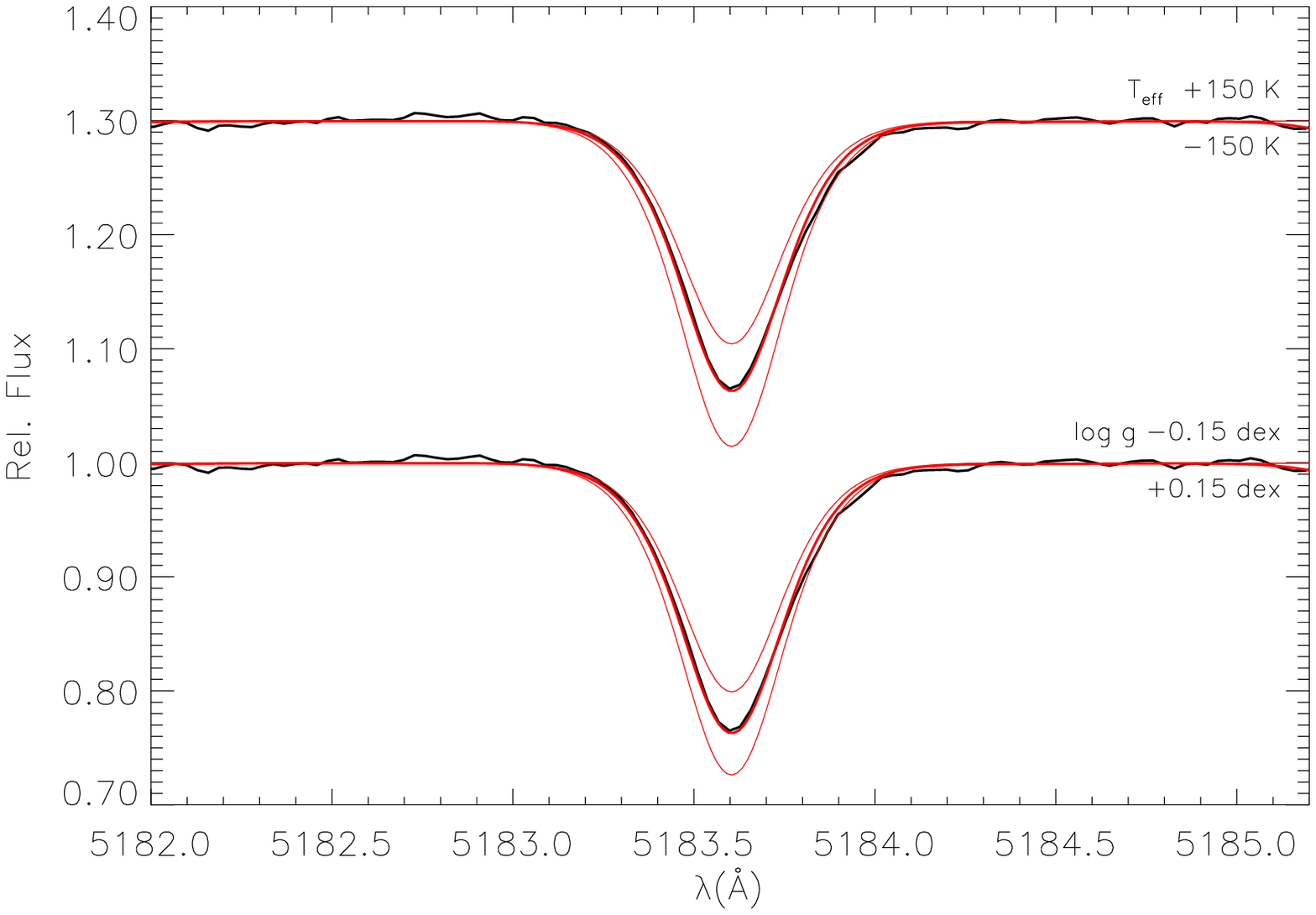}}
\hfill
\resizebox{0.497\hsize}{!}{\includegraphics{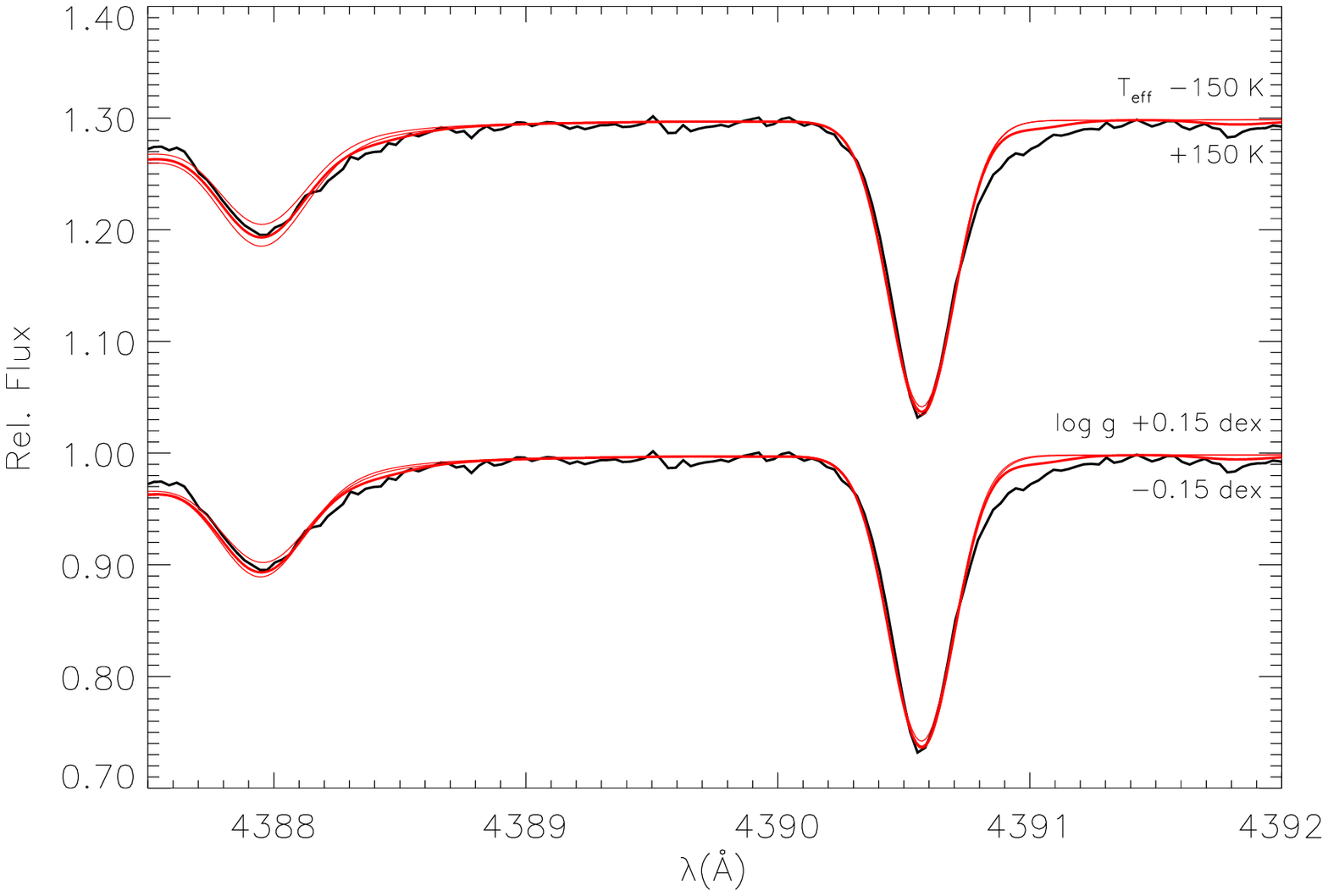}}\\[-7mm]
\caption[]{Temperature determination for $\eta$\,Leo using the \ion{Mg}{i/ii}
non-LTE ionization equilibrium. Displayed are the
observed line profiles for some of the strategic
lines, \ion{Mg}{i}
$\lambda$\,5183 (left panel) and \ion{Mg}{ii} $\lambda$\,4390
and \ion{He}{i} $\lambda$\,4387 (right panel), 
and the best non-LTE fit for stellar parameters as given in
Table~\ref{obj}
(thick red line); theoretical profiles for varied
parameters are also shown (thin lines, as indicated).
A vertical shift by 0.3\,units has been applied to the upper set of profiles.
Note the strong sensitivity of the minor ionic species, \ion{Mg}{i}, to
parameter changes, while \ion{Mg}{ii} is virtually unaffected.}
\label{ttest}
\end{figure*}

\begin{figure*}
\resizebox{0.497\hsize}{!}{\includegraphics{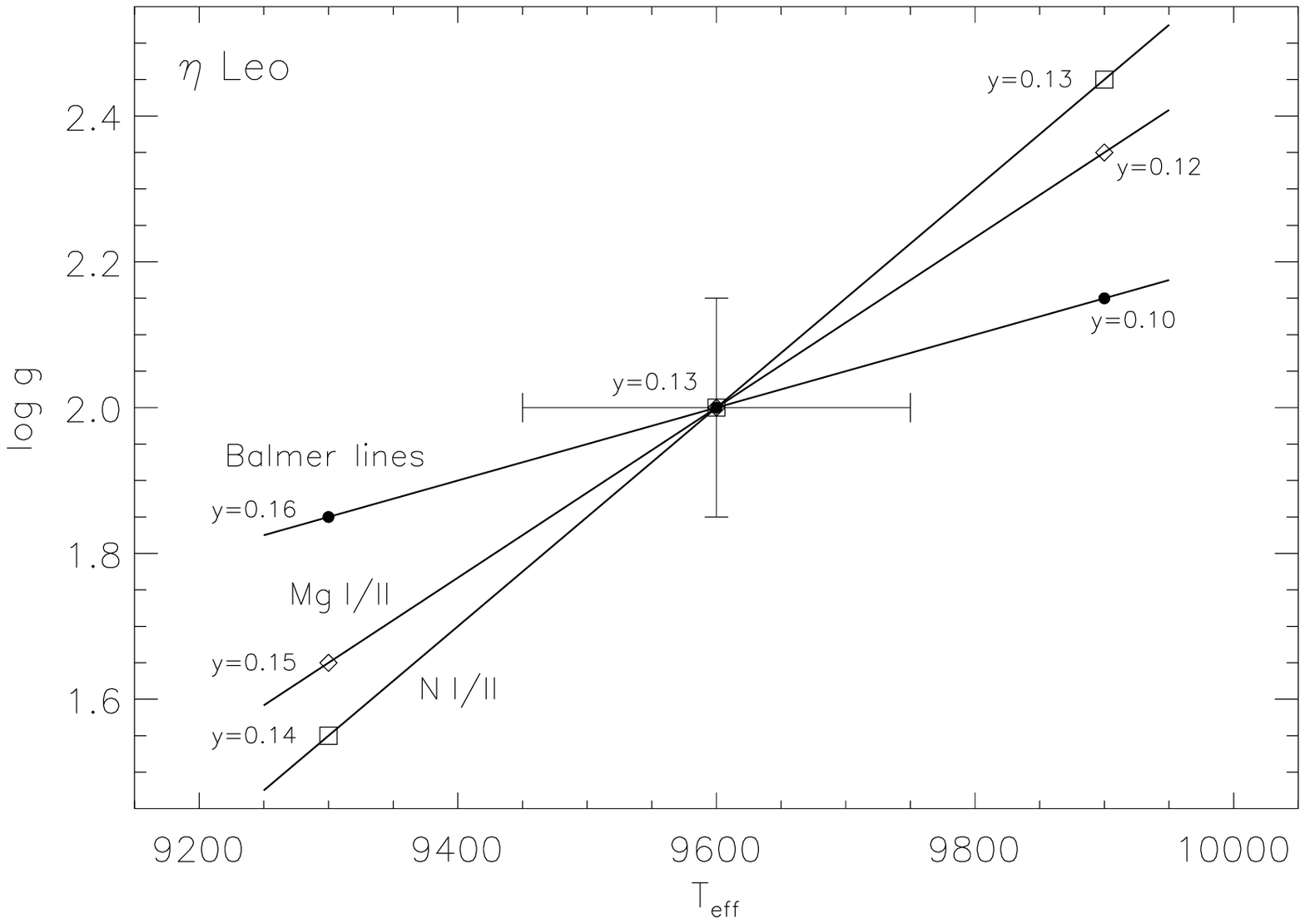}}
\hfill
\resizebox{0.497\hsize}{!}{\includegraphics{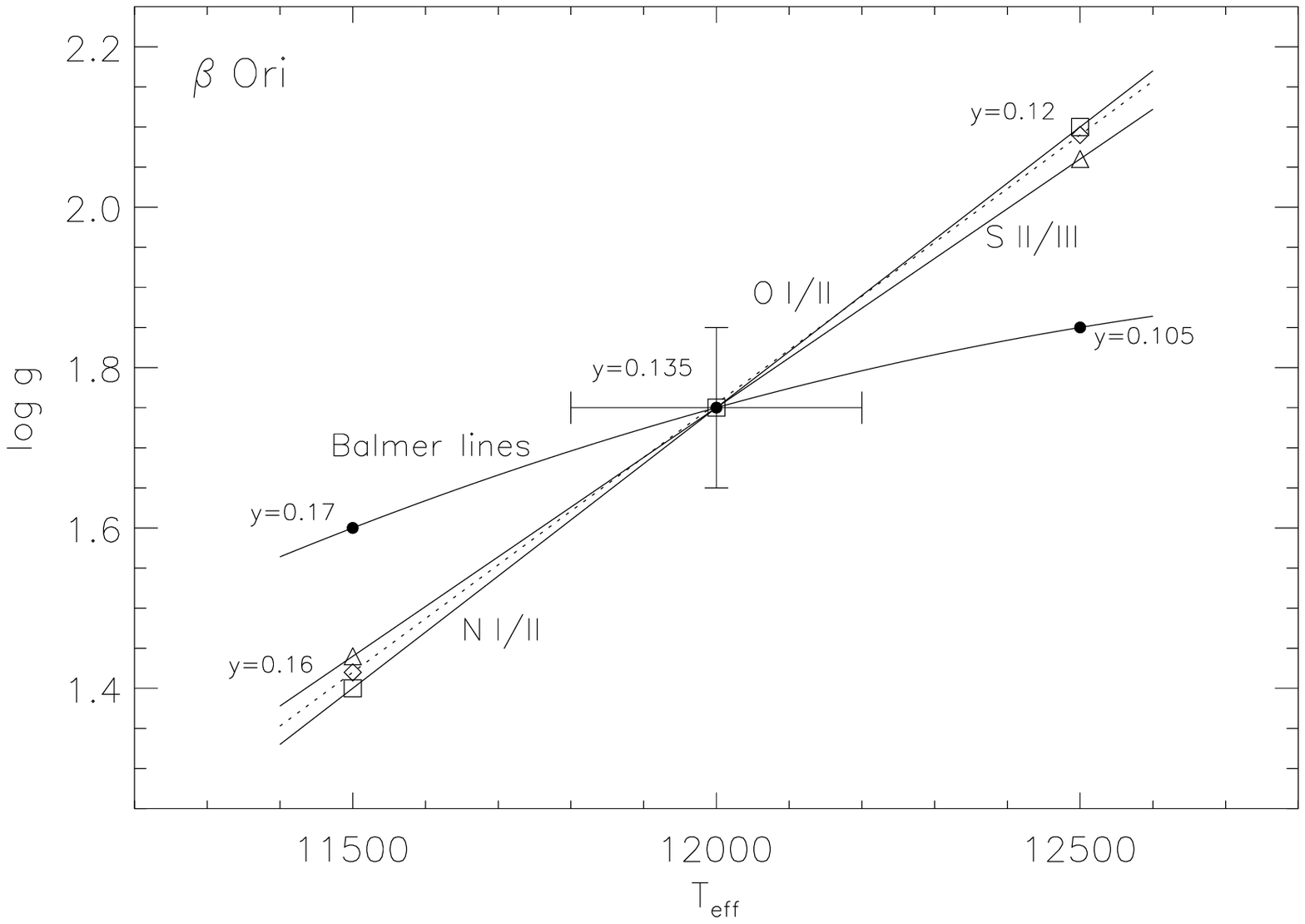}}\\[-7mm]
\caption[]{Fit diagrams of temperature- and gravity-sensitive indicators for
$\eta$\,Leo and $\beta$\,Ori in the $T_{\rm eff}$--$\log g$-plane. The
curves are parameterised by surface helium abundance $y$. The intersection
determines $T_{\rm eff}$ and $\log g$, with uncertainties as indicated.}
\label{kiel}
\end{figure*}

\subsection{Spectroscopic indicators}
The hydrogen lines are the most noticeable features in the spectra of
BA-type stars. Their broadening by the linear Stark effect gives a sensitive
surface gravity indicator in the mostly ionized atmospheric plasma.
We employ the recent Stark broadening tables 
of Stehl\'e \& Hutcheon~(\cite{SH99}, SH), which compare well 
with the classically used data from 
Vidal et al.~(\cite{VCS73}, VCS; with extensions of the grids 
by Sch\"oning \& Butler, private communication) in the BA-SG regime.
The SH tables have not only the advantage of being based on the more
sophisticated theory but they also cover the Paschen (and Lyman) series for
transitions up to principal quantum number $n$\,$=$\,30.
A detailed discussion of our hydrogen non-LTE line-formation computations
for BA-SGs can be found in Przybilla \& Butler~(\cite{PrBu04a}), where 
lines of the Brackett and Pfund series were also studied. In short, it has been shown
that except for the lower series members -- which are affected or even
dominated by the stellar
wind -- excellent consistency can be achieved from all available indicators,
if accurate data for excitation via electron collisions are accounted for.
Note that the only other study of H lines from these four series, in the
prototype A-SG \object{Deneb} (Aufdenberg et al.~\cite{Aufdenbergetal02}), fails
in this. Their non-LTE model atmosphere and line-formation
computations produce much too strong near-IR features because of inaccurate 
collisional data in their hydrogen model atom, as indicated by our findings.
The problems are resolved using the improved effective collision strengths
(P.~Hauschildt, private communication).

\begin{figure*}
\sidecaption
\includegraphics[width=12cm]{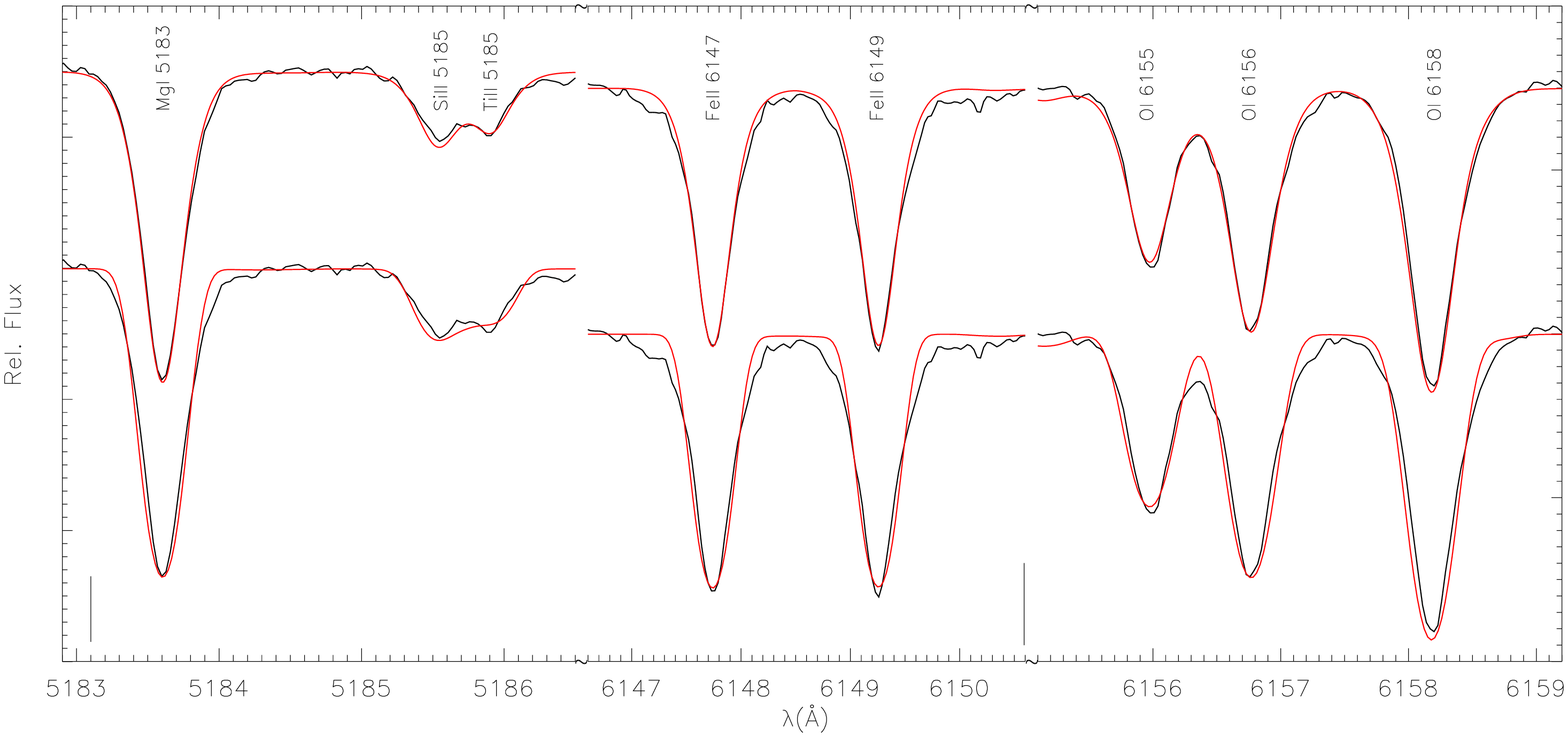}
\caption[]{Determination of projected rotational and macroturbulent
velocities, exemplarily shown for three spectral regions of $\eta$\,Leo including 
discrete and blended absorption features. The upper comparison between
observation (black line) and spectrum synthesis (red line, for finally
derived mean abundances) is made for
$\zeta$\,$=$\,16.5\,km\,s$^{-1}$ and zero $v\,\sin i$ (our finally adopted
optimum values), the lower comparison is made assuming a pure rotational profile with $v\,\sin
i$\,$=$\,13.3\,km\,s$^{-1}$. See the text for further discussion. Line
identifiers are given. The markers are vertically extended by 0.05 units.}
\label{rotmac}
\end{figure*}

Our method for deriving surface gravities from the hydrogen lines deviates
in a few details from the usual approach, which is based on the modelling of
the H$\gamma$ and H$\delta$ {\em line wings}.
The effect of variations of $\log g$ and $T_{\rm eff}$ on the H$\delta$ 
profiles of $\eta$\,Leo and HD\,92207 are displayed in Fig.~\ref{loggtest}.
We have chosen parameter offsets to our finally adopted values 
which we consider {\em conservative} estimates for the uncertainties of 
our analysis. It is clear that the {\em internal} accuracy of our
modelling is higher, amounting to $\lesssim$\,0.10\,dex at LC Ib and   
as accurate as 0.05\,dex at LC Iae. The temperature sensitivity is
smaller, but non-negligible. Note the strong decrease of the H$\delta$ line strength
in the luminosity progression. Close to the Eddington limit a
modification by 0.05\,dex in $\log g$ can amount to a change in
$W_{\lambda}$ of the Balmer lines by 15\%. Note also that in the less luminous supergiant
not only the line wings but also
the {\em entire line profile} and thus also the {\em equivalent width} 
is accurately reproduced (which becomes important for intermediate- and 
low-resolution studies), indicating that the model atmosphere is 
sufficiently realistic over all line-formation depths. At the highest
luminosities the accuracy of the spectrum synthesis degrades somewhat, because
of stellar wind and sphericity effects. However, excellent agreement can be 
restored even for the higher Balmer lines, and for the accessible Paschen
series members (see Fig.~11 in Przybilla \& Butler~\cite{PrBu04a}),
which are formed even deeper in the atmosphere. This also includes good
reproduction of the series limits and their transition into the continua.
Moreover, we also account for the effects of non-solar helium abundances on
the density structure (see Sect.~\ref{heliumetc}), which is mandatory for
achieving consistent results, but which has been ignored in all BA-SG
analyses so far. 

In order to resolve the ambiguity in $T_{\rm eff}/\log g$ another indicator 
has to be applied. In a spectroscopic approach this is typically the 
ionization equilibrium of {\em one} chemical species.
Poten\-tial ionization equilibria useful for $T_{\rm eff}$ and $\log g$
determinations in optical/near-IR spectroscopy of BA-SGs are \ion{C}{i/ii},
\ion{N}{i/ii}, \ion{O}{i/ii}, \ion{Mg}{i/ii}, \ion{Al}{i/ii}, 
\ion{Al}{ii/iii}, \ion{Si}{ii/iii}, \ion{S}{ii/iii}, \ion{Fe}{i/ii} and \ion{Fe}{ii/iii}.
Of these we can use only half because model atoms are currently 
unavailable
for the rest. The lines of the minor ionic species are highly sensitive to 
temperature and electron density changes, while the weak lines of the major ionic species 
are excellent abundance indicators. A suitable set of parameters 
$T_{\rm eff}/\log g$ is found in the case where both ionic species indicate the same 
elemental abundance within the individual error margins. However, great
care has to be practised in the modelling, using carefully selected
model atmospheres (see Sect.~\ref{sectmodels}) and sophisticated 
non-LTE techniques (see Sects.~\ref{sectlform} and~\ref{sectabus}).

An example is displayed in Fig.~\ref{ttest}, where tests on the \ion{Mg}{i/ii}
ionization balance in one of the sample objects are
performed. Results from the best fit obtained in the detailed non-LTE
analysis are compared with those from {\em conservative} parameter
studies at an unchanged elemental abundance.
The sensitivity of the predicted \ion{Mg}{i} line strengths to changes of
$T_{\rm eff}$ and $\log g$ within the error margins is high, indicating {\em
internal} uncertainties of $\Delta T_{\rm eff}$\,$<$\,100\,K and
of $\Delta \log g$\,$\lesssim$\,0.10\,dex. The profile changes by conservative parameter
variation are similar to those achieved by abundance variations of the order
0.1\,dex. On the other hand, no perceptible changes are seen in the \ion{Mg}{ii} line
for the same parameter variations. All \ion{Mg}{i/ii} lines behave in a
similar way, and the same qualitative characteristics are shared with ionization
equilibria from other elements.

Both, hydrogen profiles and ionization equilibria are affected by the helium
abundance, 
which therefore has to be {\em simultaneously} 
constrained as an additional parameter. Enhanced helium
abundances manifest in broadened hydrogen lines and shifts in ionization
equilibria -- in BA-SGs the spectral lines of the minor ionization species
are strengthened. Thus, parameter studies assuming a solar helium abundance
will tend to derive systematically higher effective temperatures and surface gravities. 
In the BA-SGs the transitions of \ion{He}{i} are typically weak. The helium abundance 
$y$ (by number) can therefore be inferred from line profile fits to all
available features. Typical uncertainties in the determination of stellar helium abundances
amount to $\sim$10\% because of the availability of excellent atomic data.

The whole process of the basic atmospheric parameter determination can be
summarised in an $T_{\rm eff}$--$\log g$ diagram, which is done exemplarily for 
$\eta$\,Leo and $\beta$\,Ori in Fig.~\ref{kiel}. The intersection of the different loci
marks the appropriate set of stellar parameters. Note that these diagnostic diagrams
differ from those typically found in the literature in two respects: first,
the different indicators lead to consistent stellar parameters, and second,
the loci are parameterised with the helium abundance, a necessary procedure
to obtain the excellent agreement. When deriving the stellar parameters one
can profit from the slow and rather predictable behaviour of the derived helium 
abundance with varying $T_{\rm eff}$/$\log g$.

The microturbulent velocity $\xi$ is determined in the usual way by
forcing elemental abundances to be independent of equivalent widths.
The difference to previous studies of BA-SGs is that this is done on the
basis of non-LTE line-formation (see Fig.~\ref{nlteabus} and the discussion in
Sect.~\ref{sectabus}). An LTE analysis would tend to find higher
values in order to compensate non-LTE line strengthening. Typically, the extensive 
iron spectrum is used for the microturbulence analysis. Because the \ion{Fe}{ii} non-LTE
computations are by far the most time-intensive, alternatives have to be
considered here. The \ion{Ti}{ii} and in particular the \ion{N}{i} spectra are
excellent replacements, the latter distinguished by more accurate atomic
data. Later, the iron lines can be used to verify the $\xi$-determination.
Note that this is the only occasion where equivalent widths are used in the
analysis process -- elsewhere line profile fits are preferred. 

\begin{table}
\caption[]{Basic properties and stellar parameters of the sample stars}
\label{obj}
\setlength{\tabcolsep}{.034cm}
\begin{tabular}{lr@{$\pm$}lr@{$\pm$}lr@{$\pm$}lr@{$\pm$}l}
\hline
& \multicolumn{2}{c}{HD\,87737} & \multicolumn{2}{c}{HD\,111613} &
\multicolumn{2}{c}{HD\,92207} & \multicolumn{2}{c}{HD\,34085}\\
\hline
Name & \multicolumn{2}{c}{$\eta$\,Leo} & \multicolumn{2}{c}{{\ldots}} &
\multicolumn{2}{c}{{\ldots}} & \multicolumn{2}{c}{$\beta$\,Ori, Rigel}\\
Association$^{\rm a}$ & \multicolumn{2}{c}{Field} & \multicolumn{2}{c}{Cen\,OB1} & 
\multicolumn{2}{c}{Car\,OB1} & \multicolumn{2}{c}{Ori\,OB1}\\
Spectral Type$^{\rm b}$ & \multicolumn{2}{c}{A0\,Ib} & \multicolumn{2}{c}{A2\,Iabe} &
\multicolumn{2}{c}{A0\,Iae} & \multicolumn{2}{c}{B8\,Iae:}\\
$\alpha$\,(J2000)$^{\rm b}$ & \multicolumn{2}{c}{10\,07\,19.95} & 
\multicolumn{2}{c}{12\,51\,17.98} &
\multicolumn{2}{c}{10\,37\,27.07} & \multicolumn{2}{c}{05\,14\,32.27}\\
$\delta$\,(J2000)$^{\rm b}$ & \multicolumn{2}{c}{$+$16\,45\,45.6} &
\multicolumn{2}{c}{$-$60\,19\,47.2} &
\multicolumn{2}{c}{$-$58\,44\,00.0} & \multicolumn{2}{c}{$-$08\,12\,05.9}\\
$l\,(\degr)$$^{\rm b}$ & \multicolumn{2}{c}{219.53} & \multicolumn{2}{c}{302.91} &
\multicolumn{2}{c}{286.29} & \multicolumn{2}{c}{209.24}\\
$b\,(\degr)$$^{\rm b}$ & \multicolumn{2}{c}{$+$50.75} & \multicolumn{2}{c}{+2.54} & 
\multicolumn{2}{c}{$-$0.26} & \multicolumn{2}{c}{$-$25.25}\\
$\pi$\,(mas)$^{\rm c}$ & 1.53 & 0.77 & 1.09 & 0.62 & 0.40 & 0.53 & 4.22 & 0.81\\
$d$\,(pc)$^{\rm d}$ & 630 & 90 & 2\,290 & 220 & 3\,020 & 290 & 360 & 40\\
$R_{\rm g}\,(\mbox{kpc})$ & 8.25 & 0.83 & 6.97 & 0.84 & 7.66 & 0.77 & 8.23 & 0.82\\
$\theta_{\rm D}$\,(mas)$^{\rm e}$ & \multicolumn{2}{c}{{\ldots}} &
\multicolumn{2}{c}{{\ldots}} & \multicolumn{2}{c}{{\ldots}} & 2.55 & 0.05\\
$v_\mathrm{rad}\,(\mbox{km\,s}^{-1})$$^{\rm b}$ & $+$3.3 & 0.9 & 
$-$21.0 & 2.0 & $-$8.5 & 5.0 & $+$20.7 & 0.9\\
$\mu_{\alpha}$\,(mas\,yr$^{-1}$)$^{\rm c}$ & $-$1.94 & 0.92 & $-$5.26 & 0.55 & $-$7.46
& 0.53 & 1.87 & 0.77\\
$\mu_{\delta}$\,(mas\,yr$^{-1}$)$^{\rm c}$ & $-$0.53 & 0.43 & $-$1.09 & 0.39 & 3.11 &
0.44 & $-$0.56 & 0.49\\
$U\,(\mbox{km\,s}^{-1})$ & \multicolumn{2}{c}{4} & \multicolumn{2}{c}{$-$49} & 
\multicolumn{2}{c}{$-$103} & \multicolumn{2}{c}{$-6$}\\
$V\,(\mbox{km\,s}^{-1})$ & \multicolumn{2}{c}{222} & \multicolumn{2}{c}{211} &
\multicolumn{2}{c}{201} & \multicolumn{2}{c}{214}\\
$W\,(\mbox{km\,s}^{-1})$ & \multicolumn{2}{c}{6} & \multicolumn{2}{c}{$-$6} &
\multicolumn{2}{c}{$-$6} & \multicolumn{2}{c}{1}\\[2mm]
Atmospheric:\\
$T_\mathrm{eff}$\,(K) & 9\,600 & 150 & 9\,150 & 150 & 9\,500 & 200 & 12\,000 & 200\\
$\log g$\,(cgs) & 2.00 & 0.15 & 1.45 & 0.10 & 1.20 & 0.10 & 1.75 & 0.10\\
$y$ & 0.13 & 0.02 & 0.105 & 0.02 & 0.12 & 0.02 & 0.135 & 0.02\\
$[$M/H$]$\,(dex) & $-$0.04 & 0.03 & $-$0.11 & 0.03 & $-$0.09 & 0.07 & $-$0.06 & 0.10\\
$\xi\,(\mbox{km\,s}^{-1})$ & 4 & 1 & 7 & 1 & 8 & 1 & 7 & 1\\
$\zeta\,(\mbox{km\,s}^{-1})$ & 16 & 2 & 21 & 3 & 20 & 5 & 22 & 5\\
$v \sin i\,(\mbox{km\,s}^{-1})$ & 0 & 3 & 19 & 3 & 30 & 5 & 36 & 5\\[2mm]
Photometric:\\
$V$\,(mag)$^{\rm f}$ & \multicolumn{2}{c}{3.52} & \multicolumn{2}{c}{5.72} & 
\multicolumn{2}{c}{5.45} & \multicolumn{2}{c}{0.12}\\
$B-V^{\rm f}$ & \multicolumn{2}{c}{$-$0.03} & \multicolumn{2}{c}{+0.38} &
\multicolumn{2}{c}{+0.50} & \multicolumn{2}{c}{$-$0.03}\\
$U-B^{\rm f}$ & \multicolumn{2}{c}{$-$0.21} & \multicolumn{2}{c}{$-$0.10} &
\multicolumn{2}{c}{$-$0.24} & \multicolumn{2}{c}{$-$0.66}\\
$E(B-V)$ & \multicolumn{2}{c}{0.02} & \multicolumn{2}{c}{0.39} &
\multicolumn{2}{c}{0.48$^{\rm g}$} & \multicolumn{2}{c}{0.05}\\
$(m-M)_0^{\rm d}$ & 9.0 & 0.3 & 11.8 & 0.2 &
12.4 & 0.2 & 7.8 & 0.2\\
$M_V$ & $-$5.54 & 0.3 & $-$7.29 & 0.2 &
$-$8.82 & 0.2 & $-$7.84 & 0.2\\
$B.C.$ & \multicolumn{2}{c}{$-$0.29} & \multicolumn{2}{c}{$-$0.23}
& \multicolumn{2}{c}{$-$0.34} & \multicolumn{2}{c}{$-$0.78}\\
$M_\mathrm{bol}$ & $-$5.83 & 0.3 & $-$7.52 & 0.2 &
$-$9.16 & 0.2 & $-$8.62 & 0.2\\[2mm]
Physical:\\
$\log L/$L$_{\odot}$ & 4.23 & 0.12 & 4.90 & 0.08 & 
5.56 & 0.08 & 5.34 & 0.08\\
$R/$R$_{\odot}$ & 47 & 8 & 112 & 12 &
223 & 24 & 109 & 12\\
$M/$M$_{\odot}^{\rm ZAMS}$ & 10 & 1 & 16 & 1 &
30 & 3 & 24 & 3\\
$M/$M$_{\odot}^{\rm evol}$ & 10 & 1 & 15 & 1 &
25 & 3 & 21 & 3\\
$M/$M$_{\odot}^{\rm spec}$ & 8 & 4 & 13 & 4 &
29 & 10 & 24 & 8\\
$\tau_{\rm evol}$\,(Myr) & 25 & 5 & 14 & 2 & 7 & 1 & 8 & 1\\
\hline
\end{tabular}\\
$^{\rm a}$~Blaha \& Humphreys~(\cite{BlHu89})\hspace{2mm}
$^{\rm b}$~adopted from the Simbad database at CDS\hspace{2mm}
$^{\rm c}$~Perryman et al.~(\cite{Perrymanetal97})\hspace{2mm}
$^{\rm d}$~see text\hspace{2mm}
$^{\rm e}$~Hanbury Brown et al.~(\cite{HanburyBrownetal74})\hspace{2mm}
$^{\rm f}$~Nicolet (\cite{Nicolet78})\hspace{2mm}
$^{\rm g}$~with $R_V$\,$=$\,3.9, see Sect.~\ref{spectrophotometry}\vspace{-4mm}
\end{table}

Good starting estimates for the microturbulent velocity in BA-SGs are
$\xi$\,$=$\,4, 6 and 8\,km\,s$^{-1}$ for objects of LC Ib, Iab
and Ia, respectively. After the determination of an improved value of $\xi$
the model atmosphere has to be recalculated in some cases, and small
corrections to $T_{\rm eff}$/$\log g$/$y$ may become applicable. Only
one iteration step is typically necessary to reestablish consistency
in the atmospheric parameters. The uncertainties amount to typically
$\pm$1\,km\,s$^{-1}$. In the present study the different microturbulence
indicators give a single value for $\xi$ within these error margins. This is 
consistent with findings from recent (LTE) work on less-luminous A-type 
supergiants (Venn~\cite{Venn95a}, \cite{Venn99}), while older studies based
on more simple model atmospheres and less accurate oscillator strengths had to invoke
different values for various elemental species, or a depth-dependent $\xi$ 
(e.g. Rosendhal~\cite{Rosendhal70}; Aydin~\cite{Aydin72}). 

Microturbulence also has to be considered as additional non-thermal broadening agent 
in the radiative transfer and statistical equilibrium computations. The
Doppler width is then given by 
$\Delta \lambda_{\rm D} = \lambda_0/c\,(v_{\rm th}^2 + \xi^2)^{1/2}$, where
$\lambda_0$ is the rest wavelength of the transition, $c$ the speed of light
and $v_{\rm th}$ the thermal velocity of the chemical species of interest.
The main effect is a broadening of the frequency bandwidth for absorption in
association with a shift in line-formation depth, typically leading to
a net strengthening of individual lines by different amounts, see e.g. Przybilla
et al.~(\cite{Przybillaetal00, Przybillaetal01a, Przybillaetal01b})
for further details.

Individual metal abundances are typically of secondary importance for the computation
of model atmospheres, as elemental ratios are remarkably constant (i.e.
close to solar) over a wide variety of stars, except for the chemically peculiar. 
We determine the stellar metallicity $[$M/H$]$\footnote{using the usual
logarithmic notations $[X]=\log \varepsilon (X)_{\star} - \log \varepsilon (X)_{\odot}$,
with $\log \varepsilon(X)=\log (N_{X} / N_{\rm H}) + 12$, the 
$\varepsilon (X)$ being the abundance of element $X$ and the $N_i$ 
number densities}, which is the more important parameter, as the arithmetic mean from five
elements, for which non-LTE computations can be done and which are
unaffected by mixing processes: $[{\rm M/H}]\equiv([{\rm O/H}]+[{\rm
Mg/H}]+[{\rm S/H}]+[{\rm Ti/H}]+[{\rm Fe/H}])/5$. High weight
is thus given to the $\alpha$--process elements, which have a different
nucleosynthesis history than the iron group elements, but 
$[\alpha$/Fe$]$\,$\sim$\,0 can be expected for these Population\,I objects.
Because of the importance of metallicity for line blanketing a further
iteration step may be required in the parameter determination. 

Finally, the projected rotational velocity $v\sin i$ and the (radial-tangential)
macroturbulent velocity
$\zeta$ are derived from a comparison of observed line profiles with
the spectrum synthesis. Single transitions as well as line blends should be
used for this, as they contain some complementary information. Both quantities are treated
as free parameters to obtain a best fit via convolution of the synthetic
profile with rotation and macroturbulence profiles (Gray~\cite{Gray92a}), 
while also accounting for a Gaussian instrumental profile. 
An example is shown in Fig.~\ref{rotmac}, where the comparison with
observation indicates pure macroturbulence broadening for $\eta$\,Leo. 
The theoretical profiles intersect the observed profiles
if the spectral lines are broadened by rotation alone, resulting in
slightly too broad line cores and insufficiently broad line wings. Typically,
both parameters are non-zero, with the macroturbulent velocity amounting to
less than twice the sonic velocity in the atmospheric plasma.
Macroturbulence is suggested to be related to surface motions caused by
(high-order) nonradial oscillations (Lucy~\cite{Lucy76}).
Weak lines should be used for the $v \sin i$ and $\zeta$-determinations 
in supergiants in order to avoid systematic uncertainties due to asymmetries 
introduced by the outflowing velocity field in the strong lines.

\subsection{Stellar parameters of the sample objects}
We summarise the basic properties and derived stellar parameters of the
sample supergiants in Table~\ref{obj}. The first block~of information
concentrates on quantities which in principle can be deduced directly,
with only little modelling involved. Besides Henry-Draper catalogue
numbers, alternative names, information on association membership
and spectral classifica\-tion and equatorial and galactic coordinates are given.
Hippar\-cos parallaxes $\pi$ have been included for completeness, as none of
the measurements is of sufficient statistical significance. The stellar 
distances $d$ are therefore deduced by other~means.

\begin{figure*}[ht!]
\sidecaption
\includegraphics[width=12cm]{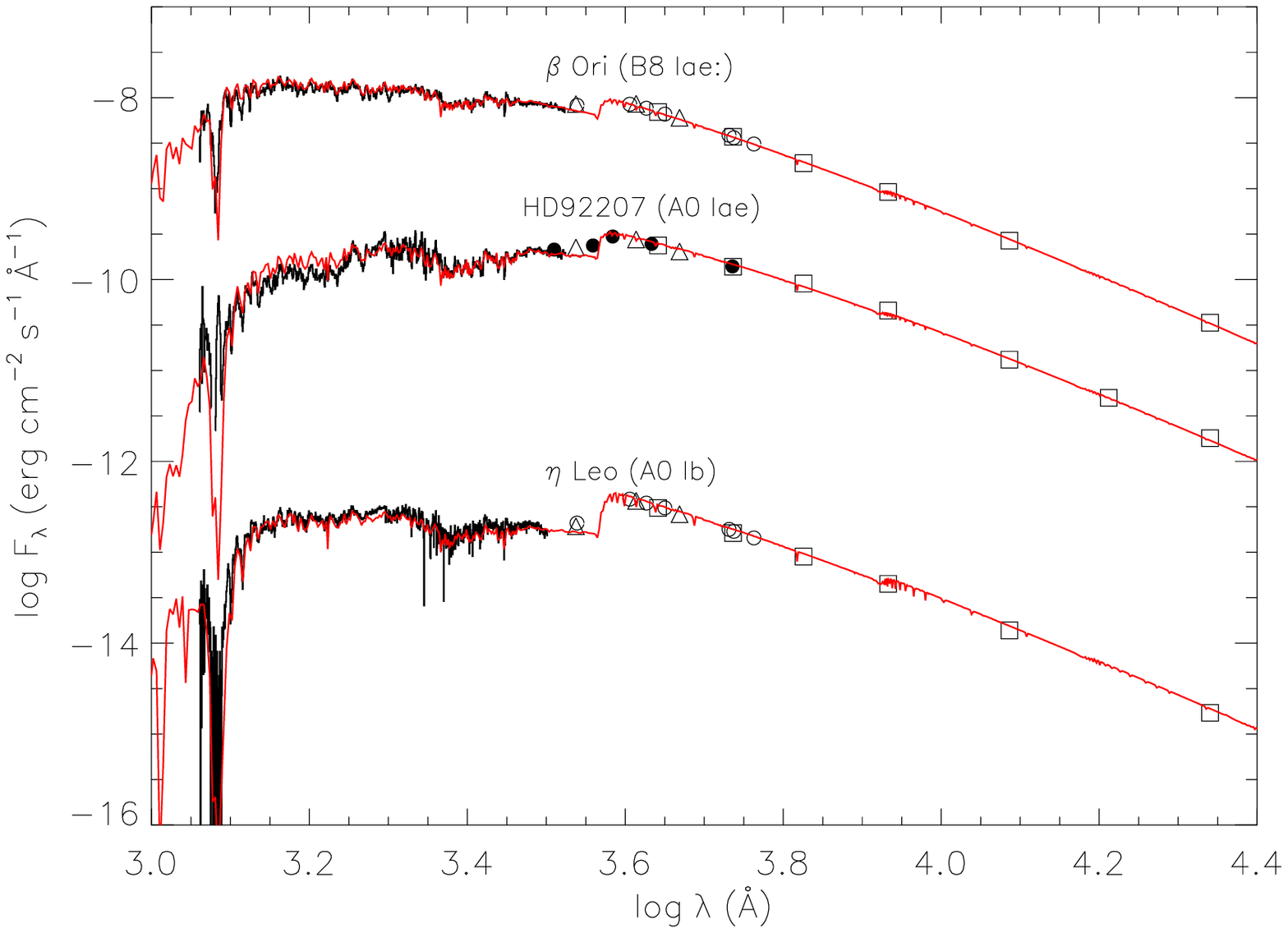}
\caption[]{Computed (red full lines) and measured spectral energy 
distributions for the sample stars from the UV to the near-IR in the
$K$-band. Overall, good to excellent agreement is obtained.
Displayed are IUE spectrophotometry
(full lines), Johnson (boxes), Str\"omgren (triangles), Walraven (filled circles)
and Geneva photometry (open circles). The observations have been dereddened
using colour excesses according to Table~\ref{obj}. A ratio
$R_V$\,$=$$A_V/E(B-V)$\,$=$\,3.1, as typical for the local ISM, is assumed for
$\beta$\,Ori and $\eta$\,Leo. In the case of the notably reddened HD\,92207 both, 
color excess and $R_V$, can be determined, implying $R_V$\,$=$\,3.9. For clarity, the data 
for $\eta$\,Leo has been shifted vertically by $-$3\,units. IUE
spectrophotometry for HD\,111613 is not available. Consequently, this star
has been omitted from the comparison.}
\label{fluxes}
\end{figure*}

A distance modulus of 8\fm5 (viz. 500\,pc) is found for the
Ori~OB\,1 association by Blaha \& Humphreys~(\cite{BlHu89}). However, 
$\beta$\,Ori shows a smaller radial velocity than the mean of the Ori OB1 
association ($\sim$30\,km\,s$^{-1}$). Therefore,
the association distance gives only an upper limit, if one assumes that the star was
formed near the centre of Ori OB1. During its lifetime $\beta$\,Ori could have moved
$\sim$100\,pc from its formation site. The finally adopted distance of 360\,pc is 
indicated by Hoffleit \& Jaschek~(\cite{HoJa82}), who associate $\beta$\,Ori with 
the $\tau$\,Ori~R1 complex.

Recent studies of the cluster NGC\,4755 found a distance modulus of 
11.6$\pm$0.2\,mag, with a~mean~value~of~$E(B$\,$-$\,$V)$ of 0.41$\pm$0.05\,mag
(Sagar \& Cannon~\cite{SaCa95}) and 0.36 $\pm$0.02\,mag (Sanner et
al.~\cite{Sanneretal01}). However, 
HD\,111613 is situated at a rather large distance from the cluster centre
and was not observed in either study. In a previous work (Dachs \&
Kaiser~\cite{DaKa84}) the object was found to be
slightly behind the cluster by 0.2\,mag. Consequently,
this difference is accounted for in the present study, otherwise using the
modern distance.

The line of sight towards the Car~OB1 association coincides with a Galactic
spiral arm, such that the star population in Car~OB1 is distributed 
in depth over 2 to 3\,kpc (Shobbrook \& Lyng{\aa}~\cite{ShLy94}). A more
decisive constraint on the distance of HD\,92207 is indicated by 
Carraro et al.~(\cite{Carraroetal01}), who
argue that the star may be associated with the cluster \object{NGC\,3324}.
The star will not be gravitationally bound because of its large 
proper motion, but appears to be spatially close to the cluster, and may
have been ejected. We therefore adopt the cluster distance modulus for
HD\,92207 with an increased error margin.
Finally, the distance to the field star $\eta$\,Leo can
only be estimated, based on its spectroscopic parallax.

Galactocentric distances of the stars are calculated from the coordinates
and $d$, using a galactocentric solar distance of
$R_0$\,$=$\,7.94$\pm$0.42\,kpc (Eisenhauer et al.~\cite{Eisenhaueretal03}).
In one case an interferometrically determined true angular diameter
$\theta_{\rm D}$ (allowing for limb darkening) is available from the
literature. Our measured radial velocities $v_{\rm rad}$ (from
cross-correlation with synthetic spectra) are compatible with the
literature values, which in combination with proper motions $\mu_{\alpha}$
and $\mu_{\delta}$ are used to calculate galactocentric velocities $U$, $V$
and $W$, assuming standard values for the solar motion
($U$\,$=$\,10.00\,km\,s$^{-1}$, $V$\,$=$\,5.23\,km\,s$^{-1}$,
$W$\,$=$\,7.17\,km\,s$^{-1}$, Dehnen \& Binney~\cite{DeBi98}) relative to 
the local standard of rest (220\,km\,s$^{-1}$, Kerr \&
Lynden-Bell~\cite{KeLy86}).

In the second block our results from the spectroscopic stellar parameter
determination are summarised. Photometric data are collected in the third block.
Observed visual magnitudes and colours in the Johnson system are used to
derive the colour excess $E(B-V)$ by comparison with synthetic colours.
Note that for HD\,111613 and HD\,92207 the derived colour excess is in
excellent agreement with literature values for their parent clusters. From
the distance moduli $(m-M)_0$ absolute visual magnitudes $M_V$ are
calculated, which are transferred to absolute bolometric magnitudes
$M_\mathrm{bol}$ by application of bolometric corrections $B.C.$ 
from the model atmosphere computations. Note also our previous comments on
the photometric variability of BA-SGs, which can introduce some systematic
uncertainty into our analysis, as the photometry and our spectra may reflect
slightly different physical states of the stellar atmospheres. The
sensitivity of the Johnson colours to $T_{\rm eff}$-changes is not high enough to
use them as an alternative indicator for a precise determination of the
stellar effective temperature. 

Finally, we derive the physical parameters luminosity $L$, stellar radius
$R$ and spectroscopic mass $M^{\rm spec}$ of the sample stars by combining
atmospheric parameters and photometry. Thus the determined stellar radius of
$\beta$\,Ori is in good agreement with that derived from the angular
diameter measurement (99$\pm$11\,R$_{\odot}$). From comparison with stellar
evolution computations (Meynet \& Maeder~\cite{MeMa03}) zero-age
main-sequence masses $M^{\rm ZAMS}$ and evolutionary masses $M^{\rm evol}$
are determined, and the evolutionary age $\tau_{\rm evol}$. The stars were
of early-B and late-O spectral type on the main sequence. The results
will be discussed in detail in Sect.~\ref{sectevol}. Note that the rather
large uncertainties in the distance determination dominate the error budget
of the physical parameters.

\begin{table*}
\caption[]{Comparison of stellar parameters of $\eta$\,Leo and $\beta$\,Ori from the literature}
\label{etaleoparams}
\setlength{\tabcolsep}{.15cm}
\begin{tabular}{lr@{$\pm$}lr@{$\pm$}lr@{$\pm$}lll}
\hline
Source & \multicolumn{2}{c}{$T_{\rm eff}$\,(K)} & \multicolumn{2}{c}{$\log g$\,(cgs)} & \multicolumn{2}{c}{$\xi$\,(km\,s$^{-1}$)} & Method & Notes\\
\hline
\\[-3mm]
\multicolumn{9}{c}{{\underline{     HD\,87737     }}}\\[1mm]
This work & 9\,600 & 150 & 2.00 & 0.15 & 4 & 1 & NLTE \ion{H}{i}, \ion{N}{i/ii}, \ion{Mg}{i/ii}, (\ion{C}{i/ii}), & {\ldots}\\
& \multicolumn{2}{c}{} & \multicolumn{2}{c}{} & \multicolumn{2}{c}{} & spectrophotometry & \\
Venn~(\cite{Venn95a}) & 9\,700 & 200 & 2.0 & 0.2 & 4 & 1 & H$\gamma$, NLTE \ion{Mg}{i/ii} & Kurucz~(\cite{Kurucz93}) atmospheres\\ 
Lobel et al.~(\cite{Lobeletal92}) & 10\,200 & 370 & 1.9 & 0.4 & 5.4 & 0.7 & LTE \ion{Fe}{i/ii} & Kurucz~(\cite{Kurucz79}) atmospheres,\\
& \multicolumn{2}{c}{} & \multicolumn{2}{c}{} & \multicolumn{2}{c}{} & & equivalent widths from Wolf~(\cite{Wolf71})\\
Lambert et al.~(\cite{Lambertetal88}) & \multicolumn{2}{c}{10\,500} & \multicolumn{2}{c}{2.2} & \multicolumn{2}{c}{{\ldots}} & Str\"omgren photometry $+$ H$\beta$ & Kurucz~(\cite{Kurucz79}) atmospheres\\
Wolf~(\cite{Wolf71}) & 10\,400 & 300 & 2.05 & 0.20 & \multicolumn{2}{c}{2{\ldots}10} & Balmer lines, Balmer jump, & early LTE atmospheres, no line-\\
& \multicolumn{2}{c}{} & \multicolumn{2}{c}{} & \multicolumn{2}{c}{} & LTE \ion{Mg}{i/ii}, \ion{Fe}{i/ii} & blanketing, photographic spectra\\[4mm]
\multicolumn{9}{c}{{\underline{     HD\,34085     }}}\\[1mm]
This work & 12\,000 & 200 & 1.75 & 0.10 & 7 & 1 & NLTE \ion{H}{i}, \ion{N}{i/ii}, \ion{O}{i/ii}, \ion{S}{ii/iii}, & {\ldots}\\
& \multicolumn{2}{c}{} & \multicolumn{2}{c}{} & \multicolumn{2}{c}{} & spectrophotometry & \\
McErlean et al.~(\cite{McErleanetal99}) & 13\,000 & 1\,000 & 1.75 & 0.2 & \multicolumn{2}{c}{10} & NLTE H$\gamma$, H$\delta$, \ion{Si}{ii/iii} & Hubeny~(\cite{Hubeny88}) H$+$He NLTE models,\\
& \multicolumn{2}{c}{} & \multicolumn{2}{c}{} & \multicolumn{2}{c}{} & & no line-blanketing\\
Israelian et al.~(\cite{Israelianetal97}) & 13\,000 & 500 & 1.6 & 0.1 &
\multicolumn{2}{c}{7} & NLTE H$\gamma$, H$\delta$, Si & Hubeny~(\cite{Hubeny88}) H$+$He NLTE models,\\
& \multicolumn{2}{c}{} & \multicolumn{2}{c}{} & \multicolumn{2}{c}{} & & no line-blanketing\\
Takeda~(\cite{Takeda94}) & 13\,000 & 500 & 2.0 & 0.3 & \multicolumn{2}{c}{7} & LTE H$\gamma$, H$\delta$, spectrophotometry & Kurucz~(\cite{Kurucz79}) atmospheres\\
Stalio et al.~(\cite{Stalioetal77}) & 12\,070 & 160 & \multicolumn{2}{c}{{\ldots}} &
\multicolumn{2}{c}{{\ldots}} & angular diameter, total flux & early
line-blanketed LTE atmospheres\\
\hline
\end{tabular}
\vspace{3mm}
\end{table*}

\subsection{Spectrophotometry}\label{spectrophotometry}
The principal aim of the present study is a self-consistent method
for the analysis of BA-type supergiants in the whole. This requires above
all the reproduction of the stellar spectral energy distribution (SED).
A comparison of our model fluxes for three of the sample supergiants with
observation is made in Fig.~\ref{fluxes}. The observational database
consists of IUE spectrophotometry (as obtained from the IUE Final Archive)
and photometry in several passbands (broad-band to small-band) from the
near-UV to near-IR, in the Johnson  (Morel \& Magnenat~\cite{MoMa78}; 
Ducati~\cite{Ducati02}, for HD\,92207), Geneva (Rufener~\cite{Rufener88}), 
Str\"omgren (Hauck \& Mermilliod~\cite{HaMe98}) and Walraven systems 
(de Geus et al.~\cite{deGeusetal90}). We have omitted HD\,111613 from the
comparison because crucial UV spectrophotometry is unavailable for this
star. The observations have been de-reddened using a reddening law according
to Cardelli et al.~(\cite{Cardellietal89}). Overall, good to excellent agreement 
is found. This {\em verifies} our spectroscopic stellar parameter determination.

Use of SED information can provide an additional $T_{\rm eff}$-indicator,
independent of weak spectral lines used in the ionization equilibria approach. This
makes SED-fitting attractive for extragalactic applications, as it
has the potential to improve on the modelling situation when only intermediate-resolution
spectra are available (see Sect.~\ref{sectmedres}). As the present work
concentrates on the spectroscopic approach to BA-SGs analyses, we will
report on a detailed investigation of this elsewhere.

\subsection{Comparison with previous analyses}\label{sectparamsother}
The two MK standards $\eta$\,Leo and $\beta$\,Ori have been analysed with model
atmosphere techniques before. A comparison of our derived stellar parameters
with literature values is made in Table~\ref{etaleoparams}. The methods used for
the parameter determination are indicated and comments on the model
atmospheres and the observational data are given.
For the comparison we have omitted publications earlier than the 1970ies.

Good agreement of the present parameters for $\eta$\,Leo with those of 
Venn~(\cite{Venn95a}) is found, as the methods employed are comparable. 
Other authors find a substantially hotter $T_{\rm eff}$ for this star,
while the values for surface gravity and microturbulence are comparable. All
these values are based on less elaborate model atmospheres, with less or no
line-blanketing, and they completely ignore non-LTE effects on the line
formation. Note that a $T_{\rm eff}$ of 9380\,K is obtained from Str\"omgren
photometry, using the calibration of Gray~(\cite{Gray92b}).

For $\beta$\,Ori basically two disjunct values for the effective temperature are found in
the literature. Model atmosphere analyses so far found a systematically
higher $T_{\rm eff}$, which is likely to be due in good part to ignored/reduced 
line-blanketing. The measurements of surface gravity and microturbulent
velocity are again similar. The other group of $T_{\rm eff}$ analyses
mostly find significantly lower values than that derived in the present work,
from measured total fluxes and interferometric radius determinations, or
from the infrared flux method: 11\,800\,$\pm$\,300/11\,700\,$\pm$\,200 (Nandy \&
Schmidt~\cite{NaSch75}), 11\,550\,$\pm$\,170 (Code et
al.~\cite{Codeetal76}), 11\,410\,$\pm$\,330
(Beeckmans~\cite{Beeckmans77}), 11\,780 (Underhill et al.~\cite{Underhilletal79}),
11\,014 (Blackwell et al.~\cite{Blackwelletal80}), 11\,380
(Underhill \& Doazan~\cite{UnDo82}) and
11\,023/11\,453\,K (Glushneva~\cite{Glushneva85}). These methods are prone
to systematic errors from inappropriate corrections for interstellar
absorption. The authors all {\em supposed} zero interstellar extinction. In
fact, the only such study that accounts for a non-zero $E(B-V)$ (Stalio et
al.~\cite{Stalioetal77}, $+$0.04 vs. $+$0.05 as derived here) finds a
temperature in excellent agreement with the present value, see
Table~\ref{etaleoparams}. However, using LTE \ion{Fe}{ii/iii} and \ion{Si}{ii/iii}
ionization equilibria and Balmer profiles they derive 
$T_{\rm eff}$\,$\gtrsim$\,13\,000\,K.


\begin{landscape}
\begin{table}
\rule{0cm}{2cm}
\setlength{\tabcolsep}{.295cm}
\caption[]{Elemental abundances in the sample stars\\[-6mm]\label{tableabu}}
\begin{tabular}{lr@{$\pm$}lr@{$\pm$}lr@{$\pm$}lr@{$\pm$}lr@{$\pm$}lr@{$\pm$}lr@{$\pm$}lr@{$\pm$}lr@{$\pm$}lr@{$\pm$}l}
\hline
Element & \multicolumn{2}{c}{Sun$^\mathrm{a}$} & 
\multicolumn{2}{c}{$\eta\,$Leo} & 
\multicolumn{2}{c}{HD\,111613} &
\multicolumn{2}{c}{HD\,92207} & 
\multicolumn{2}{c}{$\beta$\,Ori} & 
\multicolumn{2}{c}{Gal AI$^\mathrm{b}$} & 
\multicolumn{2}{c}{Gal BV$^\mathrm{c}$} & 
\multicolumn{2}{c}{Gal BV$^\mathrm{d}$} &
\multicolumn{2}{c}{Gal BV$^\mathrm{e}$} &
\multicolumn{2}{c}{Gal BV$^\mathrm{f}$}\\
\hline
\ion{He}{i} & 10.99 & 0.02 & {\em 11.18} & {\em 0.04}\,(14) & {\em 11.07} & {\em 0.05}\,(10) & {\em 11.14} & {\em 0.04}\,(10) & {\em 11.19} & {\em 0.04}\,(15) & \multicolumn{2}{c}{\ldots} & \multicolumn{2}{c}{\ldots} & {\em 11.13} & {\em 0.22} & {\em 11.05} & {\em 0.10} & \multicolumn{2}{c}{\ldots}\\
\ion{C}{i} & 8.52 & 0.06 & {\em 7.94} & {\em 0.10}\,(4) & {\em 8.10} & {\em 0.07}\,(2) & \multicolumn{2}{c}{\ldots} & \multicolumn{2}{c}{\ldots} & {\em 8.14} & {\em 0.13} & \multicolumn{2}{c}{\ldots} & \multicolumn{2}{c}{\ldots} & \multicolumn{2}{c}{\ldots} & \multicolumn{2}{c}{\ldots}\\
\ion{C}{ii} & 8.52 & 0.06 & {\em 8.10} & {\em 0.09}\,(3) & {\em 8.24} & {\em 0.06}\,(3) & \multicolumn{2}{c}{{\em 8.33}\,(1)} & {\em 8.15} & {\em 0.05}\,(3) & \multicolumn{2}{c}{\ldots} & {\em 8.27} & {\em 0.13} &{\em 8.20} & {\em 0.10} & {\em 8.22} & {\em 0.15} & 7.87 & 0.16\\
\ion{N}{i} & 7.92 & 0.06 & {\em 8.41} & {\em 0.09}\,(20) & {\em 8.40} & {\em 0.10}\,(16) & {\em 8.25} & {\em 0.04}\,(11) & {\em 8.50} & {\em 0.07}\,(11) & {\em 8.37} & {\em 0.21$^*$} & \multicolumn{2}{c}{\ldots} & \multicolumn{2}{c}{\ldots} & \multicolumn{2}{c}{\ldots} & \multicolumn{2}{c}{\ldots}\\
\ion{N}{ii} & 7.92 & 0.06 & \multicolumn{2}{c}{{\em 8.32}\,(1)} & \multicolumn{2}{c}{{\em 8.36}\,(1)} & \multicolumn{2}{c}{{\em 8.28}\,(1)} & {\em 8.51} & {\em 0.06}\,(16) & \multicolumn{2}{c}{\ldots} & {\em 7.63} & {\em 0.15} & {\em 7.75} & {\em 0.27} & {\em 7.78} & {\em 0.27} & 7.90 & 0.22\\
\ion{O}{i} & 8.83 & 0.06 & {\em 8.78} & {\em 0.05}\,(13) & {\em 8.70} & {\em 0.04}\,(9) & {\em 8.79} & {\em 0.07}\,(6) & {\em 8.78} & {\em 0.04}\,(6) & 8.77 & 0.12 & \multicolumn{2}{c}{\ldots} & \multicolumn{2}{c}{\ldots} & \multicolumn{2}{c}{\ldots} & \multicolumn{2}{c}{\ldots}\\
\ion{O}{ii} & 8.83 & 0.06 & \multicolumn{2}{c}{\ldots} & \multicolumn{2}{c}{\ldots} & \multicolumn{2}{c}{\ldots} & {\em 8.83} & {\em 0.03}\,(5) & \multicolumn{2}{c}{\ldots} & {\em 8.55} & {\em 0.14} &{\em 8.64} & {\em 0.20} & {\em 8.52} & {\em 0.16} & 8.89 & 0.14\\
\ion{Ne}{i} & 8.08 & 0.06 & 8.39 & 0.07\,(7) & 8.42 & 0.06\,(5) & 8.50 & 0.14\,(7) & 8.40 & 0.08\,(9) & \multicolumn{2}{c}{\ldots} & \multicolumn{2}{c}{\ldots} & \multicolumn{2}{c}{\ldots} & 8.10 & 0.05$^{**}$ & \multicolumn{2}{c}{\ldots}\\
\ion{Mg}{i} & 7.58 & 0.01 & {\em 7.52} & {\em 0.08}\,(7) & {\em 7.46} & {\em 0.04}\,(4) & {\em 7.60} & {\em 0.04}\,(4) & \multicolumn{2}{c}{\ldots} & 7.48 & 0.17 & \multicolumn{2}{c}{\ldots} & \multicolumn{2}{c}{\ldots} & \multicolumn{2}{c}{\ldots} & \multicolumn{2}{c}{\ldots}\\
\ion{Mg}{ii} & 7.58 & 0.01 & {\em 7.53} & {\em 0.04}\,(12) & {\em 7.43} & {\em 0.04}\,(6)& {\em 7.40} & {\em 0.05}\,(5) & {\em 7.42} & {\em 0.02}\,(4) & 7.46 & 0.17 & {\em 7.42} & {\em 0.22} & {\em 7.59} & {\em 0.22} &  {\em 7.38} & {\em 0.12} & 7.70 & 0.34\\
\ion{Al}{i} & 6.49 & 0.01 & 6.11 & 0.06\,(2) & 6.06 & 0.06\,(2) & \multicolumn{2}{c}{\ldots} & \multicolumn{2}{c}{\ldots} & \multicolumn{2}{c}{\ldots} & \multicolumn{2}{c}{\ldots} & \multicolumn{2}{c}{\ldots} & \multicolumn{2}{c}{\ldots} & \multicolumn{2}{c}{\ldots}\\
\ion{Al}{ii} & 6.49 & 0.01 & 6.39 & 0.17\,(5) & 6.55 & 0.27\,(6) & 6.38 & 0.38\,(6) & 6.14 & 0.08\,(4) & \multicolumn{2}{c}{\ldots} & \multicolumn{2}{c}{\ldots} & \multicolumn{2}{c}{\ldots} & \multicolumn{2}{c}{\ldots} & \multicolumn{2}{c}{\ldots}\\
\ion{Al}{iii} & 6.49 & 0.01 & \multicolumn{2}{c}{\ldots} & \multicolumn{2}{c}{\ldots} & \multicolumn{2}{c}{\ldots} & 7.00 & 0.38\,(4) & \multicolumn{2}{c}{\ldots} & {\em 6.08} & {\em 0.12} & {\em 6.21} & {\em 0.19} & {\em 6.12} & {\em 0.18} & 6.31 & 0.15\\
\ion{Si}{ii} & 7.56 & 0.01 & 7.58 & 0.19\,(4) & 7.45 & 0.29\,(3) & 7.33 & 0.05\,(2) & 7.27 & 0.13\,(3) & 7.33 & 0.17 & \multicolumn{2}{c}{\ldots} & {\em 7.42} & {\em 0.23} & {\em 7.19} & {\em 0.21} & \multicolumn{2}{c}{\ldots}\\
\ion{Si}{iii} & 7.56 & 0.01 & \multicolumn{2}{c}{\ldots} & \multicolumn{2}{c}{\ldots} & \multicolumn{2}{c}{\ldots} & 8.13 & 0.23\,(3) & \multicolumn{2}{c}{\ldots} & {\em 7.26} & {\em 0.19} & {\em 7.42} & {\em 0.23} & {\em 7.19} & {\em 0.21} & 7.50 & 0.15\\
\ion{P}{ii} & 5.56 & 0.06 & \multicolumn{2}{c}{\ldots} & \multicolumn{2}{c}{\ldots} & \multicolumn{2}{c}{\ldots} & 5.53 & 0.06\,(4) & \multicolumn{2}{c}{\ldots} & \multicolumn{2}{c}{\ldots} & \multicolumn{2}{c}{\ldots} & \multicolumn{2}{c}{\ldots} & \multicolumn{2}{c}{\ldots}\\
\ion{S}{ii} & 7.20 & 0.06 & {\em 7.15} & {\em 0.07}\,(14) & {\em 7.07}& {\em 0.08}\,(10) & {\em 7.12} & {\em 0.08}\,(16) & {\em 7.05} & {\em 0.09}\,(26) & \multicolumn{2}{c}{\ldots} & \multicolumn{2}{c}{\ldots} & \multicolumn{2}{c}{\ldots} & \multicolumn{2}{c}{\ldots} & \multicolumn{2}{c}{\ldots}\\
\ion{S}{iii} & 7.20 & 0.06 & \multicolumn{2}{c}{\ldots} & \multicolumn{2}{c}{\ldots} & \multicolumn{2}{c}{\ldots} & {\em 7.08} & {\em 0.04}\,(2) & \multicolumn{2}{c}{\ldots} & {\em 7.24} & {\em 0.14} & \multicolumn{2}{c}{\ldots} & 6.87 & 0.26 & \multicolumn{2}{c}{\ldots}\\
\ion{Ca}{ii} & 6.35 & 0.01 & \multicolumn{2}{c}{6.31\,(1)} & \multicolumn{2}{c}{\ldots} & \multicolumn{2}{c}{\ldots} & \multicolumn{2}{c}{\ldots} & \multicolumn{2}{c}{\ldots} & \multicolumn{2}{c}{\ldots} & \multicolumn{2}{c}{\ldots} & \multicolumn{2}{c}{\ldots} & \multicolumn{2}{c}{\ldots}\\
\ion{Sc}{ii} & 3.10 & 0.01 & 2.57 & 0.14\,(3) & 2.64 & 0.27\,(2) & \multicolumn{2}{c}{2.42\,(1)} & \multicolumn{2}{c}{\ldots} & 3.13 & 0.20 & \multicolumn{2}{c}{\ldots} & \multicolumn{2}{c}{\ldots} & \multicolumn{2}{c}{\ldots} & \multicolumn{2}{c}{\ldots}\\
\ion{Ti}{ii} & 4.94 & 0.02 & {\em 4.89} & {\em 0.13}\,(29) & {\em 4.86} & {\em 0.12}\,(25) & {\em 4.87} & {\em 0.09}\,(15) & {\em 5.01} & {\em 0.07}\,(4) & 4.86 & 0.25 & \multicolumn{2}{c}{\ldots} & \multicolumn{2}{c}{\ldots} & \multicolumn{2}{c}{\ldots} & \multicolumn{2}{c}{\ldots}\\
\ion{V}{ii} & 4.02 & 0.02 & 3.57 & 0.06\,(6) & 3.63 & 0.08\,(6) & 3.55 & 0.04\,(3) & \multicolumn{2}{c}{\ldots} & \multicolumn{2}{c}{\ldots} & \multicolumn{2}{c}{\ldots} & \multicolumn{2}{c}{\ldots} & \multicolumn{2}{c}{\ldots} & \multicolumn{2}{c}{\ldots}\\
\ion{Cr}{ii} & 5.69 & 0.01 & 5.62 & 0.08\,(29) & 5.55 & 0.08\,(24) & 5.28 & 0.08\,(21) & 5.42 & 0.11\,(8) & 5.61 & 0.23 & \multicolumn{2}{c}{\ldots} & \multicolumn{2}{c}{\ldots} & \multicolumn{2}{c}{\ldots} & \multicolumn{2}{c}{\ldots}\\
\ion{Mn}{ii} & 5.53 & 0.01 & 5.38 & 0.02\,(7) & 5.36 & 0.04\,(6) & \multicolumn{2}{c}{5.24\,(1)} & \multicolumn{2}{c}{5.33\,(1)} & 5.81 & 0.20 & \multicolumn{2}{c}{\ldots} & \multicolumn{2}{c}{\ldots} & \multicolumn{2}{c}{\ldots} & \multicolumn{2}{c}{\ldots}\\
\ion{Fe}{i} & 7.50 & 0.01 & 7.34 & 0.12\,(21) & 7.37 & 0.10\,(13) & \multicolumn{2}{c}{\ldots} & \multicolumn{2}{c}{\ldots} & 7.56 & 0.24 & \multicolumn{2}{c}{\ldots} & \multicolumn{2}{c}{\ldots} & \multicolumn{2}{c}{\ldots} & \multicolumn{2}{c}{\ldots}\\
\ion{Fe}{ii} & 7.50 & 0.01 & {\em 7.52} & {\em 0.09}\,(35) & {\em 7.42} & {\em 0.05}\,(35) & {\em 7.34} & {\em 0.07}\,(24) & {\em 7.47} & {\em 0.08}\,(20) & 7.40 & 0.11 & \multicolumn{2}{c}{\ldots} & \multicolumn{2}{c}{\ldots} & \multicolumn{2}{c}{\ldots} & \multicolumn{2}{c}{\ldots}\\
\ion{Fe}{iii} & 7.50 & 0.01 & \multicolumn{2}{c}{\ldots} & \multicolumn{2}{c}{\ldots} & \multicolumn{2}{c}{\ldots} & 7.45 & 0.07\,(10) & \multicolumn{2}{c}{\ldots} & \multicolumn{2}{c}{\ldots} & \multicolumn{2}{c}{\ldots} &  7.36 & 0.20 & \multicolumn{2}{c}{\ldots}\\
\ion{Ni}{ii} & 6.25 & 0.01 & 6.30 & 0.06\,(7) & 6.17 & 0.04\,(7) & 6.00 & 0.01\,(2) & 5.91 & 0.08\,(4) & \multicolumn{2}{c}{\ldots} & \multicolumn{2}{c}{\ldots} & \multicolumn{2}{c}{\ldots} & \multicolumn{2}{c}{\ldots} & \multicolumn{2}{c}{\ldots}\\
\ion{Sr}{ii} & 2.92 & 0.02 & 2.37 & 0.04\,(2) & 2.41 & 0.03\,(2) & 2.49 & 0.04\,(2) & \multicolumn{2}{c}{\ldots} & 2.41 & 0.21 & \multicolumn{2}{c}{\ldots} & \multicolumn{2}{c}{\ldots} & \multicolumn{2}{c}{\ldots} & \multicolumn{2}{c}{\ldots}\\
\ion{Ba}{ii} & 2.22 & 0.02 & \multicolumn{2}{c}{2.00\,(1)} & \multicolumn{2}{c}{2.13\,(1)} & \multicolumn{2}{c}{\ldots} & \multicolumn{2}{c}{\ldots} & \multicolumn{2}{c}{\ldots} & \multicolumn{2}{c}{\ldots} & \multicolumn{2}{c}{\ldots} & \multicolumn{2}{c}{\ldots} & \multicolumn{2}{c}{\ldots}\\
\hline
\end{tabular}\\
$^\mathrm{a}$ Grevesse \& Sauval (\cite{GrSa98}): meteoritic abundances for
non-volatile elements; \hspace{1mm}
$^\mathrm{b}$ Venn (\cite{Venn95a}, \cite{Venn95b}); \hspace{1mm}
$^\mathrm{c}$ Daflon \& Cunha (\cite{DaCu04}): mean values of stars from 17
clusters/associations at $R_0$\,$\pm$2\,kpc; \hspace{1mm}
$^\mathrm{d}$ Gummersbach et al. (\cite{Gummersbachetal98}): mean values of
10 stars at $R_0$\,$\pm$2\,kpc; \hspace{1mm}
$^\mathrm{e}$ mean from Kilian (\cite{Kilian92}, \cite{Kilian94}); \hspace{1mm}
$^\mathrm{f}$ Rolleston et al. (\cite{Rollestonetal00}): mean of
stars from 11 clusters/associations at $R_0$\,$\pm$2\,kpc; \linebreak
$^\mathrm{*}$~reanalysis by Venn \& Przybilla~(\cite{VePr03}) -- result from
Venn~(\cite{Venn95b}): {\em 8.05}$\pm${\em 0.19}; \hspace{1mm}
$^\mathrm{**}$ \ion{Ne}{ii}
\end{table}
\end{landscape}

To conclude, effective temperatures from our approach appear to be 
systematically lower than from previous model
atmosphere analyses and higher than from previous (spectro)photometric
studies. Not only is consistency now achieved from these indicators, but
also the uncertainties in the parameter determination have been markedly
reduced by bringing all the indicators simultaneously into agreement. This
provides a solid basis for precise metal abundance analyses, which are
discussed next.

\section{Abundance analysis}\label{sectabus}
The chemical composition of the stars is deduced from the analysis of the
weak line spectra of the individual elements. This is done by fitting
non-LTE and LTE {\em line profiles} to the observation, as this puts tighter
constraints on the modelling than is possible by reproducing equivalent widths alone.
The results from the analysis of several hundred spectral lines from 30
ionization stages of 20 chemical species are summarised in Appendix~\ref{apa}.
We have also thoroughly tested the line list for suitability in abundance
analyses of BA-stars on the main sequence (Przybilla~\cite{Przybilla02}) 
and recommend it for use in future applications.

\begin{figure*}[ht!]
\sidecaption
\includegraphics[width=12cm]{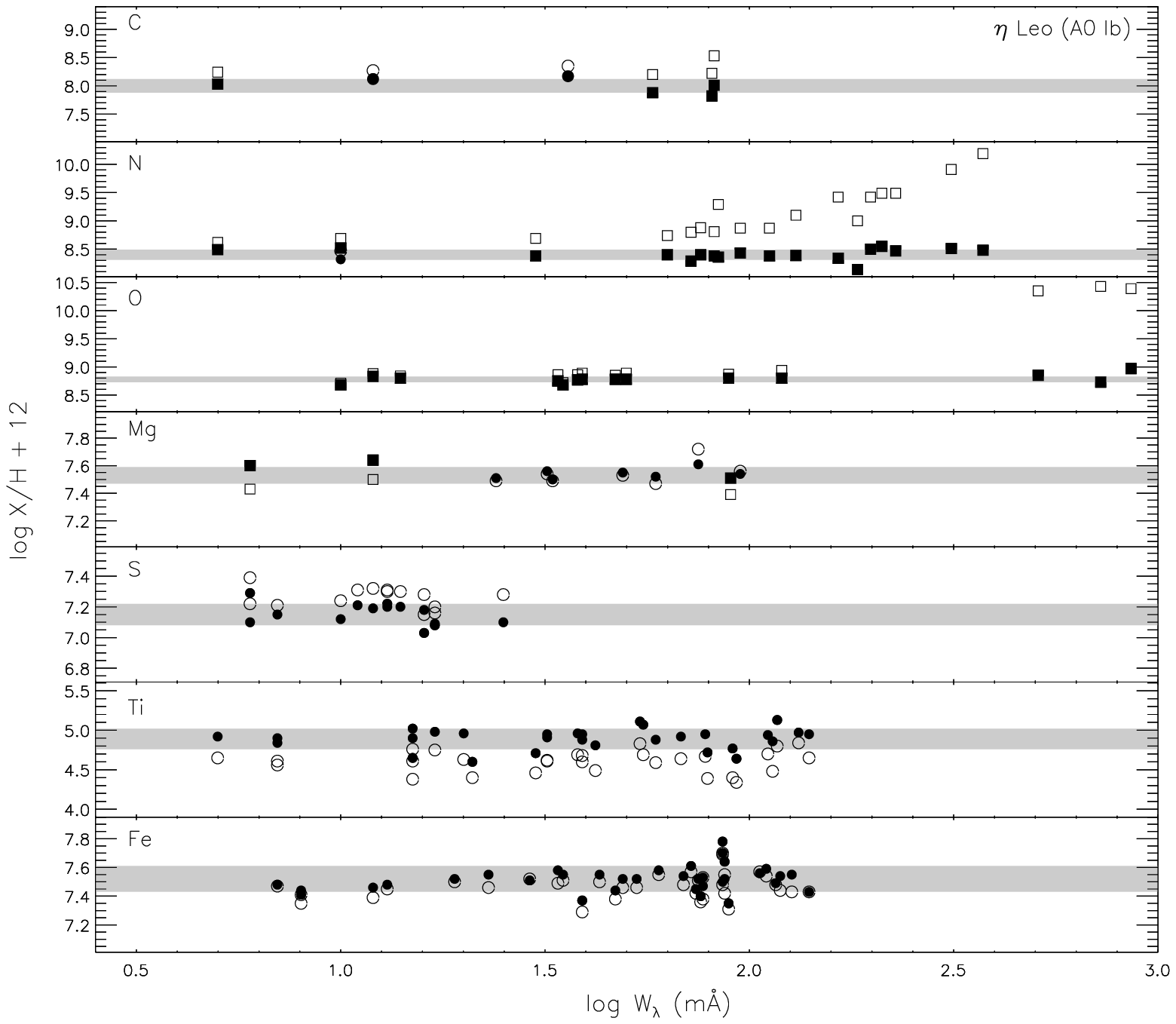}
\caption[]{Elemental abundances in $\eta$\,Leo from individual spectral
lines of elements with non-LTE calculations, plotted as a function
of equivalent width: non-LTE (solid) and LTE results (open symbols) for
neutral (boxes) and single-ionized species (circles). The grey bands cover
the 1$\sigma$-uncertainty ranges around the non-LTE mean values (see
Table~\ref{tableabu}). Note that in the determination of these some
additional spectral lines have been used in some cases, which are accessible 
to spectrum synthesis only. Data for the strongest \ion{O}{i} lines has been
adopted from Przybilla et al.~(\cite{Przybillaetal00}).
Proper non-LTE calculations reduce the
line-to-line scatter and remove systematic trends. Note that even weak lines
can show considerable departures from LTE.}
\label{nlteabus}
\end{figure*}

A comparison of non-LTE and LTE results from the individual metal lines in 
$\eta$\,Leo is made in Fig.~\ref{nlteabus}. This super\-giant shows the richest
metal-line spectrum of all our sample objects. Note that a few lines from
Table~\ref{taba1} without measured equivalent widths had to be omitted for
this, and results for a few strong \ion{O}{i} lines have been added 
from Przybilla et al.~(\cite{Przybillaetal00}). By correctly accounting for non-LTE
effects homogeneous abundances are derived from {\em all} the available
indicators, which show little scatter around the mean value (the derived
abundance). This also implies that the microturbulent velocity is correctly
chosen, and it is the same for the elements under study. The non-LTE analysis thus
avoids the systematic trends of abundance with line-strength in the approximation of 
LTE, such as for \ion{N}{i} or \ion{O}{i}. In other cases, like for
singly ionized sulphur, titanium and iron, the non-LTE abundances are
systematically shifted relative to LTE. Also the statistical scatter is
reduced in non-LTE, and contrary to common assumption significant non-LTE
abundance corrections are found even in the weak line limit. These can
exceed 0.3\,dex for lines with
$W_{\lambda}$\,$\lesssim$\,10\,m{\AA}, in particular for the more luminous
objects, where the non-LTE effects are amplified. The qualitative behaviour
of non-LTE and LTE abundances is similar in all the sample stars, which
occurs because of the reasons discussed in Sect.~\ref{sectlform}.

Mean elemental abundances for the sample stars are given in
Table~\ref{tableabu}, along with the solar standard and results from the
literature on Galactic A-supergiants and their main sequence progenitors of early
B-type at less than 2\,kpc distance (in order to avoid artefacts introduced
by the Galactic abundance gradient). 
Note that we prefer solar standard abundances as summarised by Grevesse \&
Sauval~(\cite{GrSa98}) over the more recent compilation of Asplund et
al.~(\cite{Asplundetal05}) for several reasons discussed below.
Non-LTE results are indicated by italics.
For the sample stars statistical uncertainties (1-$\sigma$ standard
deviations from the line-to-line scatter) are given, with the number of
analysed lines in parenthesis. In the case of the previous studies the 
uncertainties denote 1-$\sigma$ standard deviations from the star-to-star scatter.

The present {\em non-LTE} analysis finds abundances compatible with the LTE study 
(with non-LTE corrections for a few elements) of
Venn~(\cite{Venn95a}, \cite{Venn95b}) on less-luminous A-SGs. This
is because the objects in that study are, on the mean, far cooler and
of higher surface gravity than ours, which reduces the importance of non-LTE. In
such a case a {\em tailored} LTE analysis is meaningful for most elements. 
Consequently, we and Venn recover abundances which agree within the
uncertainties with the present-day ISM abundances (which appear to be
slightly sub-solar), as expected. Exceptions
are the carbon and nitrogen abundances, which reflect mixing of the surface layers of 
the supergiants by nuclear processed material, see below.

The B main sequence stars (B-MSs) together with \ion{H}{ii} regions define the
present-day abundances in the solar neighbourhood. Consequently, mutual agreement with
our results exists for the $\alpha$-process elements with non-LTE abundances. The B-star 
abundances of helium and nitrogen are lower and the carbon abundances higher than in the
BA-SGs, indicating little to no mixing with nuclear processed matter, which is
also expected. Note that the B-MSs oxygen abundances appear to be slightly lower than
in our BA-SGs; this may be an effect of small number statistics. Note also
that the LTE analysis by Rolleston et al.~(\cite{Rollestonetal00}) finds
systematically different B-MSs abundances than the other three B-MSs studies.
Little information is available for the iron-group
abundances, and nothing on s-process abundances in B-MSs. BA-type
supergiants and B-MSs are therefore complementary for galactochemical and
stellar evolution studies: the former contain information on the heavier element abundances
and mixing patterns, while the latter provide stellar present-day abundances
for the light elements. Consistency can be checked on the basis of the
$\alpha$-elements for which both star groups have features in common.

\begin{figure*}[ht!]
\resizebox{0.497\hsize}{!}{\includegraphics{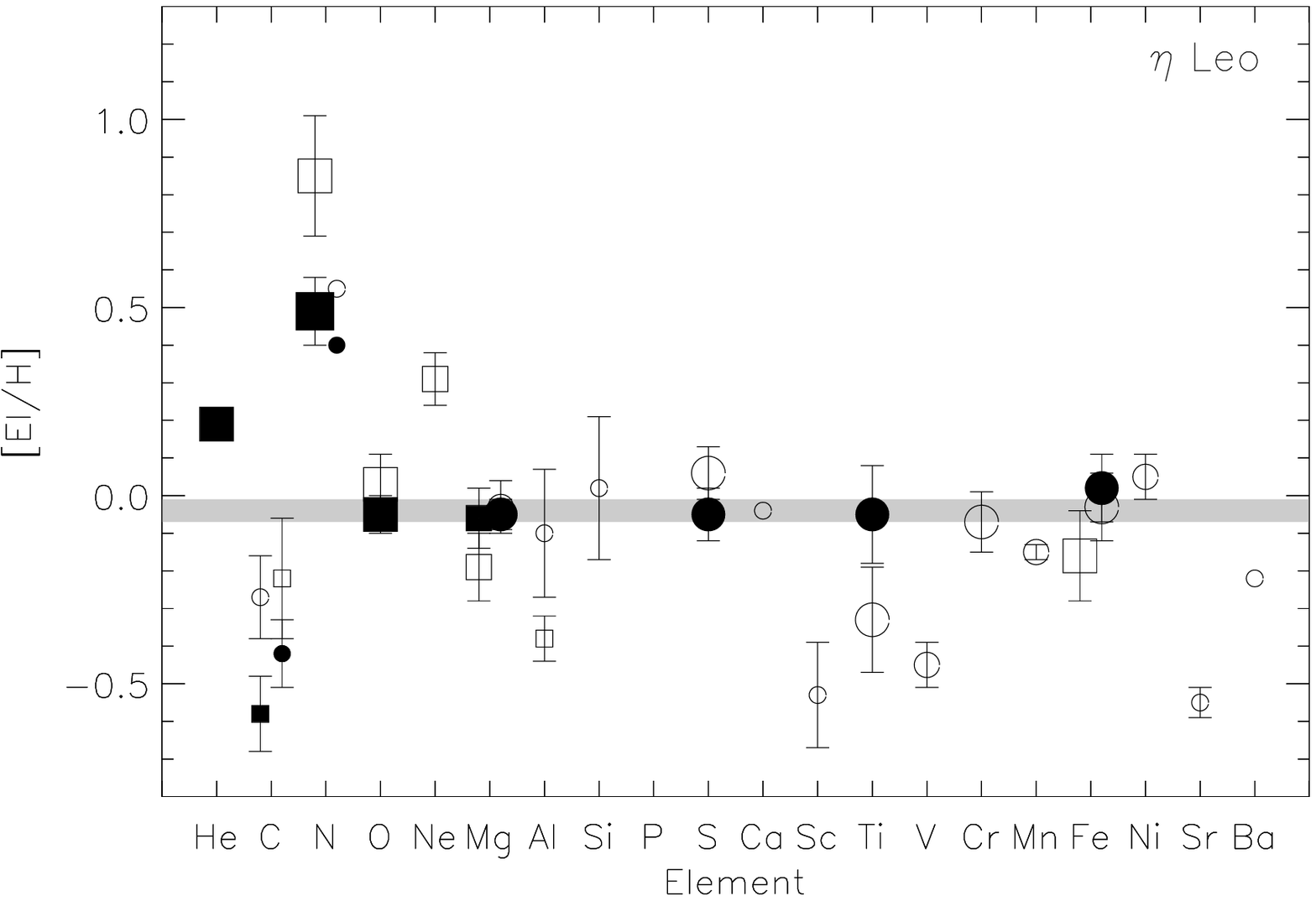}}
\hfill
\resizebox{0.497\hsize}{!}{\includegraphics{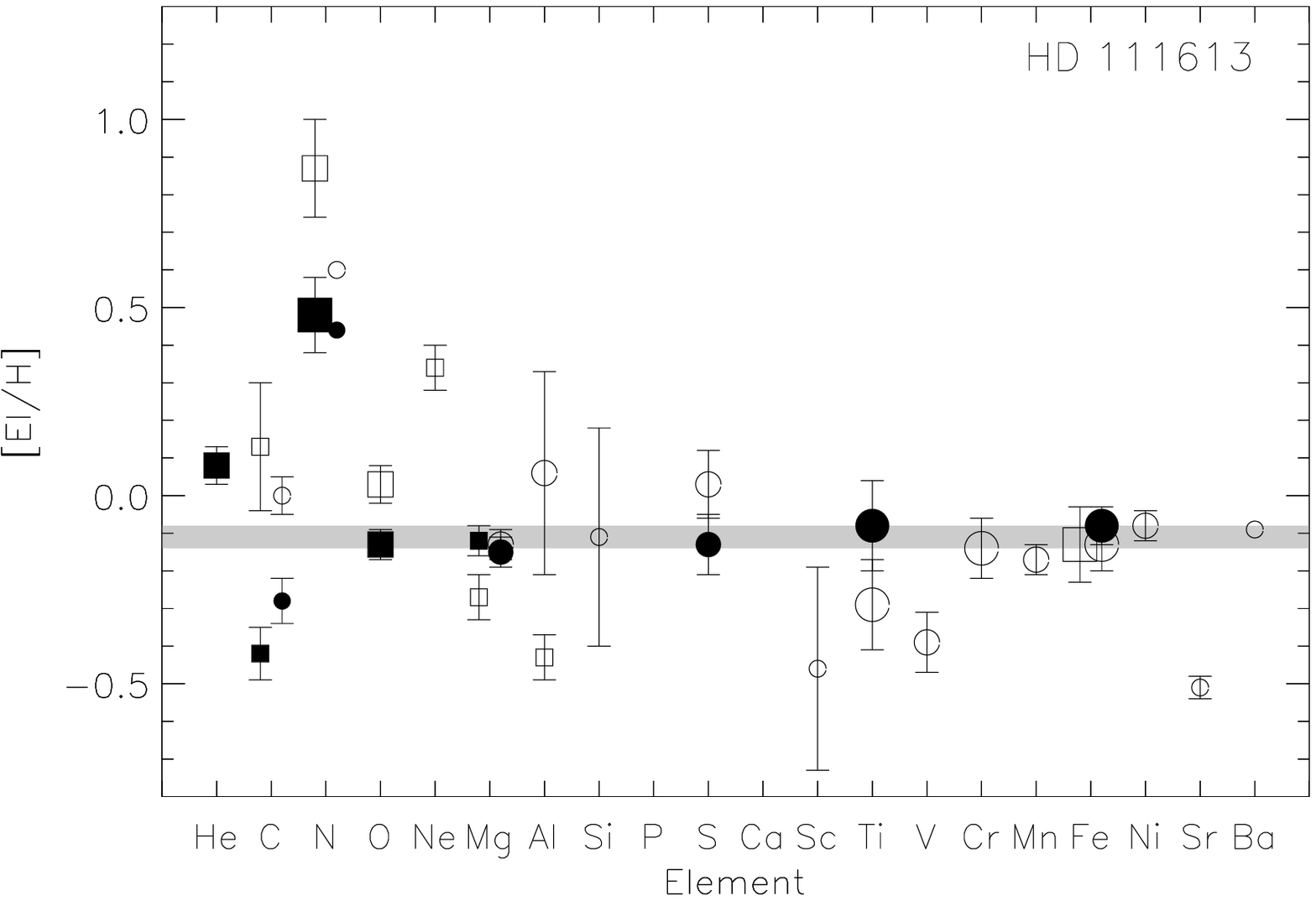}}\\[-2mm]
\resizebox{0.497\hsize}{!}{\includegraphics{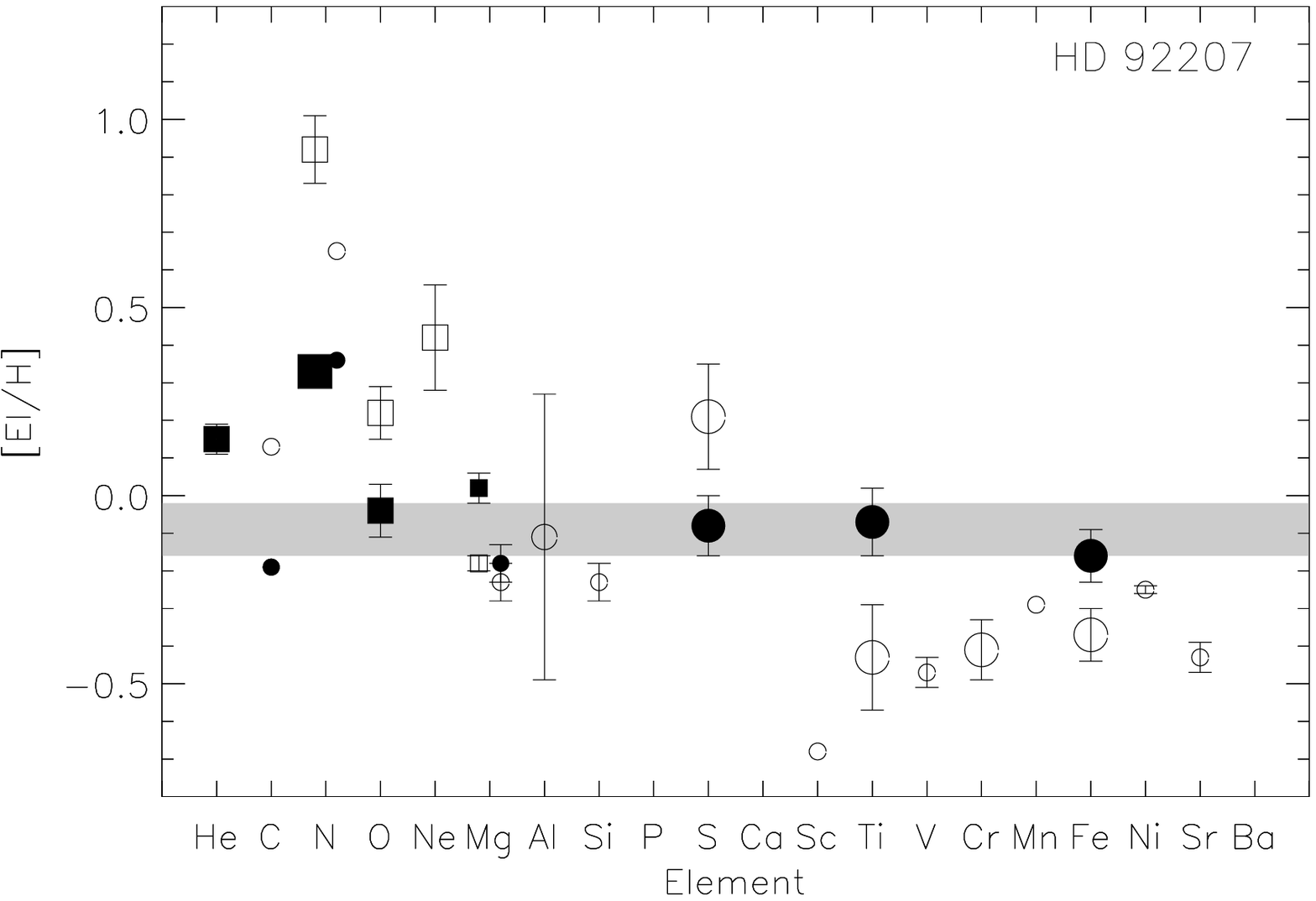}}
\hfill
\resizebox{0.497\hsize}{!}{\includegraphics{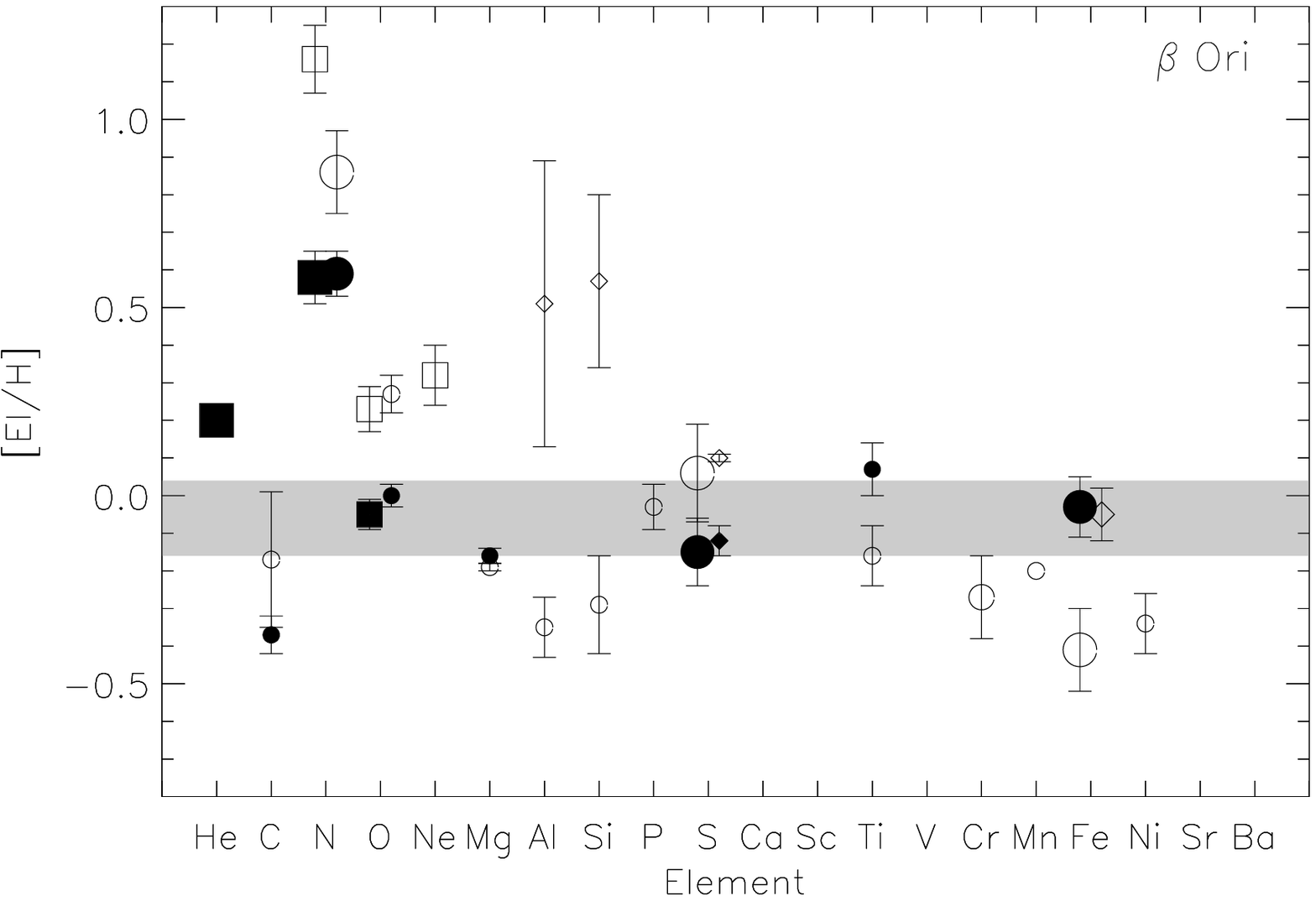}}
\caption[]{Results from the elemental abundance analysis (relative to the
solar composition) for our sample stars. Filled symbols denote non-LTE, open
symbols LTE results. The symbol size codes the number of spectral lines
analysed -- small: 1 to 5, medium: 6 to 10, large: more than 10. 
Boxes: neutral, circles: single-ionized, diamonds: double-ionized species. 
The error bars represent 1$\sigma$-uncertainties from the
line-to-line scatter. The grey shaded area marks the deduced metallicity of
the objects within 1$\sigma$-errors. 
Non-LTE abundance
analyses reveal a striking similarity to the solar abundance distribution,
except for the light elements which have been affected by mixing with
nuclear-processed matter.}
\label{abuplot}
\end{figure*}

The comparison of our sample stars with the solar standard 
(Grevesse \& Sauval~\cite{GrSa98}) is visualised in Fig.~\ref{abuplot}. 
First, we want to discuss the results from the LTE analysis, which comprises
a wider variety of elements than can be done in non-LTE at present.
For the least luminous supergiant of our sample, $\eta$\,Leo, close to solar
abundances are derived for most of the $\alpha$-process and iron group
elements. The only s-process elements accessible through their resonance
lines, \ion{Sr}{ii} and \ion{Ba}{ii}, appear underabundant, as are the
lightest iron group elements, scandium through vanadium. Neon appears to be
overabundant and a discrepancy in the \ion{Al}{i/ii} ionization equilibrium
is noteworthy, while \ion{Mg}{i/ii} is on the verge of a discrepancy. 
Of the light elements, carbon is under- and nitrogen
overabundant, with a discrepancy in the \ion{N}{i/ii} ionization balance
indicated. Note that these LTE results are valid for the finally adopted
parameters from the non-LTE analysis. A pure LTE study would indicate a slightly higher 
$T_{\rm eff}$ from most of the ionization equilibria, implying
an overall increase of the metallicity and a more pronounced \ion{N}{i/ii} discrepancy.
Similar abundance patterns are found for the LC Iab supergiant HD\,111613, with several of the
$\alpha$-processes elements (O, Ne and S) showing a tendency to being overabundant, as
also appears to be the case for carbon. For these two stars an LTE analysis
appears meaningful, in the sense that abundances are derived which are expected 
for a young star of the Galactic thin disk, with only a few exceptions. 

The situation changes drastically when considering an object of even higher
luminosity, like HD\,92207. Practically no element comes close to the solar
standard, with the entire iron group appearing to be vastly underabundant, 
and the $\alpha$-process and light elements largely overabundant. 
An LTE analysis suggests an abundance pattern resembling $\alpha$-enhancement for this star, 
which is appropriate for the old halo population but not for a Population I
object. This pattern is basically also derived for $\beta$\,Ori.
However, at the higher atmospheric temperatures features from double-ionized species
appear, which suggest drastic imbalances in several ionization equilibria.
No parameter combination can be found that brings all the available
ionization equilibria simultaneously into agreement in LTE. 

The non-LTE analysis of some of the astrophysically most interesting elements 
on the other hand reveals a rather homogeneous appearance of the abundance 
distribution in the four stars. Note that we derive {\em absolute abundances}, 
which depend on physical data and not on oscillator strengths (and line broadening data)
{\em tuned} to match observations, as is the case in many {\em differential}
studies of solar-type stars. The
$\alpha$-process and iron group elements reproduce the solar abundance
pattern, showing little scatter around slightly sub-solar mean metallicities
(the grey bands in Fig.~\ref{abuplot}). Among the light elements, helium and nitrogen 
are enhanced and carbon is underabundant, which is qualitatively well
understood in the framework of the evolution of rotating massive stars, see
the next section for details. 

The total abundance of the fusion
catalysts C, N, and O in the sample stars correlates well with the stellar metallicity. 
Use of the solar abundances of Asplund et al.~(\cite{Asplundetal05}) 
instead of the Grevesse \& Sauval~(\cite{GrSa98}) data as 
standard would imply an overabundance of CNO relative to the heavier
elements in these young stars, in particular [O/Fe]\,$>$\,0. This is inconsistent with
other astrophysical indicators. Here, the more recent solar abundance data
may imply another conflict like the recently unveiled
solar model problem, i.e. the loss of accordance between helioseismological 
measurements and the predictions of theoretical solar models 
(Bahcall et al.~\cite{Bahcalletal05}).

Note that good agreement for the available non-LTE
ionization equilibria is obtained, and that the statistical scatter in the
abundance determination is typically reduced with respect to LTE, amounting
to 0.05 to 0.10\,dex (1-$\sigma$ standard deviations). 
Nitrogen and oxygen are particularly good examples. This comes close to the accuracy 
that is usually obtained in differential analyses of solar-type stars, however on absolute
terms. 

The non-LTE abundance corrections vary from element to element and
there is a well-known tendency for increased non-LTE corrections with luminosity (i.e.
lower $\log g$), and temperature. Typically, {\em non-LTE abundance corrections
amount to a factor of two to three on the mean}. In some cases, the
corrections can amount to factors of several tens for individual
transitions, see e.g. oxygen in Fig.~\ref{nlteabus}, and in the most extreme
cases they exceed a factor of a hundred. Many strong lines of
\ion{Ti}{ii} and \ion{Fe}{ii} have been omitted in the present study,
because of the additional computational costs and their insensitivity as abundance
indicators. These also show stronger non-LTE effects as the corresponding
weak-line spectra. 
This is also the case for the prominent \ion{Mg}{ii}
$\lambda$4481 feature (Przybilla et al.~\cite{Przybillaetal01a}). 

The only exception from our general improvement due to non-LTE refinements
is the \ion{Mg}{i/ii} ionization balance in HD\,922207, which we
attribute to small insufficiencies of our LTE model atmospheres at highest
luminosity. The
\ion{Mg}{i} lines are highly sensitive to even the smallest changes in the
modelling close to the Eddington limit, see the discussion in
Sect.~\ref{sectmodels}. In this case the \ion{N}{i/ii} ionization
equilibrium is a better choice for the temperature indicator, as its
behaviour is better constrained -- much more accurate atomic data are
available for nitrogen than for magnesium, and the ground state
ionization for \ion{N}{i} is determined by the Lyman continuum instead of the 
more complicated radiation field longward of the Lyman jump, which is 
more prone to modelling uncertainties, as in the case of \ion{Mg}{i}.

The detailed non-LTE behaviour of the ionic species under study is well
explained from their atomic structure, as discussed in Sect.~\ref{sectlform}.
This allows us to estimate, at least qualitatively, the non-LTE behaviour of the large
number of elements in the supergiant spectra, for which we do not have non-LTE 
spectrum synthesis at
hand for the moment. Certainly, the entire singly-ionized iron group elements 
will show a similar non-LTE weakening like \ion{Ti}{ii} and
\ion{Fe}{ii}, as they all are subject to non-LTE overionization. This will
also be the case for \ion{Sr}{ii} and \ion{Ba}{ii}. A non-LTE strengthening
of the \ion{Ne}{i} lines was described by Sigut~(\cite{Sigut99}) in B-MSs,
which can be expected to occur to a similar extent 
also at lower $T_{\rm eff}$ in the BA-SGs. This would imply a
present-day neon abundance from the sample stars compatible with their
overall metallicity close to the solar value as given by Grevesse \&
Sauval~(\cite{GrSa98}), and agreement with the ISM abundance as derived from the Orion
nebula (Esteban et al.~\cite{Estebanetal04}), for non-LTE abundance corrections by a factor
$\sim$2. Compatibility with the value of {\rm $\log$\,Ne/H\,$+$\,12\,$=$\,8.27} 
as determined by Drake \& Testa~(\cite{DrTe05}) from nearby solar-like stars is possible 
for lower non-LTE abundance corrections. The solar value of Asplund et
al.~(\cite{Asplundetal05}) which contributes to the solar model problem
could only be restored for unrealistically high non-LTE corrections by a
factor $\sim$4--5 (the \ion{Ne}{i} lines are typically weak).
Note that an increase of the surface abundance of neon due to mixing with
nuclear-processed matter is only expected in the Wolf-Rayet phase of the evolution of 
massive stars, such that BA-SGs should reflect present-day neon abundances of the ISM.

\begin{figure*}[ht!]
\resizebox{\hsize}{!}{\includegraphics{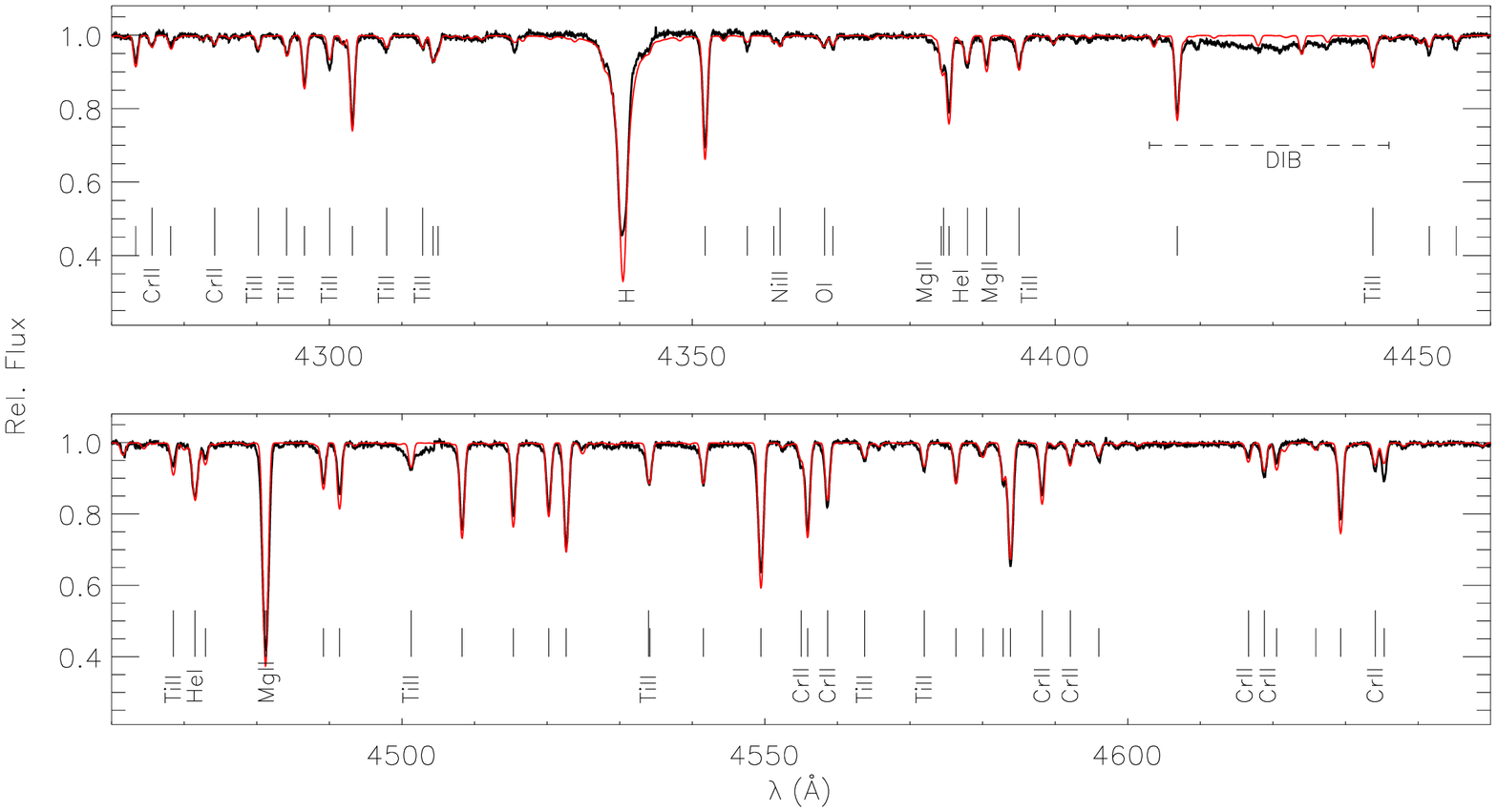}}\\[-7mm]
\caption[]{Comparison of our spectrum synthesis (thin red line) with
the high-resolution spectrum of HD\,92207 (full line). The major spectral
features are identified, short vertical marks designate \ion{Fe}{ii} lines.
The location of the diffuse interstellar band around 4430\,{\AA} is indicated. 
With few exceptions, excellent agreement between theory and observation is found
(H$\gamma$ is affected by the stellar wind).}
\label{synthesis}
\end{figure*}

We conclude that the derived abundance patterns from
our LTE study are probably an artefact of the assumption
of detailed equilibrium. It may be speculated that close to solar abundance
patterns will be recovered for the four sample stars from a full non-LTE analysis. 
We will continue our efforts to improve the status of the modelling situation, but 
note that for many ionic species practically none of the required atomic data 
needed to construct reliable non-LTE model atoms are available at present.

Another important aspect of the non-LTE abundance analysis are constraints
for the systematic uncertainties. This includes tests for the response of
the spectrum synthesis to modifications of the stellar parameters as well as
to variations of the atomic data that is used for assembling the non-LTE model 
atoms. 
Such tests have been done for \ion{C}{i/ii}, \ion{N}{i/ii}, \ion{O}{i} and
\ion{Mg}{i/ii} by Przybilla et al.~(\cite{Przybillaetal00}, \cite{Przybillaetal01a},b) 
and Przybilla \& Butler~(\cite{PrBu01}). When we scale these results for the stellar
parameter uncertainties of the present work and consider continuum placement
in our high-S/N spectra to be a negligible problem we can conclude that the
systematic uncertainties of our non-LTE abundance determination are of the
order $\sim$0.10\,dex (1-$\sigma$-error). This is from quadratic summation over 
all uncertainty contributors, assuming them to be independent. As this is not strictly the
case for the stellar parameters (e.g. $T_{\rm eff}$/$\log g$ have to be
varied simultaneously to retain consistency of a spectroscopic indicator
with observation, see e.g. Fig.~\ref{kiel}), the true systematic uncertainties may be slightly
smaller. The given uncertainty has consequently to be viewed as a
conservative estimate. Note that the main contribution to the systematic uncertainty
of abundance determinations in BA-SGs comes from the uncertainties in the stellar
parameters. The typical uncertainties in oscillator strengths,
photoionization and collisional excitation cross-sections from
quantummechanical {\em ab-initio} computations are of the order 10--20\% and
therefore contribute only little to the total error budget. 
Thoroughly constraining the systematic uncertainties is an extremely
computationally intensive task, requiring hundreds of model runs with modified
parameters, the problem thus being $\sim$2 orders of magnitude more complex
than the abundance determination itself. We therefore {\em assume} the systematic 
abundance uncertainties for \ion{S}{ii/iii}, \ion{Ti}{ii} and \ion{Fe}{ii}
also to amount to $\sim$0.10\,dex (1-$\sigma$-error). 
The atomic data used for assembling these model atoms is of similar quality
(perhaps slightly
lower because the atoms are more complex) than in the lighter elements. 
The response to stellar parameter changes can also be
expected to be similar, as deduced from a few exploratory test calculations. 

We omit a detailed comparison of our non-LTE abundance determination with previous
work from the literature. For \ion{C}{i/ii}, \ion{N}{i/ii}, \ion{O}{i} and
\ion{Mg}{i/ii} this has been done elsewhere (Przybilla et
al.~\cite{Przybillaetal00}, \cite{Przybillaetal01a},b; Przybilla \&
Butler~\cite{PrBu01}). The main conclusions are that in general the results
from the other studies are close to our findings, or origins of the 
differences have been found, e.g. different stellar parameters, see
Sect.~\ref{sectparamsother}. Major discrepancies were found only for
\ion{N}{i}, where the previous studies gave systematically lower nitrogen 
abundances, by a factor $\sim2$. This has been traced to use of
approximate collision data in the previous non-LTE model atoms, but see also
Venn \& Przybilla~(\cite{VePr03}). Contrary to previous studies, our
non-LTE analysis for these elements yields homogeneous results from all 
available spectral lines. Non-LTE studies of \ion{S}{ii/iii},
\ion{Ti}{ii} and \ion{Fe}{ii} in BA-SGs have not been reported in the
literature so far. 

\begin{figure*}
\sidecaption
\includegraphics[width=12cm]{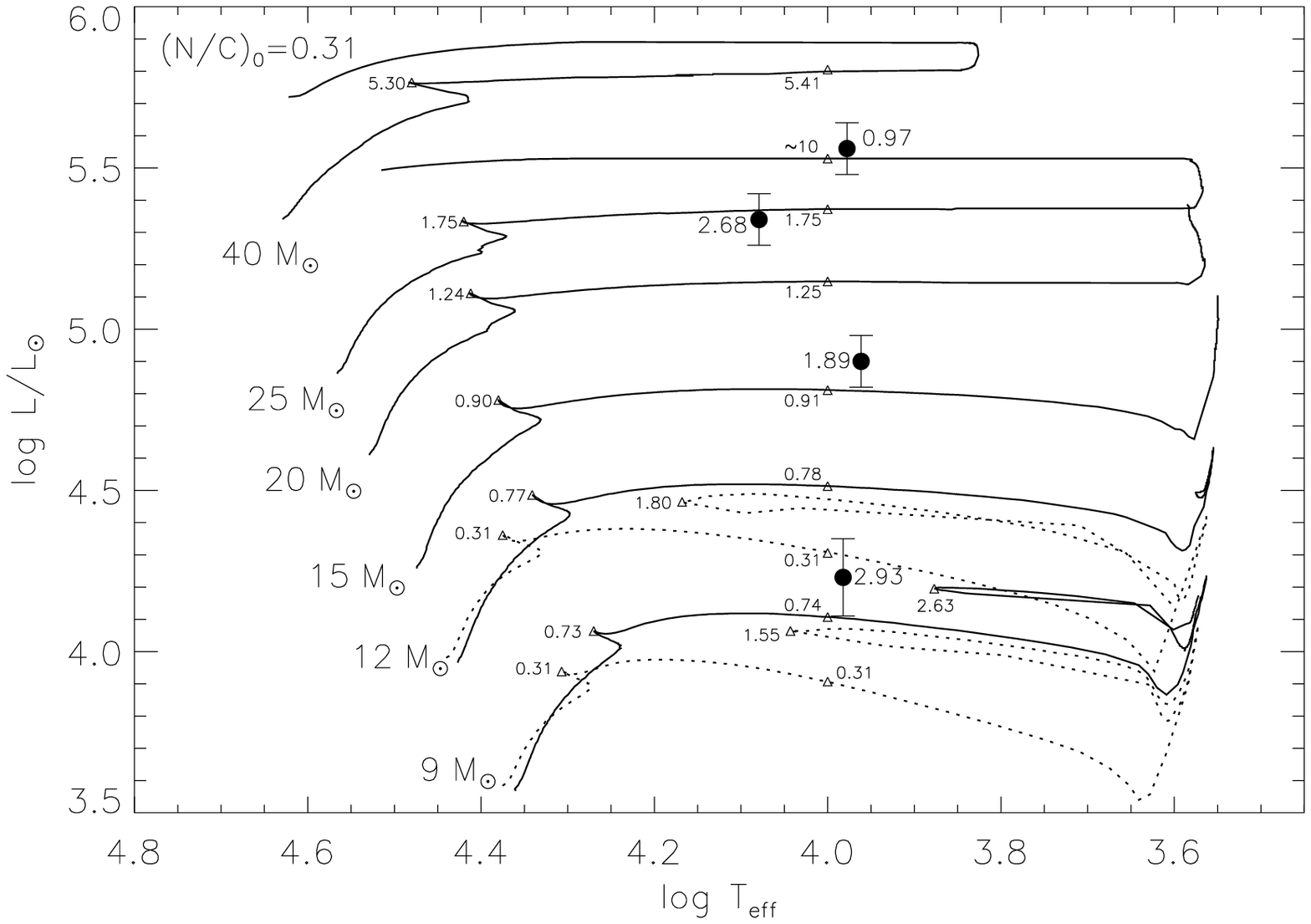}
\caption[]{The evolutionary status of the sample stars and mixing with
nuclear-processed matter. Evolutionary tracks
accounting for mass-loss and rotation from Meynet \& Maeder (\cite{MeMa03})
are displayed, for solar metallicity at an initial rotational velocity
of 300\,km\,s$^{-1}$ (full line) and the non-rotating case (dotted line). 
Numbers along the tracks give the ratio of N/C (by mass, (N/C)$_0$ is the
initial value). For the sample objects (dots) the
observed ratios are given. The three more massive supergiants
have most likely evolved directly from the main sequence, whereas $\eta$\,Leo 
appears to have already undergone the first dredge-up (as inferred from its 
high N/C ratio and also the helium enrichment) and is currently in a
blue-loop phase. Uncertainties in $\log T_{\rm eff}$ are of the order of the
symbol size.}
\label{evolplot}
\end{figure*}

Finally, a comparison of the high-resolution observations of the luminous Galactic
supergiant HD\,92207 with our spectrum synthesis is made in
Fig.~\ref{synthesis}. The model can successfully reproduce
the line spectrum with high accuracy over extended wavelength regions,
except for a few features which are strong enough to be affected by
sphericity effects and the stellar wind. An obvious example for this is
H$\gamma$. Some complications in localised spectral regions occur because of 
interstellar absorption bands, which are not modelled. Finally, CCD
defects and insufficiently corrected cosmic ray hits can render some regions 
useless for the analysis, e.g.~near \ion{Ti}{ii} $\lambda$4501 in the present case. 
In high-resolution data these are easily recognised. However at intermediate 
resolution, in particular at low S/N, special care has to be taken to identify the 
artefacts and to exclude them from the analysis.


\section{Evolutionary status}\label{sectevol}

Massive stars provide the dominant contribution to the momentum and energy
budget of the interstellar matter. They are the main drivers for the dynamical and 
chemical evolution (as sites of nucleosynthesis) of the interstellar environment,
and consequently for the evolution of galaxies. Understanding the evolution of 
massive stars is a prerequisite for such further studies. Massive star evolution 
is an active field of research (see e.g. Maeder \& Meynet~(\cite{MaMe00})
for an overview) where the overall problem has been solved in many
respects but where many details are still not conclusively explained because
of the uncertainty in several free parameters of the input physics.
These have to be constrained by observation. In particular 
{\em quantifying} the r\^ole of stellar winds and rotation as a function of
metallicity requires further efforts, as well as transport phenomena as
manifested in surface abundance anomalies, and the existence and extent of
blue loops. BA-type supergiants are of special interest in this regard, as
they allow these problems to be investigated beyond the Galaxy and its nearest
satellites, in a much wider range of different galactic environments.
In the following we wish to discuss our four sample stars in the context of
their evolutionary status, putting special emphasis on areas where 
our analysis technique allows for an improvement on previous work.

\begin{table}
\setlength{\tabcolsep}{.11cm}
\caption[]{Surface abundance data relevant for stellar evolution\\[-6mm]
\label{tableevol}}
\begin{tabular}{lr@{$\pm$}lr@{$\pm$}lr@{$\pm$}lr@{$\pm$}l}
\hline
Element & 
\multicolumn{2}{c}{$\eta\,$Leo} & 
\multicolumn{2}{c}{HD\,111613} &
\multicolumn{2}{c}{HD\,92207} & 
\multicolumn{2}{c}{$\beta$\,Ori}\\
\hline
$Y$ & 0.37 & 0.04 & 0.32 & 0.04 & 0.35 & 0.04 & 0.38 & 0.04\\
N/C & 2.93 & 1.38 & 1.89 & 0.69 & 0.97 & 0.09: & 2.68 & 0.52\\
N/O & 0.37 & 0.10 & 0.44 & 0.12 & 0.25 & 0.05 & 0.45 & 0.09\\
$[$CNO/H$]$ & $-$0.06 & 0.05 & $-$0.08 & 0.04 &
$-$0.04 & 0.05: & $\pm$0.00 & 0.04\\
$[$M/H$]$ & $-$0.04 & 0.03 & $-$0.11 & 0.03 & $-$0.09 & 0.07 & $-$0.06 & 0.10\\
\hline
\end{tabular}
\end{table}

First, reduced uncertainties in the stellar parameters (see Table~\ref{obj})
allow the objects to be located  precisely in the Hertzsprung-Russell diagram 
(HRD, Fig.~\ref{evolplot}). Here, the major remaining uncertainties arise from
the constraints of an indirect distance determination. Improvements on this
can only be expected from future astrometric satellite missions. The objects cover a
wide range in luminosity within the supergiant regime, and include two of
the most luminous and massive BA-SGs analysed so far. Evolutionary and
spectroscopic masses are in excellent agreement, indicating good
consistency between the stellar properties inferred from the present
analysis and those predicted by theory. In particular the integrated mass-loss
history of these (near-)solar metallicity objects is apparently well described.

Then, additional information on the evolutionary status of the stars comes from
abundance patterns of the light elements. The reaction rates for the
CNO-cycle result in an accumulation of nitrogen at the cost of carbon and (to lesser
extent) oxygen, with
helium being the fusion product. If transport mechanisms are active
throughout the stellar envelope, these fusion products may be mixed into the
stellar atmosphere. The observed abundance patterns are summarised in
Table~\ref{tableevol} (by mass fraction for $Y$, N/C and N/O). The 
most sensitive indicator, the N/C ratio, is also discussed in the HRD diagram in 
Fig.~\ref{evolplot}. All objects show enrichments of helium
and nitrogen, depletion of carbon and practically solar oxygen. The sum of
the CNO abundances $[$CNO/H$]$ correlates well with the stellar
metallicity $[$M/H$]$, consistent with the catalyst r\^ole of these
elements in the main fusion cycle on the main sequence.
The comparison with the predictions of evolution computations  
by Meynet \& Maeder~(\cite{MeMa03}) shows good qualitative
agreement, while discrepancies remain in the details. In particular, helium
enrichment traces the increasing N/C ratio well, while oxygen
depletion obviously does not occur to the predicted extent.

In this scenario the three more luminous supergiants appear to have evolved
directly from the main sequence, at initial rotational velocities larger
than 300\,km\,s$^{-1}$ ($\beta$\,Ori, HD\,111613) and at a lower rate, respectively, 
in the case of HD\,92207. The high helium abundance and N/C ratio (the highest of
the sample) of $\eta$\,Leo on the other hand are indicative for the first dredge-up 
and thus a blue loop scenario for this object. Note that this classification 
can only be made
for the {\em individual} objects with statistical significance because of the
reduced abundance uncertainties.
However, additional efforts are required to further improve the significance of 
the N/C ratios, which are currently still impaired by the
remaining uncertainties in the carbon abundance determination.


\section{Analyses at intermediate resolution}\label{sectmedres}
\begin{figure*}
\resizebox{\hsize}{!}{\includegraphics{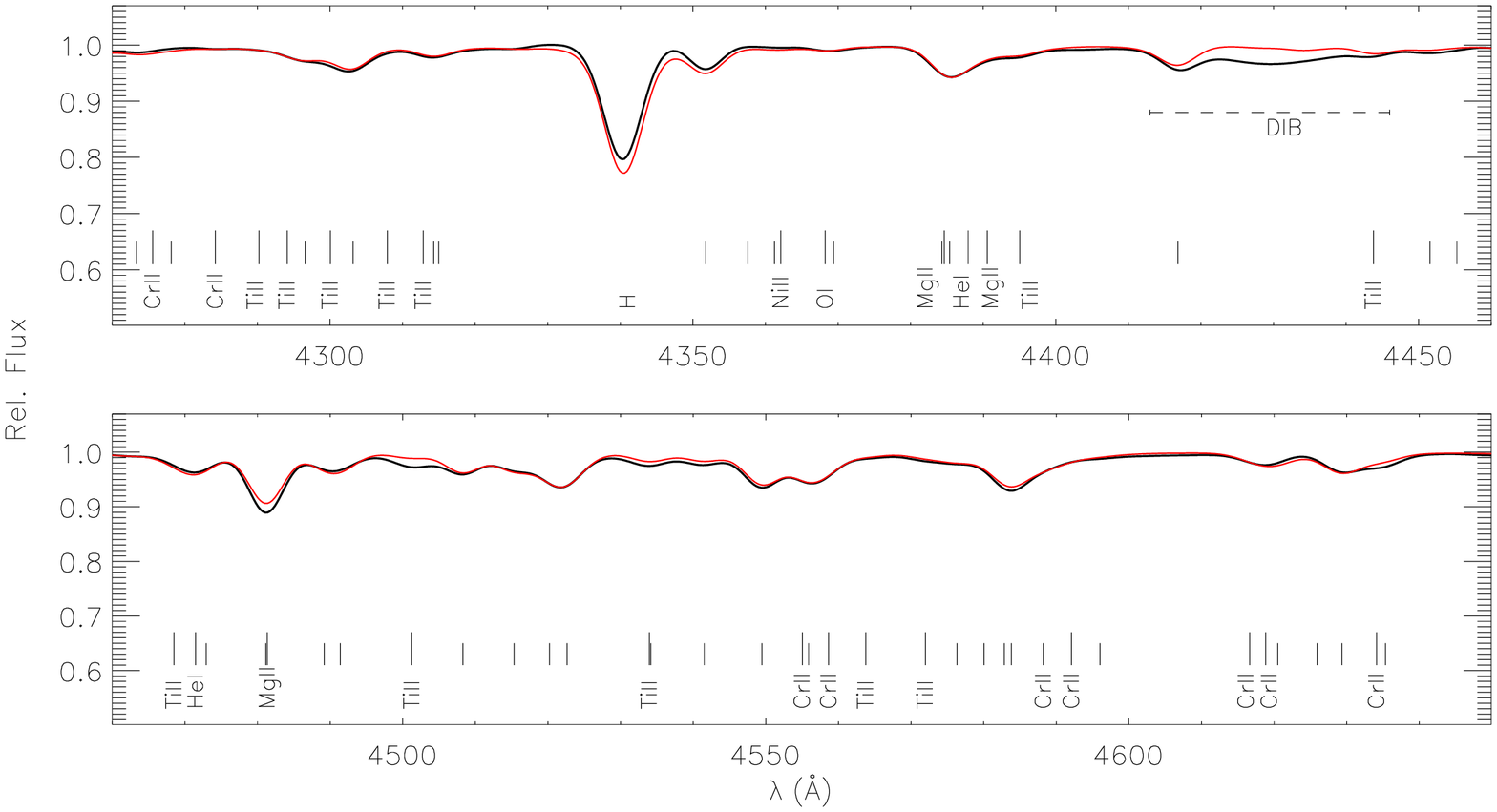}}
\caption[]{Same as Fig.~\ref{synthesis}, however artifically degraded to
5\,{\AA} resolution. The excellent agreement between theory and observation is 
preserved. Deviations are found only for H$\gamma$, \ion{Mg}{ii} $\lambda$4481 
(see text) and in the regions of the interstellar band around 4430\,{\AA} and near
\ion{Ti}{ii} $\lambda$4501 (due to a CCD defect, cf.~also
Fig.~\ref{synthesis}.}
\label{FORSsynthesis}
\end{figure*}
\begin{figure*}
\resizebox{\hsize}{!}{\includegraphics{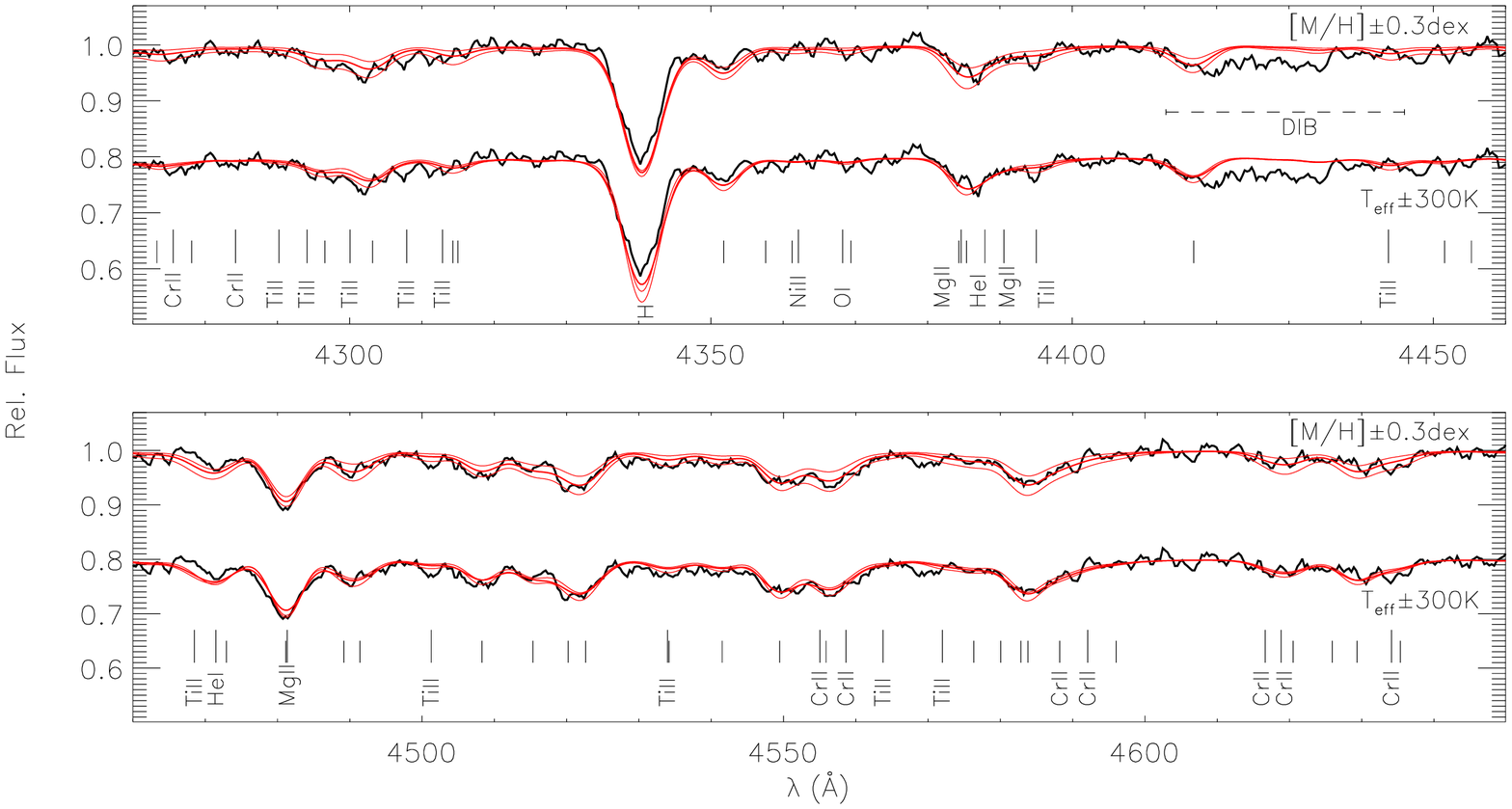}}
\caption[]{Same as Fig.~\ref{FORSsynthesis}, but artifically degraded to a
S/N\,$\simeq$\,50. The comparison with synthetic spectra for scaled abundances and
modified temperatures (thin red lines, as indicated) shows that the models react
sensitively to changes of the metal abundances only. Consequently,
stellar parameters estimated from the spectral classification
($T_{\rm eff}$) and from fitting of the higher Balmer lines ($\log g$)
suffice to constrain the stellar metallicity to $\sim$0.2\,dex.
The lower set of spectra in each panel has been shifted by $-$0.2\,units.}
\label{FORSsynthesisnoise}
\end{figure*}

In order to analyse intermediate-resolution data
($R$\,$\simeq$\,1\,000--5\,000) whole spectral
regions have to be modelled at once, as the individual spectral features are
typically no longer resolved by the instrument. The spectrum synthesis
technique implemented here allows for such analyses, as practically all 
spectral features in BA-SGs of significant strength (more than a few m{\AA}) are
included. In particular the spectral region between the Balmer jump and
$\sim$5000\,{\AA} -- the classical region for analyses of hot stars -- is well
covered.  Thus, an estimate of the overall stellar metallicity and abundances 
of individual elements from several unblended lines of intermediate strength
can be obtained from a comparison of the observed with synthesised spectra. 
The feasibility of this approach is tested in the following. Note that spectral 
regions containing lines of intermediate
strength ($W_{\lambda}$\,$\lesssim$\,300\,{m\AA}) are best suited for an
analysis, as these are strong enough to produce a noticeable signal even at
modest S/N, but are also weak enough to be of photospheric origin. 
In BA-SGs several ionic species give rise to such lines, 
typically \ion{N}{i}, \ion{O}{i}, \ion{Mg}{ii},
\ion{Si}{ii}, \ion{Ti}{ii}, \ion{Cr}{ii}, \ion{Fe}{ii} and in the hotter stars
\ion{He}{i} as well. The three iron group elements hereby dominate the line
spectrum, whereas \ion{He}{i}, \ion{N}{i} and the three $\alpha$--elements
contribute only a few distinct features.  

A test for the applicability of our spectrum synthesis at intermediate
resolution is made in the following, based on the most luminous object of
our sample, HD\,92207. The transition from the high resolution comparison
to intermediate resolution is made in Fig.~\ref{FORSsynthesis}, 
where the observed and the synthetic spectrum from Fig.~\ref{synthesis} 
have been artifically degraded to 5\,{\AA} resolution. Such data can be
expected e.g. from observations with FORS1 (FOcal Reducer and low-dispersion
Spectrograph) at the VLT. The excellent agreement between model and
observation is preserved, except for a few spectral features already
discussed above (Sect.~\ref{sectabus}). However, observational data at 
such a high S/N will typically not be available. In order to consider a 
more realistic case, the observed spectrum is further degraded to a 
S/N\,$\simeq$\,50 in Fig.~\ref{FORSsynthesisnoise}. Additionally, test 
calculations for scaled abundances and modified effective temperature are displayed.
While the surface gravity can still be determined at high accuracy from the
higher Balmer lines, testing for the temperature sensitivity of the synthetic 
spectra becomes necessary as ionization equilibria are unavailable for a direct
$T_{\rm eff}$-determination (typically, one of the ionic species shows only
weak lines). Instead, empirical (metallicity-dependent) spectral type\,--\,$T_{\rm eff}$
calibrations have to be applied (see e.g. Kudritzki et
al.~\cite{Kudritzkietal03}; Evans \& Howarth~\cite{EvHo03}), increasing the
uncertainties, or a spectrophotometric approach, which requires further
investigation (see Sect.~\ref{spectrophotometry}).
The tests show that within a considerable $T_{\rm eff}$-interval the
systematic effects on the overall spectrum synthesis are rather small. 
Modifications of the abundances on the other hand show a much larger effect,
and we conclude that our method allows the stellar metallicity to be
constrained within
approximately $\pm$0.2\,dex (1-$\sigma$ uncertainties). This is considerable larger than the
statistical uncertainties from the high-resolution analysis but sufficient
for meaningful applications in extragalactic stellar astronomy. In fact,
such uncertainties are characteristic for the status of contemporary abundance
work on BA-SGs in the literature. From a few features, like \ion{Fe}{ii}
$\lambda\lambda$4351,\,4508 and the pure \ion{Fe}{ii} blends around 4520 and
4580\,{\AA}, even the iron abundance can be derived in this wavelength
region. In a similar manner, one can use the regions around \ion{He}{i} 
$\lambda\lambda$4026 and 4471 to determine the helium abundance. More
chemical species become accessible in other wavelength regions.

Note that the microturbulent velocity can also not be directly derived from intermediate
resolution spectra, as the whole range from the weak lines to those of 
intermediate strength is not available for the analysis.
However, our high resolution analyses and data from the literature
indicate that microturbulence values of $\xi$\,$=$\,8$\pm$2\,km\,s$^{-1}$
are appropriate for BA-SGs at luminosities $\log L/L_{\sun}$\,$\gtrsim$\,5.0. 
We assume that this empirical finding holds for analogous objects in the
galaxies beyond the Local Group, for which only the next generation of
extremely large telescopes will allow for high-resolution spectroscopy and
direct microturbulence determinations.
Within the typical uncertainties in $\xi$, no marked systematic effects 
for the metallicity determination are found.

To conclude, the present spectrum synthesis technique in combination with
sophisticated non-LTE model atoms allows for reliable metallicity 
determinations for BA-type supergiants on the basis of 
intermediate-resolution spectra. With suitable spectra available,
abundances for several elements of general astrophysical relevance can also be
obtained. This opens up BA-SGs as versatile tools for extragalactic stellar
astronomy. First applications to metallicity analyses of A-type supergiants
well beyond the Local Group, in the field spiral galaxy NGC\,3621 and the Sculptor
group spiral NGC\,300, have already been discussed by 
Przybilla~(\cite{Przybilla02}) and Bresolin et al.~(\cite{Bresolinetal01}, 
\cite{Bresolinetal02}).


\section{Summary and conclusions}\label{sectsum}
We have presented a hybrid non-LTE spectrum synthesis technique for the
quantitative analysis of BA-type supergiants and related objects of lower
luminosity and tested it on
high-quality observations of four bright Galactic objects.
The statistical equilibrium and radiative transfer problem for several of the 
astrophysically most important elements is solved on the basis of classical
line-blanketed LTE atmospheres. These have been shown to be sufficient to 
obtain consistent results from the available spectroscopic
indicators. A prerequisite for this is that secondary atmospheric parameters 
such as helium abundance, metallicity and microturbulence are 
accounted for consistently, because of the high sensitivity of the model atmospheres and
line formation to the details of the computations close to the
Eddington limit. It has been shown that the stellar parameters of BA-SGs can
be determined with unprecedented precision: the internal accuracy of the
methods allows $T_{\rm eff}$ to be constrained to better than 1--2\% and $\log g$
to 0.05--0.10\,dex. Non-LTE computations with sophisticated model atoms 
reduce random errors and remove systematic trends in abundance analyses. 
The 1-$\sigma$ errors in the determination of absolute abundances amount to typically
0.05--0.10\,dex and $\sim$0.10\,dex for statistical and systematic
uncertainties, respectively. This comes close to the accuracy that is
usually achieved in differential studies of solar-type stars in the
contemporary literature. Contrary to common assumption, significant non-LTE
abundance corrections by a factor 2--3 can occur even for the weakest lines 
in BA-SGs, and considerably larger corrections for the stronger lines. 
Because of this, a purely LTE analysis of highly luminous supergiants
fails to obtain meaningful results, while the less-luminous supergiants can 
in principle be addressed in such a way, if reduced accuracy can be
tolerated. LTE analyses thereby tend to systematically overestimate the
abundances of the light and $\alpha$-process elements, and to underestimate
the abundances of the iron group species. The non-LTE analysis implies close to 
solar abundance for the heavier elements in the sample objects, and patterns
characteristic for mixing with nuclear-processed matter in the case of the
light elements. These indicate a blue-loop scenario for one of the sample
stars because of first dredge-up abundance ratios, while the other objects
were restricted to the blue part of the HRD. Using the full potential
of our spectrum synthesis, we have shown that almost the entire visual and
near-IR spectra of BA-SGs can be accurately reproduced, implying the the
technique is also suitable for analyses of intermediate-resolution
spectra. Thus BA-type supergiants have become a diagnostic tool for quantitative 
stellar spectroscopy beyond the Local Group.


\appendix
\section{Spectral line analysis}\label{apa}
In this appendix we provide details on our spectral line analysis, as a 
basis for further applications.
Table~\ref{taba1} summarises our line data and the results from the abundance 
analysis for our sample stars. It is available in electronic form only.
The first columns give the line wavelength $\lambda$ (in
{\AA}), excitation energy of the lower level $\chi$ (in eV), adopted oscillator
strength $\log gf$, an accuracy indicator and the source of the $gf$ value.
Most of the $gf$-data is retrieved from the atomic 
spectra database (V2.0) of the National Institute of Standards and Technology
(http://physics.nist.gov/cgi-bin/AtData/main\_asd).
Then, for each object the measured equivalent width $W_{\lambda}$ (in m{\AA})
is tabulated, followed by the derived abundance
$\log \varepsilon$\,=\,$\log \mathrm{(X/H)}$\,$+$\,12.
In cases with an entry for the non-LTE abundance correction
$\Delta \log \varepsilon$~=~$\log \varepsilon_\mathrm{NLTE}$~$-$~$\log
\varepsilon_\mathrm{LTE}$ this de\-notes the non-LTE abundance, else the
LTE abundance. The equivalent widths have been measured by direct
integration over the spectral lines, deviating from the usual approach
of fitting the observed profiles by a Gaussian. For high-S/N observations of 
BA-SGs as in the present study this helps to avoid systematic
uncertainties, as the line profiles are potentially subject to asymmetries
imposed by the velocity field at the base of the stellar wind. In some cases
metal lines are situated in the wings of the Balmer lines. Then their
equivalent width is measured with respect to the {\em local} continuum,
indicated by `S($W_{\lambda}$)'. Abundances are determined from a best match
of the spectrum synthesis to the observed line {\em profiles}, and not the
equivalent widths, see the discussion in Sect.~\ref{sectabus}. Equivalent widths
are used as auxiliary diagnostics for the determination of microturbulent
velocity (see e.g. Fig.~\ref{nlteabus}). Spectrum synthesis also allows
blended features to be used for abundance determinations, where a $W_{\lambda}$
measurement is hampered. These cases are marked by a sole `S' in
Table~\ref{taba1}. For \ion{He}{i} only non-LTE abundances are derived because
of the potential of this major atmospheric constituent to change atmospheric
structure and thus the stellar parameter determination, see
Sect.~\ref{heliumetc}. Entries in {\em italics} are disregarded when computing
abundance averages (Table~\ref{tableabu}).


\begin{acknowledgements}
We are indebted to the numerous atomic physicists making this research
possible by providing the vast amounts of input data required for non-LTE modelling.
It is our hope that we can
stimulate further efforts by the present work -- they would be most welcome.
We are grateful to A.~Kaufer for his help with obtaining some of the spectra 
at La Silla. 
We thank J.~Puls for numerous stimulating discussions on many
aspects of non-LTE modelling of hot stars and K.A.~Venn for many discussions
on analyses of A-type supergiants in particular.
NP would like to acknowledge the financial support of the  Max-Planck-Gesellschaft,
the BMBF through grant 05AV9WM12, and by the Insitute for Astronomy, Hawaii,
which made this project feasible throughout 
the diploma, the PhD and a postdoctoral phase.
This research has made use of the Simbad database, operated at CDS, Strasbourg, 
France.
\end{acknowledgements}



\begin{table*}
\setlength{\tabcolsep}{.0625cm}
\caption[]{Spectral line analysis\\[-6mm]\label{taba1}}
\begin{tabular}{lrrlrlrrrrrrrrrrrrrrr}
\hline
& & & & & & 
\multicolumn{3}{c}{\raisebox{-0.65ex}{$\eta\,$Leo}} & &
\multicolumn{3}{c}{\raisebox{-0.65ex}{HD\,111613}} & &
\multicolumn{3}{c}{\raisebox{-0.65ex}{HD\,92207}} & & 
\multicolumn{3}{c}{\raisebox{-0.65ex}{$\beta$\,Ori}}\\[.8mm]
\cline{7-9} \cline{11-13} \cline{15-17} \cline{19-21}
\rule[-2mm]{0mm}{1mm}{}\\[-4.mm]
$\lambda\,$(\AA) & $\chi\,(\mbox{eV})$ & $\log gf$ & Acc. & Src. & & 
$W_{\lambda}$(m\AA) & $\log \varepsilon$ &
$\Delta \log \varepsilon$ & & $W_{\lambda}$(m\AA) &
$\log \varepsilon$ & $\Delta \log \varepsilon$ & & $W_{\lambda}$(m\AA) &
$\log \varepsilon$ & $\Delta \log \varepsilon$ & & 
$W_{\lambda}$(m\AA) &
$\log \varepsilon$ & $\Delta \log \varepsilon$\\[.8mm]
\hline
\ion{He}{i}:\\
3819.60 & 20.96 & $-$0.97 & B & WSG & & S & 11.17 & {\ldots} & &
S & 11.07 & {\ldots} & & {\ldots} & {\ldots} & {\ldots} & & 
S & 11.24 & {\ldots}\\
3819.61 & 20.96 & $-$1.19 & B & WSG & & \\
3819.76 & 20.96 & $-$1.67 & B & WSG & & \\
3867.47 & 20.96 & $-$2.06 & B & WSG & & S & 11.22 & {\ldots} & &
S & 11.06 & {\ldots} & & S & 11.15 & {\ldots} & & 
86 & 11.18 & {\ldots}\\
\hline
\end{tabular}
\end{table*}

\begin{table*}
\noindent{\bf Table A1.} (cont.)\\[1mm]
\setlength{\tabcolsep}{.0625cm}
\begin{tabular}{lrrlrlrrrrrrrrrrrrrrr}
\hline
& & & & & & 
\multicolumn{3}{c}{\raisebox{-0.65ex}{$\eta\,$Leo}} & &
\multicolumn{3}{c}{\raisebox{-0.65ex}{HD\,111613}} & &
\multicolumn{3}{c}{\raisebox{-0.65ex}{HD\,92207}} & & 
\multicolumn{3}{c}{\raisebox{-0.65ex}{$\beta$\,Ori}}\\[1mm]
\cline{7-9} \cline{11-13} \cline{15-17} \cline{19-21}
\rule[-2mm]{0mm}{1mm}{}\\[-3.8mm]
$\lambda\,$(\AA) & $\chi\,(\mbox{eV})$ & $\log gf$ & Acc. & Src. & & 
$W_{\lambda}$(m\AA) & $\log \varepsilon$ &
$\Delta \log \varepsilon$ & & $W_{\lambda}$(m\AA) &
$\log \varepsilon$ & $\Delta \log \varepsilon$ & & $W_{\lambda}$(m\AA) &
$\log \varepsilon$ & $\Delta \log \varepsilon$ & & 
$W_{\lambda}$(m\AA) &
$\log \varepsilon$ & $\Delta \log \varepsilon$\\[1mm]
\hline
\ion{N}{i}:\\
8567.74 & 10.68 & $-$0.66 & B & WFD & & S(95) & 8.43 & $-$0.44 & &
S(85) & 8.41 & $-$0.47 & & S(40) & 8.18 & $-$0.52 & &
{\ldots} & {\ldots} & {\ldots}\\
8594.00 & 10.68 & $-$0.33 & B & WFD & & S(84) & 8.36 & {\em $-$0.93} & &
S & 8.32 & {\em $-$0.69} & & {\ldots} & {\ldots} & {\ldots} & &
{\ldots} & {\ldots} & {\ldots}\\
8629.24 & 10.69 & 0.08 & B & WFD & & S(184) & 8.14 & {\em $-$0.86} & &
S(209) & 8.18 & {\em $-$0.69} & & {\ldots} & {\ldots} & {\ldots} & &
{\ldots} & {\ldots} & {\ldots}\\
8655.88 & 10.69 & $-$0.63 & B & WFD & & S(72) & 8.29 & $-$0.51 & &
S(57) & 8.32 & $-$0.46 & & {\ldots} & {\ldots} & {\ldots} & &
{\ldots} & {\ldots} & {\ldots}\\
\hline
\end{tabular}
\end{table*}

\begin{table*}
\noindent{\bf Table A1.} (cont.)\\[1mm]
\setlength{\tabcolsep}{.0625cm}
\begin{tabular}{lrrlrlrrrrrrrrrrrrrrr}
\hline
& & & & & & 
\multicolumn{3}{c}{\raisebox{-0.65ex}{$\eta\,$Leo}} & &
\multicolumn{3}{c}{\raisebox{-0.65ex}{HD\,111613}} & &
\multicolumn{3}{c}{\raisebox{-0.65ex}{HD\,92207}} & & 
\multicolumn{3}{c}{\raisebox{-0.65ex}{$\beta$\,Ori}}\\[1mm]
\cline{7-9} \cline{11-13} \cline{15-17} \cline{19-21}
\rule[-2mm]{0mm}{1mm}{}\\[-3.8mm]
$\lambda\,$(\AA) & $\chi\,(\mbox{eV})$ & $\log gf$ & Acc. & Src. & & 
$W_{\lambda}$(m\AA) & $\log \varepsilon$ &
$\Delta \log \varepsilon$ & & $W_{\lambda}$(m\AA) &
$\log \varepsilon$ & $\Delta \log \varepsilon$ & & $W_{\lambda}$(m\AA) &
$\log \varepsilon$ & $\Delta \log \varepsilon$ & & 
$W_{\lambda}$(m\AA) &
$\log \varepsilon$ & $\Delta \log \varepsilon$\\[1mm]
\hline
\ion{O}{i}:\\
6156.76 & 10.74 & $-0.90$ & B+ & WFD & & \\
6156.78 & 10.74 & $-0.69$ & B+ & WFD & & \\
6158.15 & 10.74 & $-1.84$ & B+ & WFD & & 89 & 8.80 & $-$0.17 & &
89 & 8.66 & $-$0.20\\
6158.17 & 10.74 & $-1.00$ & B+ & WFD & & \\
\hline
\end{tabular}
\end{table*}

\begin{table*}
\noindent{\bf Table A1.} (cont.)\\[1mm]
\setlength{\tabcolsep}{.0625cm}
\begin{tabular}{lrrlrlrrrrrrrrrrrrrrr}
\hline
& & & & & & 
\multicolumn{3}{c}{\raisebox{-0.65ex}{$\eta\,$Leo}} & &
\multicolumn{3}{c}{\raisebox{-0.65ex}{HD\,111613}} & &
\multicolumn{3}{c}{\raisebox{-0.65ex}{HD\,92207}} & & 
\multicolumn{3}{c}{\raisebox{-0.65ex}{$\beta$\,Ori}}\\[1mm]
\cline{7-9} \cline{11-13} \cline{15-17} \cline{19-21}
\rule[-2mm]{0mm}{1mm}{}\\[-3.8mm]
$\lambda\,$(\AA) & $\chi\,(\mbox{eV})$ & $\log gf$ & Acc. & Src. & & 
$W_{\lambda}$(m\AA) & $\log \varepsilon$ &
$\Delta \log \varepsilon$ & & $W_{\lambda}$(m\AA) &
$\log \varepsilon$ & $\Delta \log \varepsilon$ & & $W_{\lambda}$(m\AA) &
$\log \varepsilon$ & $\Delta \log \varepsilon$ & & 
$W_{\lambda}$(m\AA) &
$\log \varepsilon$ & $\Delta \log \varepsilon$\\[1mm]
\hline
\ion{Al}{i}:\\
3944.01 & 0.00 & $-$0.64 & C+ & WSM & & 34 & 6.15 & {\ldots} & &
35 & 6.10 & {\ldots} & & {\ldots} & {\ldots} & {\ldots} & & 
{\ldots} & {\ldots} & {\ldots}\\
3961.52 & 0.01 & $-$0.34 & C+ & WSM & & S(43) & 6.06 & {\ldots} & &
S & 6.01 & {\ldots} & & {\ldots} & {\ldots} & {\ldots} & & 
{\ldots} & {\ldots} & {\ldots}\\[1.5mm]
\ion{Al}{ii}:\\
4663.05 & 10.60 & $-$0.29 & C & WSM & & 54 & 6.41 & {\ldots} & &
S & 6.26 & {\ldots} & & 57 & 6.16 & {\ldots} & & 38 & 6.06 & {\ldots}\\
\hline
\end{tabular}
\end{table*}

\begin{table*}
\noindent{\bf Table A1.} (cont.)\\[1mm]
\setlength{\tabcolsep}{.0625cm}
\begin{tabular}{lrrlrlrrrrrrrrrrrrrrr}
\hline
& & & & & & 
\multicolumn{3}{c}{\raisebox{-0.65ex}{$\eta\,$Leo}} & &
\multicolumn{3}{c}{\raisebox{-0.65ex}{HD\,111613}} & &
\multicolumn{3}{c}{\raisebox{-0.65ex}{HD\,92207}} & & 
\multicolumn{3}{c}{\raisebox{-0.65ex}{$\beta$\,Ori}}\\[1mm]
\cline{7-9} \cline{11-13} \cline{15-17} \cline{19-21}
\rule[-2mm]{0mm}{1mm}{}\\[-3.8mm]
$\lambda\,$(\AA) & $\chi\,(\mbox{eV})$ & $\log gf$ & Acc. & Src. & & 
$W_{\lambda}$(m\AA) & $\log \varepsilon$ &
$\Delta \log \varepsilon$ & & $W_{\lambda}$(m\AA) &
$\log \varepsilon$ & $\Delta \log \varepsilon$ & & $W_{\lambda}$(m\AA) &
$\log \varepsilon$ & $\Delta \log \varepsilon$ & & 
$W_{\lambda}$(m\AA) &
$\log \varepsilon$ & $\Delta \log \varepsilon$\\[1mm]
\hline
S\,{\sc ii}:\\
5320.72 & 15.07 & 0.50 & D & WSM & & 10 & 7.12 & $-$0.12 & &
13 & 7.09 & $-$0.12 & & 19 & 7.00 & $-$0.42 & & 57 & 6.89 & {\em $-$0.41}\\
5345.71 & 15.07 & 0.36 & D & WSM & & {\ldots} & {\ldots} & {\ldots} & &
{\ldots} & {\ldots} & {\ldots} & & {\ldots} & {\ldots} & {\ldots} & &
48 & 6.93 & $-$0.32\\
5428.66 & 13.58 & $-$0.13 & D & WSM & & 13 & 7.22 & $-$0.08 & &
14 & 7.13 & $-$0.16 & & {\ldots} & {\ldots} & {\ldots} & & 62 & 7.11 & $-$0.36\\
5432.80 & 13.62 & 0.26 & D & WSM & & {\ldots} & {\ldots} & {\ldots} & &
{\ldots} & {\ldots} & {\ldots} & & {\ldots} & {\ldots} & {\ldots} & &
108 & 7.18 & {\em $-$0.51}\\
\hline
\end{tabular}
\end{table*}

\clearpage

\begin{table*}
\noindent{\bf Table A1.} (cont.)\\[1mm]
\setlength{\tabcolsep}{.0625cm}
\begin{tabular}{lrrlrlrrrrrrrrrrrrrrr}
\hline
& & & & & & 
\multicolumn{3}{c}{\raisebox{-0.65ex}{$\eta\,$Leo}} & &
\multicolumn{3}{c}{\raisebox{-0.65ex}{HD\,111613}} & &
\multicolumn{3}{c}{\raisebox{-0.65ex}{HD\,92207}} & & 
\multicolumn{3}{c}{\raisebox{-0.65ex}{$\beta$\,Ori}}\\[1mm]
\cline{7-9} \cline{11-13} \cline{15-17} \cline{19-21}
\rule[-2mm]{0mm}{1mm}{}\\[-3.8mm]
$\lambda\,$(\AA) & $\chi\,(\mbox{eV})$ & $\log gf$ & Acc. & Src. & & 
$W_{\lambda}$(m\AA) & $\log \varepsilon$ &
$\Delta \log \varepsilon$ & & $W_{\lambda}$(m\AA) &
$\log \varepsilon$ & $\Delta \log \varepsilon$ & & $W_{\lambda}$(m\AA) &
$\log \varepsilon$ & $\Delta \log \varepsilon$ & & 
$W_{\lambda}$(m\AA) &
$\log \varepsilon$ & $\Delta \log \varepsilon$\\[1mm]
\hline
Ti\,{\sc ii}:\\
4911.18 & 3.12 & $-$0.34 & D & MFW & & 21 & 4.60 & +0.20 & &
34 & 4.64 & +0.15 & & {\ldots} & {\ldots} & {\ldots} & &
{\ldots} & {\ldots} & {\ldots}\\
5129.16 & 1.89 & $-$1.39 & D$-$ & MFW & & 20 & 4.96 & +0.33 & &
31 & 4.95 & +0.22 & & {\ldots} & {\ldots} & {\ldots} & &
{\ldots} & {\ldots} & {\ldots} \\
5188.68 & 1.58 & $-$1.21 & D$-$ & MFW & & 32 & 4.91 & +0.30 & &
50 & 4.93 & +0.24& & 12 & 4.75 & +0.48 & & {\ldots} & {\ldots} & {\ldots}\\
5336.78 & 1.58 & $-$1.70 & D$-$ & MFW & & 15 & 4.90 & +0.29 & &
S & 4.93 & +0.26 & & {\ldots} & {\ldots} & {\ldots} & &
{\ldots} & {\ldots} & {\ldots}\\
\hline
\end{tabular}
\end{table*}

\begin{table*}
\noindent{\bf Table A1.} (cont.)\\[1mm]
\setlength{\tabcolsep}{.0625cm}
\begin{tabular}{lrrlrlrrrrrrrrrrrrrrr}
\hline
& & & & & & 
\multicolumn{3}{c}{\raisebox{-0.65ex}{$\eta\,$Leo}} & &
\multicolumn{3}{c}{\raisebox{-0.65ex}{HD\,111613}} & &
\multicolumn{3}{c}{\raisebox{-0.65ex}{HD\,92207}} & & 
\multicolumn{3}{c}{\raisebox{-0.65ex}{$\beta$\,Ori}}\\[1mm]
\cline{7-9} \cline{11-13} \cline{15-17} \cline{19-21}
\rule[-2mm]{0mm}{1mm}{}\\[-3.8mm]
$\lambda\,$(\AA) & $\chi\,(\mbox{eV})$ & $\log gf$ & Acc. & Src. & & 
$W_{\lambda}$(m\AA) & $\log \varepsilon$ &
$\Delta \log \varepsilon$ & & $W_{\lambda}$(m\AA) &
$\log \varepsilon$ & $\Delta \log \varepsilon$ & & $W_{\lambda}$(m\AA) &
$\log \varepsilon$ & $\Delta \log \varepsilon$ & & 
$W_{\lambda}$(m\AA) &
$\log \varepsilon$ & $\Delta \log \varepsilon$\\[1mm]
\hline
Fe\,{\sc i}:\\
3815.84 & 1.48 & 0.30 & C+ & FMW & & S(48) & 7.19 & {\ldots} & &
73 & 7.24 & {\ldots} & & {\ldots} & {\ldots} & {\ldots} & &
{\ldots} & {\ldots} & {\ldots}\\
3859.91 & 0.00 & $-$0.71 & B+ & FMW & & 51 & 7.26 & {\ldots} & &
80 & 7.28 & {\ldots} & & {\ldots} & {\ldots} & {\ldots} & &
{\ldots} & {\ldots} & {\ldots}\\
3922.91 & 0.05 & $-$1.65 & B+ & FMW & & 7 & 7.35 & {\ldots} & &
13 & 7.44 & {\ldots} & & {\ldots} & {\ldots} & {\ldots} & &
{\ldots} & {\ldots} & {\ldots}\\
3927.92 & 0.11 & $-$1.59 & C & FMW & & 14 & 7.28 & {\ldots} & &
16 & 7.45 & {\ldots} & & {\ldots} & {\ldots} & {\ldots} & &
{\ldots} & {\ldots} & {\ldots}\\
\hline
\end{tabular}
\end{table*}

\begin{table*}
\noindent{\bf Table A1.} (cont.)\\[1mm]
\setlength{\tabcolsep}{.0625cm}
\begin{tabular}{lrrlrlrrrrrrrrrrrrrrr}
\hline
& & & & & & 
\multicolumn{3}{c}{\raisebox{-0.65ex}{$\eta\,$Leo}} & &
\multicolumn{3}{c}{\raisebox{-0.65ex}{HD\,111613}} & &
\multicolumn{3}{c}{\raisebox{-0.65ex}{HD\,92207}} & & 
\multicolumn{3}{c}{\raisebox{-0.65ex}{$\beta$\,Ori}}\\[1mm]
\cline{7-9} \cline{11-13} \cline{15-17} \cline{19-21}
\rule[-2mm]{0mm}{1mm}{}\\[-3.8mm]
$\lambda\,$(\AA) & $\chi\,(\mbox{eV})$ & $\log gf$ & Acc. & Src. & & 
$W_{\lambda}$(m\AA) & $\log \varepsilon$ &
$\Delta \log \varepsilon$ & & $W_{\lambda}$(m\AA) &
$\log \varepsilon$ & $\Delta \log \varepsilon$ & & $W_{\lambda}$(m\AA) &
$\log \varepsilon$ & $\Delta \log \varepsilon$ & & 
$W_{\lambda}$(m\AA) &
$\log \varepsilon$ & $\Delta \log \varepsilon$\\[1mm]
\hline
Fe\,{\sc ii}:\\
5362.87 & 3.20 & $-$2.74 & X & KB & & {\ldots} & {\ldots} & {\ldots} & &
{\ldots} & {\ldots} & {\ldots} & & 114 & 7.35 & +0.19 & &
{\ldots} & {\ldots} & {\ldots}\\
5425.25 & 3.20 & $-$3.36 & D & FMW & & 60 & 7.58 & +0.03 & &
87 & 7.45 & +0.05 & & 38 & 7.34 & +0.22 & &
{\ldots} & {\ldots} & {\ldots}\\
5427.83 & 6.72 & $-$1.66 & X & KB & & 29 & 7.51 & $-$0.01 & &
{\ldots} & {\ldots} & {\ldots} & & {\ldots} & {\ldots} & {\ldots} & &
{\ldots} & {\ldots} & {\ldots}\\
5534.83 & 3.24 & $-$2.93 & D & FMW & & 106 & 7.56 & $-$0.01 & &
159 & 7.52 & {\em +0.02} & & 79 & 7.37 & +0.22 & & 15 & 7.45 & +0.38\\
\hline
\end{tabular}\\
accuracy indicators -- uncertainties within: AA: 1\%; A: 3\%; B: 10\%; C:
25\%; D: 50\%; E: larger than 50\%; X: unknown\\[1mm] 
sources of $gf$-values -- BMZ: Butler et al.~(\cite{Butleretal93}); CA:
Coulomb approximation (Bates \& Damgaard~\cite{BaDa49});
D: Davidson et al.~(\cite{Davidsonetal92});
F:~Fernley et al.~(available from {\sc Topbase});
FMW: Fuhr et al.~(\cite{Fuhretal88});
KB: Kurucz \& Bell~(\cite{KuBe95});
MEL: Mendoza et al.~(available from {\sc Topbase});
MFW:~Martin et al.~(\cite{Martinetal88});
S: Sigut~(\cite{Sigut99});
T: Taylor~(available from {\sc Topbase});
WFD: Wiese et al.~(\cite{Wieseetal96});
WSG: Wiese et al.~(\cite{Wieseetal66})$^*$;
WSM:~Wiese et al.~(\cite{Wieseetal69})$^*$;
when available$^{(*)}$, improved $gf$-values from Fuhr \& Wiese~(\cite{FuWi98}) 
are favoured\\[1mm]
sources for Stark broadening parameters -- 
\ion{H}{i}: Stehl\'e \& Hutcheon~(\cite{SH99}), Vidal et al.~(\cite{VCS73});
\ion{He}{i}: Barnard et al.~(\cite{Barnardetal69}, \cite{Barnardetal74}), 
Dimitrijevi\'c \& Sahal-Br\'echot~(\cite{DiSa90}); 
\ion{C}{i}: Griem~(\cite{Griem74}),~Cowley (\cite{Cowley71});
\ion{C}{ii}: Griem~(\cite{Griem64}, \cite{Griem74}), Cowley~(\cite{Cowley71});
\ion{N}{i/ii}: Griem~(\cite{Griem64}, \cite{Griem74}), Cowley~(\cite{Cowley71});
\ion{O}{i/ii}: Cowley~(\cite{Cowley71});
\ion{Ne}{i}: Griem~(\cite{Griem74}), Cowley~(\cite{Cowley71});
\ion{Mg}{i}: Dimitrijevi\'c \& Sahal-Br\'echot~(\cite{DiSa96}), Cowley~(\cite{Cowley71});
\ion{Mg}{ii}: Griem~(\cite{Griem64}, \cite{Griem74}), Cowley~(\cite{Cowley71});
\ion{Al}{i}: Griem~(\cite{Griem74}), Cowley~(\cite{Cowley71});
\ion{Al}{ii}: Griem~(\cite{Griem64}, \cite{Griem74}), Cowley~(\cite{Cowley71});
\ion{Al}{iii}: Cowley~(\cite{Cowley71});
\ion{Si}{ii}: Lanz et al.~(\cite{Lanzetal88}), Griem~(\cite{Griem74}), Cowley~(\cite{Cowley71});
\ion{Si}{iii}: Cowley~(\cite{Cowley71});
\ion{P}{ii}: Cowley~(\cite{Cowley71});
\ion{S}{ii/iii}: Cowley~(\cite{Cowley71});
\ion{Ca}{ii}: Griem~(\cite{Griem74}), Cowley~(\cite{Cowley71});
Sc -- Ni: Cowley~(\cite{Cowley71});
\ion{Sr}{ii}: Cowley~(\cite{Cowley71});
\ion{Ba}{ii}: Dimitrijevi\'c \& Sahal-Br\'echot~(\cite{DiSa97})

\end{table*}

\end{document}